\documentclass{amsart}

\usepackage{t1enc}
\usepackage[utf8]{inputenc}

\usepackage{latexsym,pifont}
\usepackage{epsfig}
\usepackage{rotating}

\usepackage{ifpdf}
\ifpdf
\usepackage{epstopdf}
\usepackage{hyperref}
\else
\usepackage[hypertex]{hyperref}
\fi

\usepackage{amsfonts,amsma th,amssymb,mathrsfs,extarrows,MnSymbol}
\usepackage[all]{xy}

\usepackage{tikz}
\usetikzlibrary{matrix,arrows}

\usepackage[makeroom]{cancel}

\usepackage[toc,page]{appendix}
\newcommand{\stoptocwriting}{%
  \addtocontents{toc}{\protect\setcounter{tocdepth}{-5}}}
\newcommand{\resumetocwriting}{%
  \addtocontents{toc}{\protect\setcounter{tocdepth}{\arabic{tocdepth}}}}

\newcommand{\alxydim}[2]{\begin{aligned}\xymatrix#1{#2}\end{aligned}}

\newcommand{\brem}{\begin{Rem}}
\newcommand{\erem}{\end{Rem}\medskip}
\newcommand{\beg}{\begin{Eg}}
\newcommand{\eeg}{\end{Eg}}
\newcommand{\bedef}{\begin{Def}}
\newcommand{\exdef}{\begin{flushright}$\diamond$\end{flushright}
\end{Def}\vskip0.1cm}
\newcommand{\berop}{\begin{Prop}}
\newcommand{\eerop}{\end{Prop}}
\newcommand{\belem}{\begin{Lem}}
\newcommand{\elem}{\end{Lem}}
\newcommand{\bethe}{\begin{Thm}}
\newcommand{\ethe}{\end{Thm}}
\newcommand{\becor}{\begin{Cor}}
\newcommand{\ecor}{\end{Cor}}
\newcommand{\beroof}{\noindent\begin{proof}}
\newcommand{\eroof}{\end{proof}}
\newcommand{\becon}{\begin{Conv}}
\newcommand{\econ}{\begin{flushright}$\checkmark$\end{flushright}\end{Conv}}
\newcommand{\befact}{\begin{Fact}}
\newcommand{\efact}{\begin{flushright}$\checkmark$\end{flushright}\end{Fact}}
\newcommand{\bequest}{\begin{Quest}}
\newcommand{\equest}{\end{Quest}}
\newcommand{\brob}{\begin{Prob}}
\newcommand{\erob}{\end{Prob}}
\newcommand{\becj}{\begin{conj}}
\newcommand{\ecj}{\begin{flushright}$\boxtimes$\end{flushright}\end{conj}}

\newcommand{\barr}{\begin{array}}
\newcommand{\earr}{\end{array}}
\newcommand{\ben}{\begin{enumerate}}
\newcommand{\een}{\end{enumerate}}
\newcommand{\bit}{\begin{itemize}}
\newcommand{\eit}{\end{itemize}}

\newcommand{\qq}{\begin{eqnarray}}
\newcommand{\qqq}{\end{eqnarray}}

\newcommand{\nn}{\nonumber}

\newcommand{\ovl}[1]{\overline{#1}}
\newcommand{\unl}[1]{\underline{#1}}

\newcommand{\Reqref}[1]{Eq.\,\eqref{#1}}

\newcommand\void[1]{}

\newcommand{\tx}[1]{\textrm{#1}} 
\newcommand{\ciut}[1]{\tiny$#1$}

\newcommand{\gt}[1]{\mathfrak{#1}}

\def\cA{\mathcal{A}}
\def\cB{\mathcal{B}}

\def\cD{\mathcal{D}}
\def\cE{\mathcal{E}}
\def\cF{\mathcal{F}}
\def\cG{\mathcal{G}}
\def\ceH{\mathcal{H}}
\def\cI{\mathcal{I}}
\def\cJ{\mathcal{J}}
\def\cK{\mathcal{K}}
\def\ceL{\mathcal{L}}
\def\cM{\mathcal{M}}
\def\cN{\mathcal{N}}
\def\cO{\mathcal{O}}

\def\cS{\mathcal{S}}
\def\cT{\mathcal{T}}

\def\cW{\mathcal{W}}

\def\cZ{\mathcal{Z}}

\def\xcC{\mathscr{C}}

\def\xcL{\mathscr{L}}

\def\xcX{\mathscr{X}}
\def\xcY{\mathscr{Y}}
\def\xcZ{\mathscr{Z}}

\def\t{\mathbf{t}}

\def\bC{{\mathbb{C}}}
\def\bD{{\mathbb{D}}}

\def\bH{{\mathbb{H}}}
\def\bK{{\mathbb{K}}}

\def\bN{{\mathbb{N}}}

\def\bR{{\mathbb{R}}}
\def\bS{{\mathbb{S}}}

\def\bZ{{\mathbb{Z}}}

\def\a{\alpha}
\def\b{\beta}
\def\g{\gamma}
\def\G{\Gamma}
\def\d{\delta}
\def\D{\Delta}
\def\ep{\epsilon}
\def\vep{\varepsilon}

\def\Th{\Theta}

\def\kap{\kappa}
\def\la{\lambda}
\def\La{\Lambda}
\def\om{\omega}
\def\Om{\Omega}

\def\si{\sigma}
\def\Si{\Sigma}

\def\t{\tau}

\def\z{\zeta}

\def\agt{\gt{a}}

\def\ggt{\gt{g}}

\def\hgt{\gt{h}}

\def\lgt{\gt{l}}

\def\tgt{\gt{t}}

\def\zgt{\gt{z}}

\newcommand{\sfd}{{\mathsf d}}

\newcommand{\sfE}{{\mathsf E}}

\newcommand{\sfi}{{\mathsf i}}

\newcommand{\sfL}{{\mathsf L}}

\newcommand{\sfN}{{\mathsf N}}

\newcommand{\sfp}{{\mathsf p}}
\newcommand{\sfP}{{\mathsf P}}

\newcommand{\sfs}{{\mathsf s}}
\newcommand{\sft}{{\mathsf t}}
\newcommand{\sfT}{{\mathsf T}}

\newcommand{\sfY}{{\mathsf Y}}

\newcommand{\txa}{{\rm a}}
\newcommand{\txA}{{\rm A}}

\newcommand{\txB}{{\rm B}}

\newcommand{\ee}{{\rm e}}
\newcommand{\txE}{{\rm E}}

\newcommand{\txg}{{\rm g}}
\newcommand{\txG}{{\rm G}}

\newcommand{\txH}{{\rm H}}

\newcommand{\Lx}{{\rm L}}
\newcommand{\txm}{{\rm m}}

\newcommand{\txT}{{\rm T}}

\def\vH{\check{H}}

\def\exp{{\rm exp}}
\def\id{{\rm id}}
\newcommand{\pr}{{\rm pr}}

\def\too{\longrightarrow}
\def\ev{{\rm ev}}

\def\obj{{\rm Ob}}

\def\Hom{{\rm Hom}}

\def\1morf{1{\rm -Mor}}
\def\2morf{2{\rm -Mor}}
\def\dim{{\rm dim}}
\def\im{{\rm im}}
\def\ker{{\rm ker}}

\def\End{{\rm End}}

\newcommand{\Man}{{\rm {\bf Man}}}

\newcommand{\sMan}{{\rm {\bf sMan}}}

\newcommand{\sLieGrp}{{\rm {\bf sLieGrp}}}

\def\Vol{{\rm Vol}}

\newcommand{\pLie}[1]{\,{-\hspace{-8pt}\xcL}_{#1}}
\def\p{\partial}

\def\con{\lrcorner}

\def\emb{\hookrightarrow}

\def\bd1{{\boldsymbol{1}}}
\def\brd0{{\boldsymbol{0}}}

\def\det{{\rm det}}

\def\diag{\textrm{diag}}

\def\ad{{\rm ad}}
\def\Ad{{\rm Ad}}

\def\Cliff{{\rm Cliff}}

\newcommand{\uj}{{\rm U}(1)}

\def\x{\times}
\def\ox{\otimes}

\def\lx{{\hspace{-0.04cm}\ltimes\hspace{-0.05cm}}}
\def\rx{\rtimes}

\def\lact{\vartriangleright}

\def\rstr{\mathord{\restriction}}

\newcommand{\corr}[1]{\left\langle #1 \right\rangle}

\newtheorem{Thm}{Theorem}
\newtheorem{Prop}[Thm]{Proposition}
\newtheorem{Lem}[Thm]{Lemma}
\newtheorem{conj}{Conjecture}
\newtheorem{Cor}[Thm]{Corollary}
\theoremstyle{definition}
\newtheorem{Rem}[Thm]{Remark}
\newtheorem{Def}[Thm]{Definition}
\newtheorem{Eg}[Thm]{Example}
\newtheorem{Conv}[Thm]{Convention}
\newtheorem{Fact}[Thm]{Fact}
\newtheorem{Quest}[Thm]{Question}
\newtheorem{Prob}[Thm]{Problem}

\numberwithin{equation}{section} 

\newcount\hour\newcount\minute
        \hour=\time \divide\hour by60 \minute=\time
        {\multiply\hour by60 \global\advance\minute by-\hour}
        \edef\militarytime{\number\hour:\ifnum\minute<10 0\fi\number\minute}

\begin{document}

\title{Equivariant Cartan--Eilenberg supergerbes,\\ the Kosteleck\'y--Rabin defect\\ and descent to the Rabin--Crane superorbifold}

\author{Rafa\l ~R.\ ~Suszek}
\address{R.R.S.:\ Katedra Metod Matematycznych Fizyki,\ Wydzia\l ~Fizyki
Uniwersytetu Warszawskiego,\ ul.\ Pasteura 5,\ PL-02-093 Warszawa,
Poland} \email{suszek@fuw.edu.pl}

\begin{abstract}
A concrete geometrisation scheme is proposed for the Green--Schwarz cocycles in the supersymmetric de Rham cohomology of the super-minkowskian spacetime,\ which determine standard super-$\sigma$-model dynamics of super-$p$-branes.\ The scheme yields higher-supergeometric structures with supersymmetry akin to those known from the un-graded setting -- distinguished (Murray-type) $p$-gerbe objects in the category of Lie supergroups.\ These are shown to carry a canonical equivariant structure for the action of the discrete Kosteleck\'y--Rabin subgroup of the target supersymmetry group on the target,\ and thus to resolve the `topology' of the corresponding Rabin--Crane soul superorbifold of the super-minkowskian spacetime.\ The equivariant structure is seen to effectively define a novel $\si$-model of super-$p$-brane dynamics in the super-orbifold through the construction of the corresponding (gauge-)symmetry defect.
\end{abstract}


\maketitle

\tableofcontents

\section{Introduction}

The enduring relevance,\ in physics and mathematics alike,\ of $\si$-models with topological Wess--Zumino (WZ) terms rests safely on the great variety of contexts in which they appear naturally and constructively:\ From -- at the physical end -- an effective description of nuclear interactions \cite{GellMann:1960np},\ through the critical field theory of collective excitations of the quantum spin chain \cite{Haldane:1983a,Affleck:1985crit},\ to classical string \cite{Curtright:1984dz,Friedan:1985phd,Callan:1985sibf} and superstring \cite{Green:1983wt} theory and beyond,\ and -- at the mathematical end -- from the theory of infinite dimensional Lie algebras \cite{Belavin:1984vu,Witten:1983ar},\ through topological quantum field theory \cite{Witten:1988hf} and the theory of quantum groups \cite{Gawedzki:1990jc},\ to noncommutative \cite{Frohlich:1993es} and higher geometry \cite{Carey:1997xm} and cohomology \cite{Gawedzki:1987ak},\ to name only few .

The naturality and adequacy of the language of gerbe theory \cite{Giraud:1971,Gawedzki:1987ak,Murray:1994db} in a rigorous formulation and canonical description \cite{Gawedzki:1987ak,Suszek:2011hg}, symmetry and duality analysis \cite{Gawedzki:2007uz,Gawedzki:2008um,Gawedzki:2010rn,Suszek:2011hg,Gawedzki:2012fu,Suszek:2012ddg} and constructive geometric quantisation \cite{Gawedzki:1987ak,Gawedzki:1999bq,Gawedzki:2002se,Gawedzki:2003pm,Gawedzki:2004tu,Suszek:2011hg} of field theories from this distinguished class,\ has,\ by now,\ attained the status of a well-documented fact.\ Introduced,\ in the pioneering works of Alvarez \cite{Alvarez:1984es} and Gaw\k{e}dzki \cite{Gawedzki:1987ak},\ in the disguise of the Beilinson--Deligne hypercohomology which built the Cheeger--Simons differential characters \cite{Cheeger:1985} into the lagrangean formulation,\ the language has found ample structural applications in this domain upon its geometrisation by Murray {\it et al.} \cite{Murray:1994db,Murray:1999ew},\ and in particular in a cohomological classification of quantum-mechanically consistent models,\ in a concrete formulation of a universal gauge principle and in the resulting classification of gauge anomalies and of inequivalent gaugings \cite{Gawedzki:2010rn,Suszek:2011,Gawedzki:2012fu,Suszek:2012ddg,Suszek:2013},\ and -- finally -- in a straightforward geometric description of defects and their fusion in the said theories in terms of bicategories of gerbes over stratified target spaces \cite{Fuchs:2007fw,Runkel:2008gr,Suszek:2011hg}.

The $\si$-models in which the ideas and methods of gerbe theory have proven particularly robust and constructive are those with a rich configurational symmetry induced from isometries of the target space,\ to wit, the Wess--Zumino--Witten (WZW) $\si$-models of loop dynamics on compact Lie groups \cite{Witten:1983ar,Gawedzki:1990jc,Gawedzki:1999bq,Gawedzki:2001rm} and their gauged Gaw\k{e}dzki--Kupiainen variants \cite{Goddard:1984vk,Gawedzki:1988hq,Gawedzki:1988nj,Karabali:1988au,Hori:1994nc,Gawedzki:2001ye},\ defining that dynamics on homogeneous spaces.\ Here,\ the key results of the Diracesque programme of a (higher-)geometric study of the two-dimensional lagrangean field theories,\ authored and developed,\ jointly with his collaborators,\ by Gaw\k{e}dzki \cite{Gawedzki:1987ak,Gawedzki:2002se} and stimulated by Murray's findings \cite{Murray:1994db,Murray:1999ew},\ are:\ an explicit equivariant geometric quantisation \cite{Gawedzki:1987ak,Gawedzki:1999bq,Gawedzki:2002se,Gawedzki:2003pm,Gawedzki:2004tu} in the spirit of Segal's functorial quantisation \cite{Segal:2002,Gawedzki:1999bq};\ a systematic reconstruction of $\si$-models for mappings of worldsheets (also unoriented ones) into orbispaces of group actions in terms of gerbes with an (twisted) equivariant structure \cite{Gawedzki:2003pm,Gawedzki:2004tu,Schreiber:2005mi,Gawedzki:2007uz,Gawedzki:2010rn,Gawedzki:2012fu},\ the latter determining the data of the gauge-symmetry defect of \cite{Suszek:2011,Suszek:2012ddg,Suszek:2013} ({\it cp.}\ also \cite{Runkel:2008gr});\ a construction of the maximally symmetric WZW defects \cite{Fuchs:2007fw,Runkel:2009sp} and their in-depth study \cite{Runkel:2009sp,Suszek:2022lpf},\ revealing intricate novel relations to the Moore--Seiberg data of the WZW model \cite{Moore:1988qv} and to the ubiquitous Chern--Simons topological gauge field theory \cite{Witten:1988hf,Alekseev:1993rj};\ an elucidation of the peculiar structure of the emergent spectral noncommutative geometry of the maximally symmetric D-branes on the target Lie group \cite{Recknagel:2006hp}.

The advent of supersymmetry \cite{Miyazawa:1966mfa,Gervais:1971ji,Golfand:1971iw,Volkov:1972jx,Volkov:1973ix,Akulov:1974xz,Wess:1973kz,Wess:1974tw} and supergeometry \cite{Berezin:1975,Kostant:1975,Gawedzki:1977pb,Batchelor:1979a,Rogers:1980,DeWitt:1984,Schwarz:1984,Voronov:1984} marks the beginning of a new chapter in the study of the distinguished field theories under consideration,\ in which the main protagonists are the Green--Schwarz (GS) \emph{super}-$\si$-models \cite{Green:1983wt,Green:1983sg,Achucarro:1987nc,Metsaev:1998it,Arutyunov:2008if,Gomis:2008jt,Fre:2008qc,DAuria:2008vov,Bandos:1997ui,deWit:1998yu,Claus:1998fh,Bergshoeff:1985su,Bergshoeff:1987cm} of inner-$\Hom$ `mappings' \cite{Freed:1999} into a target supermanifold,\ the latter replacing smooth mappings into the target manifold of the original non-graded formulation.\ The underlying physics requires the existence of an additional structure on the supergeometric targets of these models,\ to wit,\ an action of a Lie supergroup \cite{Kostant:1975,Fioresi:2007zz,Carmeli:2011} of supersymmetries,\ playing the r\^ole of the rigid symmetries of the superfield theory.\ Finally, the pragmatic criterion of irreducibility fixes the choice of the target in the form of a homogeneous space of the supersymmetry group.\ With the action `functional' composed of a tensorial (metric) `kinetic' term and an intrinsically cohomological WZ term,\ just as in the non-graded setting,\ the super-$\si$-models beg for an application of the higher-algebraic and -geometric methods \emph{along the lines} of the successful treatment of their non-graded counterparts.\ \emph{\bf This is the immediate (super)geometric context and the ultimate goal,\ motivated amply in the previous paragraphs,\ of the approach advocated in the present work.}\ The state of the art in this field of mathematical and physical research prompts a careful explanation of the last declaration,\ to be followed by its concretisation.\smallskip

The super-$\si$-model on a homogeneous space $\,\txG/\txH\equiv\txT\,$ of the supersymmetry Lie supergroup $\,\txG\,$ comes with an obvious choice of cohomology in which to place the usual analysis of its WZ term:\ The theory is founded on the assumption of a quantum-mechanically consistent realisation of supersymmetry,\ and so the relevant cohomology is the $\txG$-invariant de Rham cohomology of $\,\txT$.\ Whenever the latter supermanifold itself is a Lie supergroup,\ we arrive at the Cartan--Eilenberg cohomology $\,{\rm CaE}^\bullet(\txT)\equiv H_{\rm dR}^\bullet(\txT)^\txT\,$ of $\,\txT$,\ which can be mapped to the Le\"ites super-variant \cite{Leites:1975} of the Chevalley--Eilenberg cohomology $\,{\rm CE}^\bullet(\tgt)\,$ of the tangent Lie superalgebra $\,\tgt\,$ of $\,\txT$,\ and otherwise the tools of the theory of non-linear realisations of (super)symmetry \cite{Schwinger:1967tc,Weinberg:1968de,Coleman:1969sm,Callan:1969sn,Salam:1969rq,Salam:1970qk,Isham:1971dv,Volkov:1972jx,Volkov:1973ix,Ivanov:1978mx,Lindstrom:1979kq,Uematsu:1981rj,Ivanov:1982bpa,Samuel:1982uh,Ferrara:1983fi,Bagger:1983mv} can be employed to `embed' it in the same cohomology of the full supersymmetry Lie superalgebra $\,\ggt={\rm sLie}\txG\,$ with a \emph{reductive} decomposition $\,\ggt=\tgt\oplus{\rm Lie}\txH\,$.\ The problem with this purely algebraic presentation of the $(p+2)$-form backgrounds of the GS super-$\si$-models of super-$p$-branes is twofold:
\bit 
\item The higher-algebraic structure encoded by cohomology groups $\,{\rm CE}^{p+2}(\tgt)\,$ with $\,p>0$,\ relevant to the $(p+1)$-dimensional GS super-$\si$-models,\ does not admit a \emph{simple} interpretation in terms of central extensions of $\,\tgt$,\ \emph{conditionally} amenable to \emph{straightforward} integration to the Lie-(super)group level as is the case for $\,{\rm CE}^2(\tgt)\,$ \cite{Chevalley:1948,Tuynman:1987ij}.\ An interpretation of classes in the higher cohomology groups in terms of Lie-$(p+1)$-(super)algebras based on $\,\tgt\,$ was given by Baez {\it et al.} in \cite{Baez:2004hda6,Baez:2010ye},\ whereupon the Getzler--Henriques scheme of $L_\infty$-integration \cite{Getzler:2009Linf,Henriques:2008Linft} was adapted to the setting of interest \cite{Fiorenza:2010mh,Huerta:2011aa}.\ This led to a \emph{formal} definition of higher geometric objects (the so-called Lie $(p+1)$-supergroups) over $\,\txT\,$ which were then proposed to provide a `realisation' of the GS classes in $\,{\rm CE}^{p+2}(\tgt)\,$ \cite{Fiorenza:2013nha}.\ The formal nature of the definition may,\ at times,\ effectively obstruct investigation of a number of physically fundamental issues ({\it cp.}\ below).
\item The existing formal definition does not resolve,\ in any obvious manner,\ the \emph{topology} of the target supermanifold of the super-$\si$-model,\ as \emph{probed by the embedded worldvolume} of the charged fundamental object (the super-$p$-brane),\ in the same (or similar) manner as an abelian $p$-gerbe gives a resolution of the homology class (of the $\si$-model target) dual to the cohomology class of the curvature of that $p$-gerbe \cite{Gajer:1996}.\ The said target supermanifold is simply \emph{assumed} to be $\,\txT$,\ which is the topologically featureless super-Minkowski space in the simplest case under consideration\footnote{It is altogether unclear what the formal approach tells us about the non-flat supergeometries \emph{of physical relevance},\ such as,\ {\it e.g.},\ the homogeneous superspaces of \cite{Metsaev:1998it,Park:1998un,Zhou:1999sm}.},\ and so there seems to be no topology to speak of.\ Here,\ the word `seems' is crucial.
\eit

\noindent The former aspect could well be igonored,\ were it not for the significance of the following mutually entangled questions that one ought to be able to ask and then answer in any attempt at explaining the \emph{physics} of the super-$\si$-model:
\ben
\item \emph{How does the $\kap$-symmetry of the super-$\si$-model manifest itself in the higher-supergeometric formulation?}\ The gauged supersymmetry of the superfield theory is responsible for restoring equibalance between Gra\ss mann-odd and -even degrees of freedom in the vacuum \cite{deAzcarraga:1982dhu,Siegel:1983hh,Siegel:1983ke,Suszek:2020xcu},\ and so it is central to the \emph{consistency} of the physical model -- it fixes the WZ term,\ and with it,\ the structure of the superfield theory.\ As any gauge symmetry,\ it reflects an intrinsic redundancy of the latter,\ and furnishes a consistent description of the underlying dynamics on the symmetry orbispace of the target.\ The gauging scheme in the presence of higher-geometric structures was worked out and elucidated in \cite{Gawedzki:2010rn,Gawedzki:2012fu,Suszek:2012ddg,Suszek:2013}.\ The problem here is that the symmetry does \emph{not} preserve the metric and topological terms in the GS action functional \emph{separately},\ so a direct application of the standard techniques is precluded,\ but some manifestation is nevertheless to be found if the higher supergeometry is to determine a (pre)quantisation of the superfield theory.
\item \emph{What is the higher-supergeometric representation of supersymmetric defects of the super-$\si$-model compatible with the bulk $\kap$-symmetry?}\ The fundamental r\^ole of these defects in the implementation -- along the lines of \cite{Fuchs:2007fw,Runkel:2008gr,Suszek:2011hg} -- of the generating loop-group symmetries,\ and so also in an in-depth understanding of the the non-graded counterparts of the super-minkowskian super-$\si$-model of the superstring,\ {\it i.e.},\ of the bosonic WZW $\si$-models,\ was demonstrated in \cite{Runkel:2008gr,Runkel:2009sp,Suszek:2022lpf}.\ These results do more than justify the question,\ and furnish a ready-to-use methodology of work towards an answer.
\item \emph{What are the higher-supergeometric mechanisms which lift the asymptotic correspondences between the known curved super-$\si$-model backgrounds and the flat ones?}\ The correspondences are a \emph{constitutive} element ({\it cp.}\ \cite[Sec.\,3]{Metsaev:1998it}) of the Metsaev--Tseytlin construction of the super-$\si$-models for the super-cosets with curved bodies of the general form $\,{\rm AdS}_p\x\bS^q\,$ \cite{Metsaev:1998it,Park:1998un,Zhou:1999sm},\ and as such they promise to shed light on any future higher-geometric attempt at elucidating the celebrated AdS/CFT holography.\ Rooted firmly in the purely Lie-(super)algebraic mechanism of the \.In\"on\"u--Wigner contraction,\ they are known to admit a rather subtle transcription into the corresponding Cartan--Eilnberg cohomology \cite{Hatsuda:2002hz},\ which clearly indicates that an essential refinement of a na\"ive approach to `gerbification' is requisite \cite{Hatsuda:2001pp}.
\een

\noindent It is not even clear how to formulate these questions without \emph{concrete and tractable models} of the higher-supergeometric realisations of the GS classes in the supersymmetric de Rham cohomology of the physically relevant supertargets.\ \emph{\bf These we construct explicitly in the present work} for the super-minkowskian GS $p$-cocycles with $\,p\in\{0,1,2\}\,$ in what can be regarded as a Lie-integrated variant of a method,\ originally devised by de Azc\'arraga {\it et al.} \cite{Chryssomalakos:2000xd} and amenable to straightforward generalisation for $p\geq 3$,\ of a stepwise trivialisation of the GS cocycles on $H^2$-runged ladders of (super)central extensions of superspaces.\ For a given $p$,\ the method essentially yields a \emph{\bf $p$-gerbe object in the category of Lie supergroups},\ which we call the {\bf Cartan--Eilenberg $p$-supergerbe}.\ The latter is manifestly $\txT$-invariant.\ In the case of the supermembrane ({\it i.e.},\ $\,p=2$),\ the super-2-gerbe arises naturally over the fully extended 11$d$ superpoint of Def.\,\ref{def:full-ext-spoint11} rather than over the 11$d$ super-Minkowski space.\ The construction of the Cartan--Eilenberg $p$-supergerbes paves the way for an equally concrete and tractable treatment of questions (1)--(3),\ which shall be reported in upcoming publications \cite{Suszek:2018bvx,Suszek:2018ugf,Suszek:2019cum,Suszek:2020xcu,Suszek:2021hjh,Suszek:2022lpf}.\ Prior to discussing its details,\ though,\ let us address the second aspect,\ revealing its deeper meaning along the way.\smallskip

In its most basic form,\ a $p$-gerbe -- a principal $\bC^\x$-bundle ($p=0$),\ a bundle gerbe ($p=1$),\ a 2-gerbe {\it etc.} -- is a geometric structure surjectively submersed onto the support $\,X\,$ of an integral de Rham $(p+2)$-cocycle $\,\chi\,$ in which the class of homology cycles in $\,X\,$ dual to $\,[\chi]\,$ is resolved.\ The associated Cheeger--Simons differential character determines,\ \`a la Dirac \cite{Dirac:1933pi},\ quantum-mechanical amplitudes for the elementary dynamical object -- a material point,\ a tense loop,\ a membrane {\it etc.} -- charged under $\,\chi$,\ modelling,\ {\it i.a.},\ Aharonov--Bohm-type interferences between Feynman histories of the physical object which wrap noncontractible cycles in $\,X\,$ in an interplay between its dynamics and intrinsic topology.\ Thus,\ a $p$-gerbe encodes information on the topology of $\,X\,$ \emph{and} translates that information into the lagrangean dynamics of the $p$-loop.\ The algebraic nature of the (CE) cohomology relevant to our considerations obscures or even suggests inadequacy of this line of thought in the present context,\ and our pragmatic choice of the Lie supergeometry to be considered in what follows -- that of the topologically trivial super-Minkowski space -- appears to seal its fate.\ \emph{\bf In our work,\ we demonstrate the fallacy of the latter reasoning by identifying the \emph{purely} topological content of our definition of the supergerbe.}\ The idea underlying our demonstration,\ and hence the inspiration and conceptual foundation of the present work,\ comes from an old and largely overlooked series of articles \cite{Rabin:1984rm,Rabin:1985tv} by Rabin and Crane.\ Without going into the technical details,\ which shall be provided in Sec.\,\ref{sec:RCorb},\ we can summarise the idea as a consistent interpretation of the supersymmetric refinement of the de Rham cohomology on the super-minkowskian target $\,{\rm sMink}(d,1|D_{d,1})\,$ of the GS super-$\si$-models as (a model of) \emph{the} de Rham cohomology of an orbifold $\,{\rm sMink}(d,1|D_{d,1})//\G_{{\rm KR}}\,$ \cite{Rabin:1984rm},\ to be referred to as the {\bf Rabin--Crane} ({\bf RC}) {\bf orbifold},\ of $\,{\rm sMink}(d,1|D_{d,1})\,$ with respect to a natural action of the Kosteleck\'y--Rabin discrete supersymmetry group $\,\G_{{\rm KR}}\subset{\rm sMink}(d,1|D_{d,1})\,$ \cite{Kostelecky:1983qu}.\ The identification is attained in the direct limit over nested skeletal Rogers--DeWitt presentations \cite{Rogers:1980,DeWitt:1984} (or,\ more formally,\ over a nested family of superpoints in $\,{\rm sMink}(d,1|D_{d,1})$) -- anticipated by and coherent with the Freed functorial interpretation \cite{Freed:1999} of the superfield theory on Gra\ss mann-even spacetimes -- of the target supermanifold of the super-$\si$-model.\ What makes the identification possible is the polynomial dependence of the super-vielbeins on the global supercoordinates on $\,{\rm sMink}(d,1|D_{d,1})$.\ The orbifold,\ which does not seem\footnote{To the best of the Author's knowledge,\ no attempt has been made to date to analyse the Rabin--Crane construction within the Jadczyk--Pilch framework of \cite{Jadczyk:1980xp},\ which is not excluded by Rothstein's axioms.} to fit into Rothstein's axiomatics for supermanifolds given in \cite{Rothstein:1986ax},\ is a soul fibration over the body $\,{\rm Mink}(d,1)\,$ of the target with a \emph{topologically nontrivial fibre}.\ In particular,\ it has \emph{compact} Gra\ss mann-odd fibres.\ \emph{\bf We demonstrate that the CaE super-$p$-gerbe ensures a quantum-mechanically consistent descent of the superfield theory to the RC orbifold.}\ The goal is attained,\ in conformity with  \cite{Gawedzki:2010rn,Gawedzki:2012fu},\ through identification of a suitable $\G_{{\rm KR}(L)}$-equivariant structure on the $p$-gerbe,\ and -- equivalently \cite{Runkel:2008gr,Suszek:2012ddg,Suszek:2013} -- through construction of the corresponding gauge-symmetry defect,\ and the upshot seems to be the first concrete definition of a superfield theory with a nontrivial (non-Rothstein) superorbifold as a model for its internal degrees of freedom.\ After this,\ necessarily general,\ introduction,\ we are ready to enter the formal discourse of the work.\smallskip 

\noindent\emph{A chronological note:} The present work is a revised,\ conceptually elaborated and extended version of the Author's original (unpublished) arXiv report \cite{Suszek:2017xlw},\ in which some later findings of \cite{Suszek:2019cum,Suszek:2018bvx,Suszek:2020xcu} have been integrated,\ and in which the fully fledged interpretation of the main construction in terms of the super-orbifold superfield theory has been provided.\bigskip

\noindent{\bf Acknowledgements:}  This work is a humble tribute to the memory of a Friend and Teacher,\ Professor Krzysztof Gaw\k{e}dzki ($\ast$ 1947 -- $\dagger$ 2022),\ whose rigorous yet imaginative approach to the higher algebra and geometry of field theory has always been and remains an inexhaustible source of intuition and a guidepost in the Author's research.\ Many of the ideas underlying the work reported herein stem from the Author's fruitful collaboration and discussions with Professor Gaw\k{e}dzki in the years 2005--19.

\section{The gerbe theory behind the bosonic $\si$-model \& its symmetries}\label{sec:sigrb}

Realisation of the goals set up in the Introduction requires,\ on the one hand,\ a working knowledge of the structure and properties of the $\si$-model,\ with special emphasis on its configurational symmetries,\ and,\ on the other hand,\ a sound understanding of the general idea behind and some technicalities of the scheme of geometrisation of integral classes in the de Rham cohomology of a manifold $\,M\,$ proposed in and inspired by the works of Murray {\it et al.},\ alongside the corresponding higher-geometric rendering of the symmetries of $\,M$.\ In this section,\ we review those aspects of the two classes of problems which are central to the supergeometric adaptation of the constructions from the non-graded category advanced in the supergeometric setting,\ to be undertaken in Sec.\,\ref{sec:dAzcladder},\ and in its conceptual elucidation in the spirit of superfield theory on (super)orbifolds,\ given in Sec.\,\ref{sec:RCorb}.\ In so doing,\ we leave,\ wherever possible,\ technical details aside for the sake of conciseness,\ refering the Reader to the rich literature on the subject.\smallskip

The (mono-phase) two-dimensional non-linear $\si$-model with the topological WZ term is a lagrangean theory of smooth mappings $\,x\in[\Si,M]\,$ of a closed connected oriented two-dimensional manifold $\,\Si\,$ (the {\bf worldsheet}) into a metric manifold $\,(M,\txg)\,$ (the {\bf target space}) endowed with a 1-gerbe with connection $\,\cG\,$ of curvature $\,\chi\equiv{\rm curv}(\cG)\in Z^3_{\rm dR}(M)\,$ with periods $\,{\rm Per}(\chi)\subset 2\pi\bZ$.\ The theory is determined by the principle of least action for the Dirac--Feynman amplitude
\qq\label{eq:DFampl}
\cA^\Si_{\rm DF}\ :\ [\Si,M]\too\uj\ :\ x\longmapsto\exp\left(\sfi\int_\Si{\rm Vol}\bigl(\Si,x^*\txg\bigr)\right)\cdot{\rm Hol}_\cG\bigl(x(\Si)\bigr)\,,
\qqq
in which the WZ factor returns the value attained by the Cheeger--Simons differential character $\,{\rm Hol}_\cG\in\Hom_{{\rm {\bf AbGrp}}}(Z_2(M),\uj)$,\ known as the {\bf surface holonomy of} $\,\cG$,\ on the 2-cycle $\,x(\Si)\in Z_2(M)$.\ The holonomy can be understood abstractly as the image of the class $\,[x^*\cG]\,$ of the pullback 1-gerbe in the group $\,\cW^3(\Si;0)\,$ of isoclasses of flat 1-gerbes over $\,\Si\,$ under the composite isomorphism $\,\cW^3(\Si;0)\cong\vH^2(\Si,\unl{\uj})\cong\uj$,\ but it also admits an explicit presentation in terms of the sheaf-theoretic data of $\,\cG$,\ the latter composing a 2-cocycle in the 2${}^{\rm nd}$ (real) Beilinson--Deligne cohomology group $\,\bH^2(M,\cD(2)^\bullet)$ of $\,M\,$ for the Deligne complex of sheaves $\,\cD(2)^\bullet: \unl{\uj}{}_M\xrightarrow{\ -\sfi\,\sfd\log\ }\unl{\Om}^1(M)\xrightarrow{\ \sfd\ }\unl{\Om}^2(M)\,$ \cite{Gawedzki:1987ak,Murray:1994db,Murray:1999ew}.\ Before we recall the essential implications of the above definition of the field theory of interest,\ let us outline the structural underpinnings of Murray's geometrisation scheme which yields the higher-geometric object $\,\cG\,$ upon application to $\,\chi$.

The inherently hierarchical nature of the geometrisation scheme \cite{Johnson:2003} is most succinctly represented by the `Murray diagram' of a {\bf $(p+1)$-gerbe $\,\cG^{(p+1)}$}:
{\scriptsize\qq\nn\hspace{-1.25cm}
\alxydim{@C=.65cm@R=1.25cm}{ {\tiny\D^{(p+3)}\Phi^{(p)}_{p+1}=\id} \ar@{->>}[d] & {\tiny\Phi^{(p)}_{p+1}:\D^{(p+2)}\Phi^{(p)}_p\cong\id} \ar@{->>}[d] & \cdots & {\tiny\Phi^{(p)}_2:\D^{(3)}\Phi^{(p)}_1\cong\id} \ar@{->>}[d] & {\tiny\Phi^{(p)}_1:\D^{(2)}\cG^{(p)}\cong\cI^{(p)}_0} \ar@{->>}[d] & \cG^{(p)} \ar@{->>}[d] & \cI^{(p+1)}_\b \ar@{=}[d] \\ \sfY^{[p+4]}M \ar@<1.ex>[r]^{d^{(p+3)}_\cdot} \ar@<.5ex>[r] \ar@{.}@<0ex>[r] \ar@<-.5ex>[r] \ar@<-1.ex>[r] & \sfY^{[p+3]}M \ar@<1.ex>[r]^{d^{(p+2)}_\cdot} \ar@{.}@<.5ex>[r] \ar@{.}@<0ex>[r] \ar@<-.5ex>[r] \ar@<-1.ex>[r] & \cdots \ar@<1.ex>[r]^{d^{(4)}_\cdot} \ar@<.5ex>[r] \ar@<0ex>[r] \ar@<-.5ex>[r] \ar@<-1.ex>[r] &  \sfY^{[4]}M \ar@<.75ex>[r]^{d^{(3)}_\cdot} \ar@<.25ex>[r] \ar@<-.25ex>[r] \ar@<-.75ex>[r] & \sfY^{[3]}M \ar@<.5ex>[r]^{d^{(2)}_\cdot\qquad} \ar@<0ex>[r] \ar@<-.5ex>[r] & (\sfY^{[2]}M,\D^{(1)}\b) \ar@<.25ex>[r]^{\qquad d^{(1)}_\cdot} \ar@<-.25ex>[r] & (\sfY M,\b) \ar@{->>}[d]_{\pi_{\sfY M}} \\ &  &  &  &  & & (M,\chi)}\,,
\qqq}
~\vspace{-25pt}
\qq\label{diag:Murray}
\qqq
which defines $\,\cG^{(p+1)}\,$ in terms of a $0$-cell $\,\cG^{(p)}\,$ and a collection of $k$-cells $\,\Phi^{(p)}_k\,$ (the so-called $k$-isomorphisms) of the weak $(p+2)$-category $\,\gt{Grb}_\nabla^{(p)}(M)\,$ (with a tensor product and duality) of $p$-gerbes with connective structure over a given base $\,M\,$ \cite{Murray:1999ew,Stevenson:2001grb2,Waldorf:2007mm} ({\it cp.}\ also \cite{Stevenson:2000wj,Waldorf:2007phd} for an introduction into the subject).\ The diagram is to be read according to the following key:\label{p:Murray}
\bit
\item[(i)] the diagram is written in the category $\,\Man\,$ of smooth manifolds;
\item[(ii)] all its arrows are \emph{epimorphism} in that category,\ {\it i.e.},\ surjective submersions;
\item[(iii)] the root of the diagram is an object $\,M\in\obj\Man\,$ together with a $(p+3)$-\emph{cocycle} $\,\chi\in Z^{p+3}_{\rm dR}(M)\,$ in the differential complex naturally associated with the underlying category,\ {\it i.e.},\ in the de Rham complex;
\item[(iv)] over the root,\ there is an object $\,\sfY M\in\obj\Man\,$ together with a $(p+2)$-cochain $\,\b\in\Om^{p+2}(\sfY M)\,$ of the same complex (for $\,\sfY M$),\ which jointly \emph{resolve} $\,(M,\chi)$,\ as expressed by $\,\pi_{\sfY M}^*\chi=\sfd\b\,$ -- this is the so-called {\bf trivial $(p+1)$-gerbe} $\,\cI^{(p+1)}_\b$;
\item[(v)] objects $\,\sfY^{[n]}M\in\obj\Man,\ n\in\ovl{1,p+4}\,$ in the middle row are components of the nerve $\,\sfN_\bullet{\rm Pair}_M(\sfY M)\equiv\sfY^{(\bullet+1)}M\,$ of the $M$-fibred pair groupoid $\,{\rm Pair}_M(\sfY M): \alxydim{@C=2.cm}{\sfY^{[2]}M \ar@<.25ex>[r]^{\sft\equiv d^{(1)}_0=\pr_1\qquad} \ar@<-.25ex>[r]_{\sfs\equiv d^{(1)}_1=\pr_2\qquad} & \sfY M\equiv\sfY^{[1]}M}\,$ with the arrow object $\,\sfY^{[2]}M=\sfY M{}_{\pi_{\sfY M}}\hspace{-3pt}\x_{\pi_{\sfY M}}\hspace{-1pt}\sfY M=\{\ (y_1,y_2)\in\sfY M^{\x 2}\quad\vert\quad \pi_{\sfY M}(y_1)=\pi_{\sfY M}(y_2)\ \}\,$ and the face morphisms $\,d^{(n)}_i : \sfY^{[n+1]}M\too\sfY^{[n]}M,\ i\in\ovl{0,n}\,$ given by the canonical projections according to the (semi-)simplicial scheme laid out in \cite{Segal:1968}; 
\item[(vi)] the face maps induce coboundary operators $\,\D^{(n)}=\sum_{k=0}^n\,(-1)^{k-1}\,d^{(n)\,*}_k\,$ \cite{Dupont:1976};
\item[(vii)] over the arrow object,\ there is a $p$-gerbe $\,\cG^{(p)}\,$ of curvature $\,\D^{(1)}\b$;
\item[(viii)] the action of the $\,\D^{(n)}\,$ on higher-geometric objects is to be understood in terms of the monoidal structure and duality on $\,\gt{Grb}_\nabla^{(p)}(M)$,\ {\it e.g.},\ $\,\D^{(2)}\cG^{(p)}\equiv\pr_{2,3}^*\cG^{(p)}\ox\pr_{1,3}^*\cG^{(p)\,*}\ox\pr_{1,2}^*\cG^{(p)}\,$ (and further $\ox$-augmented with suitable identity cells for $\,n>2$); 
\item[(ix)] at the penultimate stage of the (hierarchical) definition of each ($k$-)isomorphism,\ there appears an \emph{isomorphism} in the (1-)category of $0$-gerbes,\ with which the hierarchy starts\footnote{With a little imagination,\ one may also regard functions as $(-1)$-gerbes,\ {\it cp.}\ \cite{Johnson:2003}.} -- this is the groupoid $\,{\rm Bun}^\nabla_{\bC^\x}(M)\,$ of principal $\bC^\x$-bundles with compatible connection over a given base $\,M$;
\item[(x)] over $\,\sfY^{[p+4]}M$,\ there is an equality of the composition of face-map pullbacks of the isomorphism $\,\Phi^{(p)}_{p+1}\,$ with the identity morphism in $\,{\rm Bun}^\nabla_{\bC^\x}(M)$.
\eit

Upon application of the key at $\,p=1$,\ we recover a 1-gerbe $\,\cG\equiv\cG^{(1)}\,$ over $\,M\,$ of curvature $\,\chi\in Z^3_{\rm dR}(M)$,\ defined by a quintuple $\,(\sfY M,\pi_{\sfY M},\b,\cG^{(0)},\Phi^{(0)}_1)\,$ in which $\,\cG^{(0)}=(L,\pi_L,\cA_L)\,$ is a principal $\bC^\x$-bundle over $\,\sfY^{[2]}M\,$ with a principal connection 1-form $\,\cA_L\in\Om^1(L)$,\ and in which $\,\mu_L\equiv\Phi^{(0)}_1:\pr_{1,2}^*L\ox\pr_{2,3}^*L\xrightarrow{\ \cong\ }\pr_{1,3}^*L\,$ is a connection-preserving isomorphism of principal $\bC^\x$-bundles over $\,\sfY^{[3]}M$,\ satisfying the identity $\,\pr_{1,2,4}^*\mu_L\circ(\id\ox\pr_{2,3,4}^*\mu_L)=\pr_{1,3,4}^*\mu_L\circ(\pr_{1,2,3}^*\mu_L\ox\id)\,$ over $\,\sfY^{[3]}M$,\ whence also its name:\ the {\bf grupoid structure of} $\,\cG^{(1)}$.\smallskip

In the canonical description of the loop dynamics described by the two-dimensional $\si$-model,\ we encounter the single-loop configuration space $\,\sfL M\equiv[\bS^1,M]\,$ and the corresponding space $\,\sfT^*\sfL M\xrightarrow{\ \pi_{\sfT^*\sfL M}\ }\sfL M\,$ of Cauchy data of a critical embedding,\ localised on a Cauchy contour $\,(\cong)\bS^1\subset\Si$.\ The latter single-loop space of states is endowed with a (pre)symplectic form \cite{Gawedzki:1972ms,Gawedzki:1990jc,Suszek:2011hg} 
\qq\nn
\Om_\si=\d\vartheta_{\sfT^*\sfL M}+\pi_{\sfT^*\sfL M}^*\int_{\bS^1}\,\ev^*\chi\,,
\qqq 
in which $\,\vartheta_{\sfT^*\sfL M}\,$ is the canonical (action) 1-form on $\,\sfT^*\sfL M$,\ and $\,\ev:\sfL M\x\bS^1\too M:(x,\varphi)\longmapsto x(\varphi)\,$ is the standard evaluation map.\ In the first-order formalism of \cite{Gawedzki:1972ms},\ the 2-form is used to compute Poisson brackets of function(al)s $\,h_\xcX\,$ on $\,\sfT^*\sfL M\,$ associated with hamiltonian vector fields $\,\xcX\in\G(\sfT\sfT^*\sfL M)\,$ in the usual manner,\ $\,\d h_\xcX=-\xcX\con\Om_\si$.\ We have 
\qq\nn
\{h_{\xcX_1},h_{\xcX_2}\}_{\Om_\si}=\xcX_2\con\xcX_1\con\Om_\si\,.
\qqq

The above is the point of departure for the prequantisation of the $\si$-model,\ in which one erects,\ after \cite{Gawedzki:1987ak},\ a line bundle $\,\ceL_\si\,$ with connection over $\,\sfT^*\sfL M\,$ with curvature $\,\Om$.\ The reconstruction proceeds along the lines of the classic clutching theorem for fibre bundles,\ with the local data of $\,\ceL_\si\,$ induced from those of $\,\cG\,$ {\it via} transgression $\,\t:\bH^2(M,\cD(2)^\bullet)\too\bH^1(\sfL M,\cD(1)^\bullet)$.\ In this framework,\ quantum states of the theory are identified with (suitably polarised) sections of $\,\ceL_\si$,\ and properties of the thus obtained Hilbert space $\,\ceH_\si=\G_{\rm pol}(\ceL_\si)\,$ of the $\si$-model (the spectrum,\ correlators,\ realisation of symmetries,\ anomalies {\it etc.}) are read off directly from the behaviour of the disc DF amplitudes $\,\cA^\Si_{\rm DF},\ \Si\cong\bD^2\,$ at the boundary $\,\p\Si\cong\bS^1\,$ \cite[Sec.\,4.1]{Gawedzki:1999bq}.\ The existence of the transgression map $\,\t\,$ decides the central r\^ole of the higher-geometric object $\,\cG\,$ in a quantum-mechanically consistent description of the loop dynamics of the $\si$-model.\smallskip

The canonical and geometrically-(pre)quantum descriptions are particularly well-adapted to the analysis of configurational symmetries of the $\si$-model.\ Here,\ we restrict our attention to symmetries induced by isometries of the metric target space $\,(M,\txg)$.\ Let $\,\txG_\si\,$ be a Lie group of those of them which preserve the DF amplitude \eqref{eq:DFampl},\ with the action $\,\la:\txG_\si\x M\too M$,\ and let $\,\ggt_\si\,$ be its tangent Lie algebra.\ The latter is mapped homomorphically to the (commutator) Lie algebra $\,(\G(\sfT M),[\cdot,\cdot]_{\G(\sfT M)})\,$ by the fundamental vector field $\,\cK_\cdot:\ggt\too\G(\sfT M)$.\ Preservation of $\,\cA^\Si_{\rm DF}\,$ implies the existence of a {\bf comomentum 1-form} $\,\kap_\cdot:\ggt\too\Om^1(M)\,$ with the defining property $\,\sfd\kap_X=-\cK_X\con\chi,\ X\in\ggt$.\ These determine the noetherian currents $\,\jmath_X[\psi]=\cK{}_X^\mu(x)\sfp_\mu+(x_*\widehat t)^\mu\kap_{X\,\mu}(x)\,$ (written in terms of the standard loop/momentum (local) coordinates $\,\psi\equiv(x^\mu,\sfp_\nu)\,$ on $\,\sfT^*\sfL M$,\ and of the normalised tangent vector field $\,\widehat t\,$ on $\,\bS^1$) with the \emph{anomalous} equal-time ($t$) Poisson brackets 
\qq\nn
\{\jmath_{X_1}[\psi](t,\varphi_1),\jmath_{X_2}[\psi](t,\varphi_2)\}_{\Om_\si}=\jmath_{[X_1,X_2]_\ggt}[\psi](t,\varphi_1)\,\d(\varphi_2-\varphi_1)\cr\cr
+\a_{X_1,X_2}[x](t,\varphi_1)\,\d(\varphi_1-\varphi_2)-2\La_{X_1,X_2}[x]\bigl(t,\tfrac{1}{2}\,(\varphi_1+\varphi_2)\bigr)\,\d'(\varphi_1-\varphi_2)
\qqq 
in which there appears the {\bf wrapping-charge 1-cocycle} $\,\a_{\cdot,\cdot}:\ggt\wedge\ggt\too Z^1_{\rm dR}(M)\,$ given by the formula $\,\a_{X_1,X_2}=\tfrac{1}{2}\,(\pLie{\cK_{X_1}}\kap_{X_2}-\pLie{\cK_{X_2}}\kap_{X_1})-\kap_{[X_1,X_2]_\ggt}$,\ alongside the {\bf Leibniz-anomaly 1-cochain} $\,\La_{\cdot,\cdot}:\ggt\ox_{\rm sym}\ggt\too C^\infty(M,\bR)\,$ defined as $\,\La_{X_1,X_2}=\cK_{X_1}\con\kap_{X_2}+\cK_{X_2}\con\kap_{X_1}$,\ {\it cp.}\ \cite{Alekseev:2004np,Gawedzki:2012fu,Suszek:2012ddg} where the algebroidal structure over the target space underlying the Poisson bracket was elucidated.\ Whenever there exist non-contractible loops in $\,M$,\ the corresponding {\bf winding states} $\,x(\bS^1)\in Z_1(M)\,$ contribute to the wrapping charge.\ These give rise to central extensions of $\,\ggt_\si\,$ which reflect an interplay between the nontrivial topology of the target space and that of the spacetime $\,\Si\,$ of the $\si$-model.\ The Leibniz anomaly,\ on the other hand,\ can occur even in the absence of such topological `defects' in the target space -- a classic example is the central term in the chiral current algebra of the WZW model for a compact \emph{1-connected} target Lie group \cite{Gawedzki:1990jc,Falceto:1992bf}.

Physical considerations distinguish the $\txG_\si$-invariant refinement $\,H_{\rm dR}^\bullet(M)^{\txG_\si}\,$ of the standard de Rham cohomology $\,H_{\rm dR}^\bullet(M)\,$ as \emph{the} classifying algebraic structure in the discussion of the topological properties of the field theory.\ As noted in \cite{deAzcarraga:1991ch} and analysed at length in \cite{deAzcarraga:1995},\ this r\^ole of the refinement becomes particularly pronounced when $\,\chi\,$ only has global (de Rham) primitives,\ $\,\chi=\sfd\b$,\ which are {\bf \emph{quasi}-$\ggt_\si$-invariant},\ {\it i.e.},\ whenever there exists a non-zero $\ggt$-indexed family of 1-forms $\,J_\cdot:\ggt\too\Om^1(M)\,$ with the property $\,\pLie{\cK_X}\b=\sfd J_X$,\ so that we may take $\,\kap_X=\cK_X\con\b-J_X$,\ {\it cp.}\ \cite{deAzcarraga:1991ch,deAzcarraga:1995}.\ In the light of the classic result:\ $\,H_{\rm dR}^\bullet(M)^{\txG_\si}\cong H_{\rm dR}^\bullet(M)\,$ for $\,\txG_\si\,$ compact \cite{Chevalley:1948},\ such quasi-$\ggt_\si$-invariance manifests itself only in the case of \emph{non}-compact symmetry groups.\ Supersymmetry,\ built into superfield theory in its form investigated in the present paper as its constitutive element,\ is modelled on an intrinsically non-compact supermanifold structure.\ Hence,\ the discrepancy between the two cohomologies is bound to and does,\ indeed,\ become relevant in the study of super-$\si$-models.\smallskip

The above symmetry analysis,\ which is readily integrated to the group level by examining the behaviour of the disc amplitudes under symmetry transformations of the global primitive $\,\b\,$ (whenever it exists),\ was carried out on the field-theory side.\ There is,\ however,\ a possibility to work directly with the higher-geometric object $\,\cG\,$ over the target space,\ taking \eqref{eq:DFampl} as the point of departure.\ Here,\ the obvious question is:\ What is a structural (bicategorial) manifestation of a configurational symmetry of the $\si$-model in $\,\gt{Grb}_\nabla^{(1)}(M)$?\ A very general answer to this question was worked out by the Author {\it et al.}\ in a series of papers \cite{Runkel:2008gr,Gawedzki:2010rn,Suszek:2011hg,Suszek:2012ddg,Gawedzki:2012fu}.\ In its simplest and most direct form,\ proven in \cite[Sec.\,2.2]{Gawedzki:2010rn},\ the answer is:\ The action $\,\la\,$ of $\,\txG_\si\subset{\rm Isom}(M,\txg)\,$ preserves the DF amplitude iff there exists a comomentum 1-form $\,\kap_\cdot\,$ as above and a $\txG_\si$-indexed family of gerbe 1-isomorphisms\label{p:rig-qmorph}
\qq\label{eq:sym-inv-1isos}
\Phi_g\ :\ \cG\xrightarrow{\ \cong\ }\wp_g^*\cG\,,\qquad\quad\wp_\cdot\ :\ M\x\txG_\si\too M\ :\ (m,g)\longmapsto\la_{g^{-1}}(m)\,.\qquad
\qqq 
These transgress to automorphisms of the prequantum bundle \cite{Suszek:2011hg,Suszek:2012ddg}.\smallskip

\begin{figure}[htb]\label{fig:def-corresp}
\centerline{\includegraphics[width=3.in]{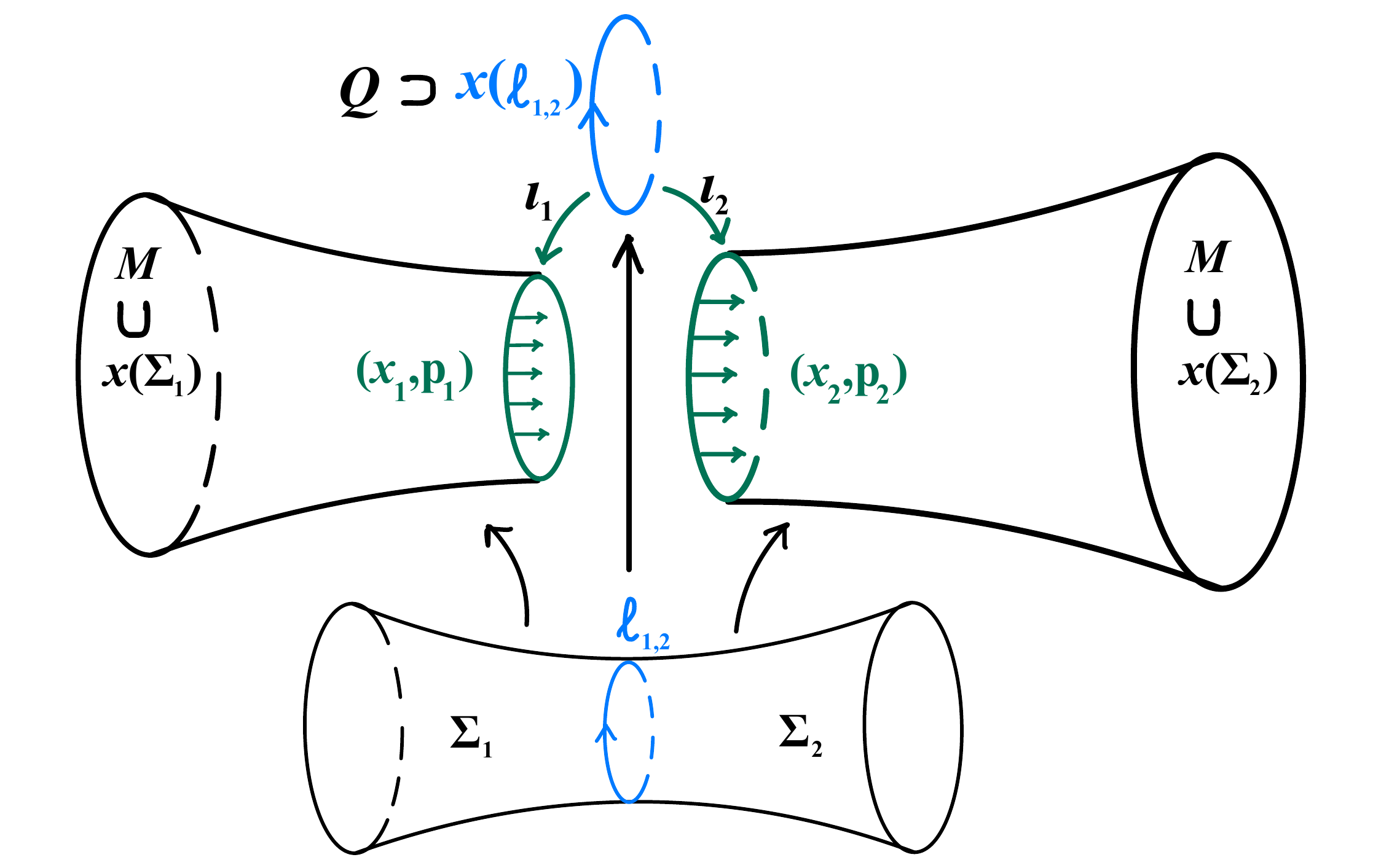}}
\vspace*{8pt}
\caption{A correspondence between states $\,(x_1,\sfp_1)\,$ and $\,(x_2,\sfp_2)\,$ across a defect line $\,\ell_{1,2}\,$ separating the two phases of the $\si$-model,\ inhabiting the spacetime patches $\,\Si_1\,$ and $\,\Si_2$,\ respectively.}
\end{figure}

In fact,\ there is an important alternative way in which the above $\,\Phi_g\,$ can be employed to \emph{implement} the symmetry,\ as discovered and elaborated in \cite{Fuchs:2007fw,Runkel:2008gr,Suszek:2011hg,Suszek:2012ddg}.\ Indeed,\ $\,\Phi_g\,$ is a particular example of a {\bf $\cG$-bi-module},\ {\it i.e.},\ of a gerbe 1-isomorphism $\,\Phi\ :\ \iota_1^*\cG\xrightarrow{\ \cong\ }\iota_2^*\cG\ox\cI^{(1)}_\om$,\ defined in terms of a smooth 2-form $\,\om\in\Om^2(Q)$,\ termed the {\bf bi-brane curvature},\ on a {\bf correspondence space} (a.k.a.\ the {\bf bi-brane worldvolume}) $\,Q\,$ which comes with a pair of smooth maps $\,\iota_A\ :\ Q\too M,\ A\in\{1,2\}$.\ Just as a gerbe $\,\cG\,$ defines a $\si$-model for the mapping space $\,[\Si,M]$,\ a $\cG$-bi-module renders meaningful its counterpart for \emph{patchwise} smooth mappings into $\,M$,\ with the patches separated by a {\bf line defect} $\,\ell\subset\Si$,\ to which the (sheaf-theoretic) data of $\,\Phi\,$ are to be pulled back along $\,x\rstr_\ell\ :\ \ell\too Q$.\ The patches,\ which may,\ in general,\ be mapped to disjoint components of a stratified (bulk) target space $\,M$,\ are to be thought of as inhabited by phases of the ensuing poly-phase $\si$-model.\ It was demonstrated in \cite{Suszek:2011hg} that the Defect Gluing Condition \cite{Runkel:2008gr} satisfied by the limiting values of the bulk embedding field $\,x\rstr_{\Si\setminus\ell}\,$ and its normal derivatives at the defect line determines an isotropic subspace,\ and hence a symplectic relation,\ in the two-phase space of states (equipped with the (pre)symplectic form given by the weighted sum of the (pre)symplectic forms of the adjacent phases),\ and $\,\Phi\,$ transgresses to a local isomorphism between the two mono-phase prequantum bundles.\ It is in this beautifully structured manner that a gerbe bi-module is seen to implement a \emph{local} state correspondence across a defect (between the limiting Cauchy data on either side of the defect line),\ as illustrated in Fig.\,\ref{fig:def-corresp}.
Circumstances under which the latter acquires the status of a (quantum-mechanically consistent) \emph{hamiltonian automorphism} were studied at length in \cite{Runkel:2008gr,Suszek:2011hg,Suszek:2012ddg},\ where it was established that a sufficient condition is the so-called \emph{topologicality} of the worldsheet defect decorated by the data of the bi-brane $\,\cB=(Q,\iota_1,\iota_2,\om,\Phi)$,\ {\it cp.}\ the original papers for details.\ The important lesson from that study is the manifest topologicality of a distinguished class of {\bf $\txG_\si$-jump defects} for the $\si$-model with symmetry $\,\txG_\si$,\ with data ($\wp_g\equiv\la_{g^{-1}}$,\ as above)
\qq\nn
&\cB_{\txG_\si}=\bigsqcup_{g\in\txG_\si}\cB_g\,,\qquad\qquad\cB_g=\bigl(Q_g\equiv M\x\{g\},\iota_\cdot,0,\Phi_g\bigr)\,,&\cr\cr
&\iota_1=\pr_1\,,\quad\iota_2=\wp\,.&
\qqq
A bulk field configuration `jumps' across a defect line mapped to the component $\,\cB_g\,$ uniformly by the action of the corresponding symmetry-group element $\,g$.

The peculiarity of the $\txG_\si$-jump defect for $\,\txG_\si\,$ \emph{discrete} lies in the fact that whenever there exists a choice of gerbe 2-isomorphisms
\qq\label{eq:discr-sym-equiv-2isos}
\varphi_{g_1,g_2}\ :\ \wp_{g_1}^*\Phi_{g_2}\circ\Phi_{g_1}\overset{\cong}{\Longrightarrow}\Phi_{g_1\cdot g_2}\,,
\qqq
making up (the $2\to 1$-component of) a {\bf $(\cG,\cB_{\txG_\si})$-inter-bi-brane}
\qq\nn
&\cJ_{\txG_\si}^{2\to 1}=\bigsqcup_{g_1,g_2\in\txG_\si}\cJ_{g_1,g_2}\,,\qquad\cJ_{g_1,g_2}=\bigl(T_{g_1,g_2}^{(3)}\equiv M\x\{g_1\}\x\{g_2\},\pi^{(3)}_{\cdot,\cdot},\Phi_g\bigr)\,,&\cr\cr
&\pi^{(3)}_{1,2}=\pr_{1,2}\,,\quad\pi^{(3)}_{2,3}=\wp\x\id_{\txG_\si}\,,\quad\pi^{(3)}_{1,3}=\id_M\x\txm\,,&
\qqq
\emph{for which the cohomology class of the associated $\uj$-valued} {\bf associator 3-cocycle} \emph{on $\,\txG_\si$},\ derived in \cite[Sec.\,2.8.1]{Runkel:2008gr},\ \emph{vanishes},\ the defect \emph{defines} a $\txG_\si$-orbifold of the original $\si$-model,\ {\it i.e.},\ a $\si$-model with the target given by the orbispace $\,M//\txG_\si$,\ descended canonically from its precursor on $\,M\,$ through a specialisation of the universal gauge principle established in \cite{Gawedzki:2010rn,Gawedzki:2012fu}.\ Indeed,\ the collection $\,(\Phi_g,\varphi_{g_1,g_2})_{g,g_1,g_2\in\txG_\si}\,$ then determines a (discrete,\ and hence automatically descendable) {\bf $\txG_\si$-equivariant structure on} $\,\cG$. \label{p:jump-inv-equiv}

\begin{figure}[htb]\label{fig:Dante}
\centerline{\includegraphics[width=2.75in]{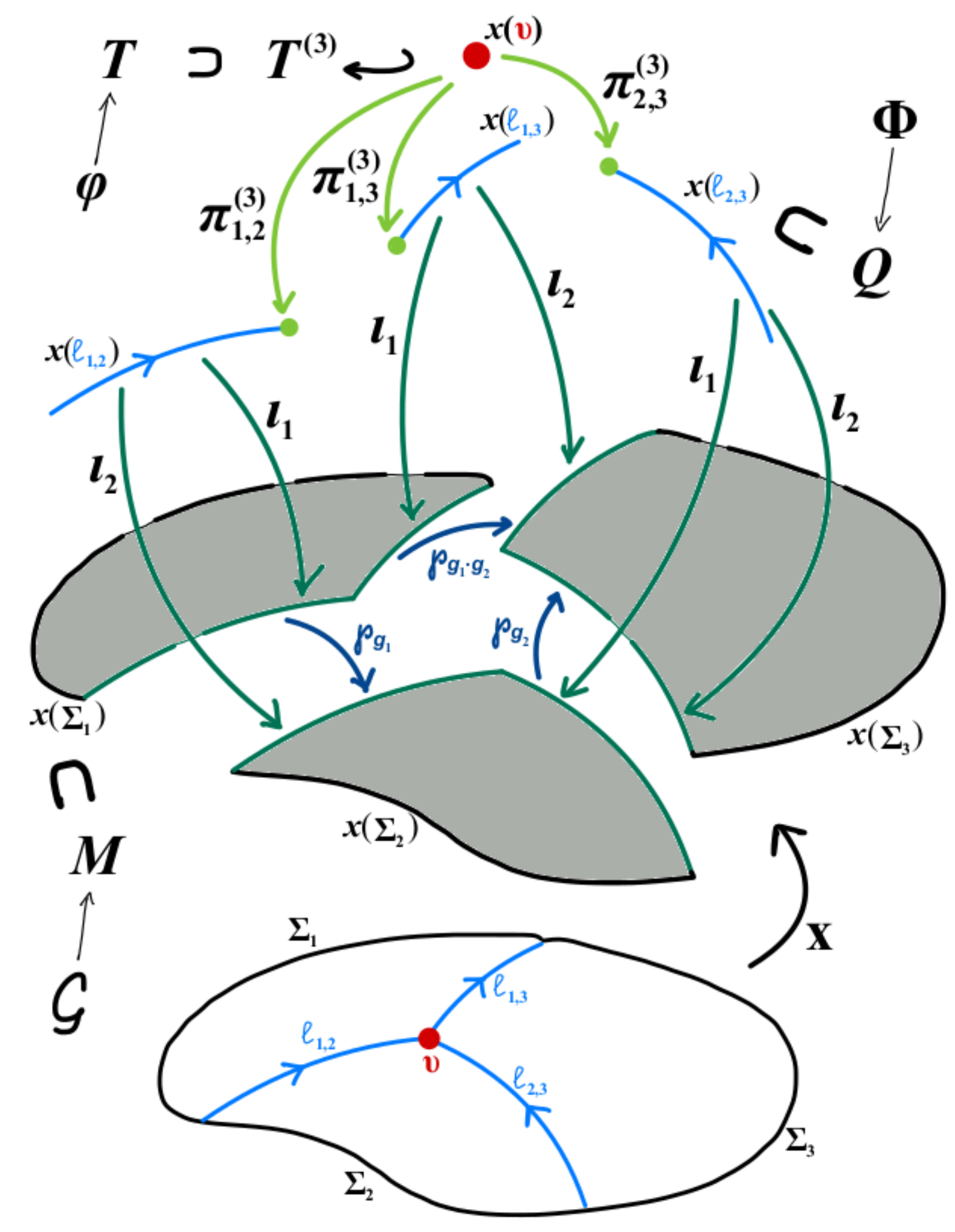}}
\vspace*{8pt}
\caption{A $\txG_\si$-twisted field configuration $\,x\,$ near a $2\to 1$ junction $\,\upsilon\,$ of (oriented) edges $\,\ell_{i,j},\ (i,j)\in\{(1,2),(2,3),(1,3)\}\,$ of a $\txG_\si$-jump defect.\ The edges separate patches $\,\Si_A,\ A\in\{1,2,3\}\,$ of the worldsheet,\ embedded by $\,x\,$ in connected components of the metric bulk target space $\,M\,$ and defined by the respective restrictions of the metric $\,\txg\,$ and of the gerbe $\,\cG$.\ The field $\,x\,$ maps the $\,\ell_{i,j}\,$ to the bi-brane worldvolume $\,Q$,\ from which it pulls back the respective restrictions of the gerbe bimodule $\,\Phi$.\ Similarly,\ the junction $\,\upsilon\,$ is mapped to the component $\,T^{(3)}\,$ of the stratified inter-bi-brane worldvolume $\,T$,\ from which the 2-isomorphism $\,\varphi\,$ is pulled back.}
\end{figure}

Equivalently,\ the data of the 1- and 2-isomorphisms can be pulled back to an arbitrary (oriented) defect graph,\ or defect quiver $\,\G\,$ embedded in $\,\Si$,\ whereby the {\bf $\txG_\si$-twisted sector} of the orbifold $\si$-model is recovered,\ as discussed in \cite{Runkel:2008gr} and generalised in \cite{Suszek:2012ddg}:\ The $\si$-model field  maps edges of the graph $\,\G\,$ to the connected components of the stratified bi-brane worldvolume $\,Q=\bigsqcup_{g\in\txG_\si}\,Q_g$,\ and its vertices of valence $\,n+1\,$ -- to the corresponding components $\,T^{(n+1)}\,$ of the stratified inter-bi-brane worldvolume,\ whose definition extends that of $\,T^{(3)}=\bigsqcup_{g_1,g_2\in\txG_\si}\,T_{g_1,g_2}^{(3)}\,$ in a self-evident manner,\ and whose higher-geometric data are \emph{induced} from those of the fusion 2-isomorphisms $\,\varphi_{g_1,g_2}\,$ in a limiting procedure for binary-tree `resolutions' of the $(n+1)$-valent vertices detailed {\it ibidem} ({\it cp.}\ also \cite{Suszek:2022lpf,Suszek:2022gtmp} for a recent study).\ Thus,\ the $\txG_\si$-twisted sector consists of the patchwise continuous field configurations with discontinuities at intersections with a worldsheet defect described earlier and so vanishing in the passage $\,M\to M//\txG_\si$,\ {\it cp.}\ \cite{Dixon:1985jw,Runkel:2008gr,Frohlich:2009gb,Suszek:2011hg,Suszek:2012ddg},\ as depicted in Fig.\,\ref{fig:Dante}.\ This is none other than a (Cartan--Borel) field-theoretic pullback of the universal homotopy-quotient model of the space of orbits $\,M//\txG_\si$,\ {\it cp.}\ \cite{Cartan:1950mix,Tu:2020}.\ We shall encounter this manifestation of the orbifold structure in the supergeometric setting discussed in the remainder of the present work. 

\section{The super-minkowskian background}\label{sec:sMink}

In the present section,\ we set the stage for a constructive adaptation of the ideas gathered in the previous section to a class of $(p+1)$-dimensional \emph{super}field theories modelling propagation of extended distributions of \emph{super}charge in an ambient (super)geometry permeated by background fields,\ known as super-$p$-branes \cite{Green:1983wt,Green:1983sg,Achucarro:1987nc,Metsaev:1998it,Arutyunov:2008if,Gomis:2008jt,Fre:2008qc,DAuria:2008vov,Bandos:1997ui,deWit:1998yu,Claus:1998fh,Bergshoeff:1985su,Bergshoeff:1987cm}.\ The superfield theories of interest are the so-called {\bf super-$\si$-models},\ whose structure resembles closely that of the two-dimensional $\si$-model with the WZ term,\ given in \eqref{eq:DFampl}.\ In these theories,\ the mapping space $\,[\Si,M]\,$ for the closed (oriented two-dimensional) manifold $\,\Si\,$ and a smooth manifold $\,M\,$ is replaced by the inner-$\Hom$ functor $\,[\Si_p,\cM]\equiv\unl\Hom{}_\sMan(\Si_p,\cM)\,$ for a closed (oriented) $(p+1)$-dimensional (non-graded) manifold $\,\Si_p$,\ termed the ({\bf super-$p$-brane}) {\bf worldvolume},\ and a \emph{super}manifold $\,\cM$.\ Whenever mentioned without a qualifier,\ the latter is to be understood as a ringed space in the sense of Berezin,\ Le\"ites and Kostant \cite{Berezin:1975,Kostant:1975}.\ The functors are to be evaluated on the nested $\bN$-indexed family of superpoints $\,\bR^{0|N},\ N\in\bN$,\ {\it cp.}\ \cite{Freed:1999}.\ The {\bf supertarget} $\,\cM\,$ comes with an action $\,\la\ :\ \txG\x\cM\too\cM\,$ of a Lie supergroup $\,\txG\,$ \cite{Kostant:1975,Carmeli:2011},\ interpreted as the {\bf supersymmetry group} of the superfield theory,\ and with a number of \emph{even} $\txG$-invariant tensor fields:\ a (pseudo-)metric field\footnote{Here,\ $\,\cT^{(*)}\cM\,$ is the (co)tangent sheaf of $\,\cM$,\ and $\,\widehat\ox\,$ denotes the $\bZ/2\bZ$-graded tensor product.} $\,\txg\in\G(\cT^*\cM\widehat\ox{}_{\cM,\bR}^{\rm sym}\cT^*\cM)_0\,$ and superdifferential-form fields coupling to the supercurrents determined by the propagation of the super-$p$-brane.

An important special class of super-$\si$-models consists of those with supertargets given by homogeneous spaces $\,\cM=\txG/\txH\,$ of the supersymmetry group \cite{Kostant:1975,Carmeli:2011} relative to even subgroups $\,\txH\subset|\txG|\,$ associated with \emph{reductive} decompositions $\,\ggt=\tgt\oplus\hgt\,$ of the tangent Lie superalgebra $\,\ggt={\rm sLie}\txG\,$ into the tangent Lie algebra $\,\hgt={\rm Lie}\txH\,$ and its $\bZ/2\bZ$-graded $\ad$-module direct-sum complement $\,\tgt$.\ In this class,\ background fields are customarily modelled on pullbacks of $\txH$-basic tensors on $\,\txG\,$ along sections of the principal $\txH$-bundle $\,\txG\too\txG/\txH$.\ The existence of the latter supermanifold morphisms can,\ with a little effort,\ be read off \cite{Suszek:2020xcu} from Kostant's seminal work \cite{Kostant:1975},\ very lucidly deciphered in this context by Fioresi {\it et al.} in \cite{Fioresi:2007zz,Carmeli:2011}.\ The tensors of interest are linear combinations of the $\tgt$-components $\,\theta_{\rm L}^{\unl a},\ \unl a\in\ovl{1,\dim\,\tgt}\,$ of the left-invariant (LI) Maurer--Cartan 1-form $\,\theta_{\rm L}\in\Om^1(\txG)\ox_\bR\ggt\,$ with $\sfT_e\Ad_\txH$-invariant tensors as (constant) coefficients.\ In this approach,\ supersymmetry is automatically ensured by the left-invariance of $\,\theta_{\rm L}$,\ which in the supergeometric setting is encoded by the conditions $\,|\ell|_g^*\theta_{\rm L}=\theta_{\rm L}\,$ and $\,\pLie{R_X}\theta_{\rm L}=0\,$ written,\ for arbitrary $\,(g,X)\in|\txG|\x\ggt$,\ in terms of the left action $\,|\ell|_g=\txm\circ(\widehat g\x\id_\txG)\,$ of the body Lie group induced from the left regular action $\,\ell\equiv\txm\,$ of $\,\txG\,$ on itself with the help of the topological points $\,\widehat g:\bR^{0|0}\too\txG$,\ and in terms of the right-invariant (RI) vector fields $\,R_X=(X\ox\id_{\cO_\txG})\circ\txm^*\,$ on $\,\txG$.\ This class of super-$\si$-models comprises the original super-$\si$-models with the super-minkowskian supertargets \cite{Green:1983wt,Green:1983sg,Achucarro:1987nc},\ as well as those with the super-${\rm AdS}_p\x\bS^q\,$ supertargets \cite{Metsaev:1998it,Park:1998un,Zhou:1999sm}.\ In all these models,\ the relevant even $p$-cocycle $\,\chi\,$ admits a de Rham primitive,\ so that the symmetry analysis developed in the previous section can be adapted directly.\smallskip

In what follows,\ we restrict our attention to the topologically trivial super-minkowskian backgrounds,\ which allows us to identify the novel features and establish a \emph{higher}-supergeometric interpretation of the supersymmetric refinement of the de Rham cohomology.\ An additional simplification comes from the fact that $\,\bR^{d,1|D_{d,1}}\equiv{\rm sMink}(d,1|D_{d,1})={\rm sISO}(d,1|D_{d,1})/{\rm Spin}(d,1)\,$ is itself a Lie supergroup with \emph{global} generators of the structure sheaf (super-coordinates),\ to be denoted as $\,x^a,\ a\in\ovl{0,d}\,$ (even) and $\,\theta^\a,\ \a\in\ovl{1,D_{d,1}}\,$ (odd).\ The odd generators carry indices of a Majorana--Weyl representation of (the group $\,{\rm Spin}(d,1)\,$ in) the Clifford algebra $\,\Cliff(\bR^{d,1})\,$ of the body Lie group $\,\bR^{d,1}\equiv{\rm Mink}(d,1)\,$ described at length in App.\,\ref{app:sISO},\ with generators  $\,\G_a\,$ assumed to satisfy the {\bf Fierz identities} \eqref{eq:Fierz},\ which imposes constraints on the admissible values of $\,p\,$ and $\,d\,$ (these compose the so-called `brane scan' of \cite{Achucarro:1987nc}).\ The latter identity ensures de Rham closedness of the distinguished {\bf Green--Schwarz} ({\bf GS}) {\bf super-$(p+2)$-cocycle}
\qq\label{eq:GS-cocyc}
\chi_{(p+2)}=\left\{ \barr{cl} \si^\a\wedge\bigl(\ovl\G{}_{a_1 a_2\ldots a_p}\bigr)_{\a\b}\,\si^\b\wedge e^{a_1 a_2\ldots a_p} & \ \tx{if}\ p\in\ovl{1,10}\\ \si^\a\wedge\bigl(\ovl\G{}_{11}\bigr)_{\a\b}\,\si^\b & \ \tx{if}\ p=0\earr \right.\,,
\qqq 
which determines the {\bf Green--Schwarz super-$\si$-model} together with the (pseudo-)metric field 
\qq\nn
\txg=\eta_{ab}\,e^a\ox e^b\,,
\qqq
both having been expressed in terms of the basis ($\bR^{d,1|D_{d,1}}$-)LI 1-forms $\,\si^\a\,$ and $\,e^a$,\ explicited in App.\,\ref{app:sISO}.\ In the presence of the global generators of the structure sheaf of $\,\bR^{d,1|D_{d,1}}$,\ it is customary (and convenient) to pass to the so-called $\cS$-point picture \cite{Carmeli:2011} in the description of left-invariance ({\it i.e.},\ supersymmetry),\ whereby the former conditions -- separate for the body Lie group and for the tangent Lie superalgebra (adapted to Kostant's formulation of Lie-supergroup theory in terms of the super-Harish--Chandra pairs) -- are replaced by the familar ones:\ 
\qq\nn
\ell_{(\vep,t)}^*\chi_{(p+2)}=\chi_{(p+2)}\,,\qquad\ell_{(\vep,t)}^*\txg=\txg\,.
\qqq 
Here,\ $\,\vep\,$ resp.\ $\,t\,$ represents the Gra\ss mann-odd resp.\ -even generators of the structure sheaf of the acting supersymmetry group $\,\txT$,\ and it is with this understanding that we write $\,(\vep,t)\in\sfT$.\ For the sake of notational economy,\ we employ the notation $\,\txT\equiv{\rm sMink}(d,1|D_{d,1})\,$ and $\,\tgt\equiv\gt{smink}(d,1|D_{d,1})$,\ and stick to the $\cS$-point picture (with $\,\cS\,$ implicit) in the remainder of the present paper.

It is not hard to convince oneself that the GS super-$(p+2)$-cocycles define nontrivial classes in $\,{\rm CE}^{p+2}(\tgt)\,$ -- an assumption to the contrary readily leads to a contradiction,\ an example of such a reasoning can be found in \cite{deAzcarraga:1989vh}.\ The standard de Rham cohomology of $\,\txT$,\ on the other hand,\ is trivial by Kostant's Theorem \cite[Thm.\,4.7]{Kostant:1975},\ and so we have
\berop
The GS super-$(p+2)$-cocycle $\,\chi_{(p+2)}\,$ of \Reqref{eq:GS-cocyc} admits a manifestly $|\txT|$-invariant de Rham primitive given by the coordinate expression
\qq\hspace{-.25cm}
\b_{(p+1)}(\theta,x)=\left\{ \barr{cl} \tfrac{1}{p+1}\,\sum_{k=0}^p\,\ovl B{}_{a_1 a_2\ldots a_p}(\theta)\wedge\sfd x^{a_1 a_2\ldots a_k}\wedge e^{a_{k+1} a_{k+2}\ldots a_p}(\theta,x) & \tx{if}\ p\in\ovl{1,10}\\ \ovl\theta\,\G_{11}\,\si(\theta) & \ \tx{if}\ p=0
\earr \right.\cr\label{eq:GSprim}
\qqq
in which $\,\ovl B{}_{a_1 a_2\ldots a_p}(\theta)=\theta\,\ovl\G_{a_1 a_2\ldots a_p}\,\si(\theta)$,\ and $\,\sfd x^{a_1 a_2\ldots a_k}\equiv\sfd x^{a_1}\wedge\sfd x^{a_2}\wedge\cdots\wedge\sfd x^{a_k}\,$ for $\,k>0\,$ (and is set equal to $\,1\,$ for $\,k=0$).
\eerop
\beroof
The case $\,p=0\,$ is trivial.\ Hence,\ assume $\,p>0\,$ to obtain
\qq\nn
\chi_{(p+2)}(\theta,x)=\sfd S_0(\theta,x)+\tfrac{p}{2}\,\bigl(\ovl B{}_{a_1 a_2\ldots a_p}\wedge\si\wedge\ovl\G{}^{a_1}\,\si\bigr)(\theta)\wedge e^{a_2 a_3\ldots a_p}(\theta,x)\,,
\qqq
with $\,S_0(\theta,x)=\ovl B{}_{a_1 a_2\ldots a_p}(\theta)\wedge e^{a_1 a_2\ldots a_p}(\theta,x)$.\ We may next use the Fierz identity \eqref{eq:Fierz} in the form $\,(\ovl\G_{a_1 a_2\ldots a_p})_{\a(\b}\,\ovl\G{}^{a_1}_{\g\d)}=-\ovl\G{}^{a_1}_{\a(\b}\,(\ovl\G_{a_1 a_2\ldots a_p})_{\g\d)}\,$ to rewrite the last equality as
\qq\nn
\chi_{(p+2)}(\theta,x)=\sfd S_0(\theta,x)+p\,\sfd\ovl B{}_{a_1 a_2\ldots a_p}(\theta)\wedge\sfd x^{a_1}\wedge e^{a_2 a_3\ldots a_p}(\theta,x)-p\,\chi_{(p+2)}(\theta,x)\,,
\qqq
whence
\qq\nn
\chi_{(p+2)}(\theta,x)=\tfrac{1}{p+1}\,\sfd S_0(\theta,x)+\tfrac{p}{p+1}\,\sfd x^{a_1}\wedge\D^1_{a_1}(\theta,x)
\qqq
with $\,\D^1_{a_1}(\theta,x)=\sfd\ovl B{}_{a_1 a_2\ldots a_p}(\theta)\wedge e^{a_2 a_3\ldots a_p}(\theta,x)$.\ Reasoning as in the previous step,\ we find
\qq\nn
\D^1_{a_1}(\theta,x)=\tfrac{1}{p}\,\sfd\unl S{}_1(\theta,x)+\tfrac{p-1}{p}\,\sfd x^{a_2}\wedge\D^2_{a_1 a_2}(\theta,x)
\qqq
with $\,\unl S{}_1(\theta,x)=\ovl B{}_{a_1 a_2\ldots a_p}(\theta)\wedge e^{a_2 a_3\ldots a_p}(\theta,x)\,$ and $\,\D^2_{a_1 a_2}(\theta,x)=\sfd\ovl B{}_{a_1 a_2\ldots a_p}(\theta)\wedge e^{a_3 a_4\ldots a_p}(\theta,x)$.\ Repeating the above reduction procedure $p$ times,\ we eventually arrive at the equality $\,\D^1_{a_1}=\sfd\unl\b{}_{a_1}\,$ with $\,\unl\b{}_{a_1}(\theta,x)=\tfrac{1}{p}\,\sum_{k=1}^p\,\ovl B{}_{a_1 a_2\ldots a_p}(\theta)\wedge\sfd x^{a_2}\wedge\cdots\wedge\sfd x^{a_k}\wedge e^{a_{k+1} a_{k+2}\ldots a_p}(\theta,x)$,\ from which the claim of the proposition follows.
\eroof

The above yields the following explicit form of the GS super-$\si$-model (inner-$\Hom$) action functional for the super-minkowskian super-$p$-brane:
\qq\nn
S_{\rm GS}[\xi]=\int_{\Si_p}\,\Vol(\Si_p)\,\sqrt{\bigl|\det\,\bigl(\eta_{ab}\,\xi^*e^a\,\xi^*e^b\bigr)\bigr|}+q_p\,\int_{\Si_p}\,\xi^*\b_{(p+1)}\,,
\qqq
with $\,\xi\equiv(\theta,x)\,$ from $\,[\Si_p,\txT]\,$ and with the charge $\,q_p\,$ fixed \cite{McArthur:1999dy} by the mechanism of restoration of supersymmetry in the vacuum of the theory,\ discovered in \cite{deAzcarraga:1982dhu,Siegel:1983hh} and known under the name of $\kap$-symmetry,\ {\it cp.}\ \cite{Suszek:2020xcu} for a supergeometric elucidation in the dual topological Hughes--Polchinsky model of \cite{Hughes:1986dn,Gauntlett:1989qe}.\ The super-$\si$-model thus defined is supersymmetric:\ The metric term of the lagrangean `density' is manifestly invariant under $\,\ell$,\ whereas the topological term is properly \emph{quasi}-invariant.\ Indeed,\ the primitive $\,\b_{(p+1)}\,$ is non-LI,\ but its exterior derivative is,\ whence $\,\sfd(\ell_{(\vep,t)}^*\b_{(p+1)}-\b_{(p+1)})=\ell_{(\vep,t)}^*\chi_{(p+2)}-\chi_{(p+2)}=0$,\ and so invoking Kostant's Theorem once more,\ we infer the existence of $p$-forms $\,\jmath_{(p)\,(\vep,t)}\,$ such that 
\qq\label{eq:quasi-beta}
\ell_{(\vep,t)}^*\b_{(p+1)}-\b_{(p+1)}=\sfd\jmath_{(p)\,(\vep,t)}\,.
\qqq
The latter determine an action of the supersymmetry group $\,\txT\,$ on the states of the quantised theory in the geometric approach,\ which is readily demonstrated for\footnote{For $\,p>1$,\ we should simply replace $\,\bD^2\,$ by a $(p+1)$-dimensional `cap' $\,D^{p+1}\,$ with the boundary $\,\p D^{p+1}=\xcC_p\,$ given by a single-$p$-brane Cauchy hypersurface $\,\xcC_p\subset\Si_p$.} $\,p=1$:\ The behaviour of the relevant disc DF amplitude under a supersymmetry transformation,\ $\,\cA_{\rm DF}^{\bD^2}[\ell_{(\vep,t)}\circ\xi]=\cA_{\rm DF}^{\bD^2}[\xi]\cdot\ee^{\sfi\,\int_{\bS^1}\,\xi^*\jmath_{(1)\,(\vep,t)}}$,\ gives rise to a `phase' correction 
\qq\nn
\bigl(R(\vep,t)\Psi\bigr)[\xi]:=c_{(\vep,t)}[\xi]\cdot\Psi\bigl[\ell_{(\vep,t)^{-1}}\circ\xi\bigr]\,,\qquad\qquad c_{(\vep,t)}[\xi]:=\ee^{\sfi\,\int_{\bS^1}\,(\la_{(\vep,t)^{-1}}\circ\xi)^*\jmath_{(1)\,(\vep,t)}}
\qqq
of the standard (pullback) realisation of $\,\txT\,$ on (disc) states in the position polarisation ($\xi\in[\bS^1,\txT]$).\ Accordingly, we find, for arbitrary $\,(\vep_1,t_1),(\vep_2,t_2)\in\txT$,\ the identity $\,R(\vep_1,t_1)\circ R(\vep_2,t_2)=(\d_\txT c)_{(\vep_1,t_1),(\vep_2,t_2)}[\cdot]\cdot R((\vep_1,t_1)\cdot(\vep_2,t_2))\,$ with the homomorphicity 2-cocycle $\,(\d_\txT c)_{(\vep_1,t_1),(\vep_2,t_2)}\,$ on $\,\txT\,$ given by the exponentiated integral over $\,\bS^1\,$ of the pullback of the current 2-cocycle $\,(\d_{\txG_\si}\jmath)_{(\vep_1,t_1),(\vep_2,t_2)}:=\la_{(\vep_2,t_2)}^*\jmath_{(\vep_1,t_1)}-\jmath_{(\vep_1,t_1)\cdot(\vep_2,t_2)}+\jmath_{(\vep_2,t_2)}\,$ along $\,\xi$.\ The latter is de Rham closed by \eqref{eq:quasi-beta},\ and hence exact,\ which implies strict homomorphicity of $\,R$.\ Thus,\ up to now,\ there seems to be no nontrivial field-theoretic effect of the non-exactness of the class of the GS super-$(p+2)$-cocycle in the relevant cohomology $\,{\rm CaE}^{p+2}(\txT)$.\ In the next section,\ we refine our cohomological analysis of the GS background,\ and discover thereby a higher-geometric structure over $\,\txT\,$ naturally associated with $\,\chi_{(p+2)}$.\ The structure shall subsequently be demonstrated  to effectively encode the topology of the soul of a generalised supermanifold `underlying' the target superspace $\,\sfT\,$ of the GS super-$\si$-model.

\section{Supergerbes from $H^2$-ladder Lie-superalgebra extensions}\label{sec:dAzcladder}

In the present section,\ we delineate and carry out in detail the central higher-geometric \emph{construction} of the paper,\ to wit,\ a manifestly supersymmetric geometrisation of the distinguished GS super-$(p+2)$-cocycles in the Cartan--Eilenberg cohomology $\,{\rm CaE}^\bullet(\txT)\,$ of the super-Minkowski space $\,\txT\equiv{\rm sMink}(d,1|D_{d,1})$.\ These super-$(p+2)$-cocycles can be mapped to the Chevalley--Eilenberg cohomology $\,{\rm CE}^\bullet(\tgt)\,$ of the corresponding Lie superalgebra $\,\tgt=\gt{smink}(d,1|D_{d,1})$,\ which enables us to perform their \emph{sequential} trivialisation through recursive application,\ to cohomologically nontrivial rank-2 $\wedge$-subfactors of the super-$(p+2)$-cocycles,\ of the classic 1-1 correspondence between -- on the one hand -- classes in the cohomology group $\,H^2(\ggt,\agt)\,$ of a Lie superalgebra $\,\ggt\,$ with values in its trivial supercommutative module $\,\agt$,\ and -- on the other hand -- equivalence classes of supercentral extensions of $\,\ggt\,$ through $\,\agt$,\ {\it cp.}\ Thm.\,\ref{thm:H2ext}.\ The ensuing trivialising extension of $\,\tgt$,\ generically no longer supercentral,\ may next,\ under propitious circumstances,\ be integrated to a Lie-supergroup extension $\,\sfY\txT\too\txT$,\ {\it i.e.},\ an epimorphism in the category $\,\sLieGrp\,$ of Lie supergroups suggested by our physically motivated choice of the Cartan--Eilenberg cohomology in which we work.\ The total space of the epimorphism then becomes the carrier of a primitive of the pullback of the original GS super-$(p+2)$-cocycle in that cohomology,\ (the binary operation of) the Lie-supergroup structure on $\,\sfY\txT\,$ being essentially \emph{fixed} by the assumed left-invariance of that primitive.\ To the thus derived {\bf trivial CaE super-$p$-gerbe} (over $\,\sfY\txT$) we may subsequently apply the standard reconstruction algorithm over $\,\sfN_\bullet{\rm Pair}_\txT(\sfY\txT)\,$ in which appropriate cells of the $p$-category of CaE super-$(p-1)$-gerbes are to be built successively,\ once again with the help of the recursive application of Thm.\,\ref{thm:H2ext} and integration to the Lie-supergroup level.\ Prior to delving into the technicalities of the geometrisation scheme,\ let us summarise its very general description presented hereabove by stating that its end product -- the {\bf Cartan--Eilenberg super-$p$-gerbe} -- is,\ morally,\ a $p$-gerbe object in $\,\sLieGrp$,\ represented by a Murray diagram with 
\bit
\item all nodes given by Lie supergroups;
\item all arrows representing Lie-supergroup epimorphisms;
\item all tensorial components of the connective structure (super-$k$-forms,\ with $\,k\in\ovl{1,p+2}$,\ the top-rank one being the initial GS super-$(p+2)$-cocycle) LI with respect to the regular action of their Lie-supergroup support on itself,
\eit
{\it cp.}\ the key to the original Murray diagram given on p.\,\pageref{p:Murray}.

Just like their non-graded prototypes (in the standard de Rham cohomology),\ CaE super-$p$-gerbes form a hierarchy (relative to the rank of the underlying curvature super-$p$-cocycle),\ which starts with geometrisations of super-2-cocycles.\ Putting together the above considerations and the results of \cite{Tuynman:1987ij},\ we arrive at
\bedef\label{def:CaEs0g}
Let $\,\txG\,$ be a Lie supergroup and let $\,\ggt\,$ be its tangent Lie superalgebra (of left-invariant vector fields). Assume given a de Rham super-2-cocycle $\,\chi_{(2)}\,$ on $\,\txG\,$ representing a class in $\,{\rm CaE}^2(\txG)$.\ A {\bf Cartan--Eilenberg super-0-gerbe} over $\,\txG\,$ of curvature $\,\chi_{(2)}\,$ is a triple  
\qq\nn
\cG^{(0)}_{\rm CaE}:=\bigl(\sfY\txG,\pi_{\sfY\txG},\b_{(1)}\bigr)
\qqq
composed of 
\bit
\item a principal $\bC^\x$-bundle\footnote{For principal bundles in $\,\sMan$,\ {\it cp.}\ \cite{Kessler:2019bwp,Bartocci:sm2012}.} $\,\pi_{\sfY\txG}:\sfY\txG\too\txG$;
\item a principal connection 1-form $\,\b_{(1)}\,$ on it,\ trivialising the pullback of the curvature super-2-form $\,\chi_{(2)}\,$ along the projection to the base,\ $\,\pi_{\sfY\txG}^*\chi_{(2)}=\sfd\b_{(1)}$,
\eit
in which the total space $\,\sfY\txG\,$ of the bundle carries the structure of a Lie supergroup (with the binary operation $\,\sfY\txm$) that extends that on $\,\txG$,\ as captured by the short exact sequence of Lie supergroups
\qq\nn
\bd1\too\bC^\x\xrightarrow{\ \sfY\jmath\ }\sfY\txG\xrightarrow{\ \pi_{\sfY\txG}\ }\txG\too\bd1
\qqq
which integrates the short exact sequence of Lie superalgebras 
\qq\nn
\bd1\too\bR\too\sfY\ggt\too\ggt\too\bd1
\qqq
determined by $\,\chi_{(2)}\,$ along the lines of Thm.\,\ref{thm:H2ext},\ and such that the connection 1-form $\,\b_{(1)}\,$ is invariant under the left regular action of the extended Lie supergroup $\,\sfY\txG\,$ upon itself,\ $\,\sfY\ell\equiv\sfY\txm\ :\ \sfY\txG\x\sfY\txG\too\sfY\txG$.

Given CaE super-0-gerbes $\,\cG^{(0)\,A}_{\rm CaE}=(\sfY_A\txG,\pi_{\sfY_A\txG},\b_{(1)}^A),\ A\in\{1,2\}\,$ over a common base $\,\txG$,\ an {\bf isomorphism} between them is an isomorphism of the principal $\bC^\x$-bundles $\,\varphi:\sfY_1\txG\xrightarrow{\ \cong\ }\sfY_2\txG\,$ which preserves the connection,\ $\,\varphi^*\underset{\tx{\ciut{(1)}}}{\txa_2}=\underset{\tx{\ciut{(1)}}}{\txa_1}$,\ and is,\ at the same time,\ an isomorphism of the respective Lie supergroups which defines an equivalence between the two extensions,\ as described by the commutative diagram in $\,\sLieGrp$:
\qq\nn
\alxydim{@C=1cm@R=1cm}{ & & \sfY_1\txG \ar[dd]^{\varphi}_{\cong} \ar[dr]^{\pi_{\sfY_1\txG}} & & \\ \bd1 \ar[r] & \bC^\x \ar[ur]^{\sfY_1\jmath} \ar[dr]_{\sfY_2\jmath} & & \txG \ar[r] & \bd1 \\ & & \sfY_2\txG \ar[ur]_{\pi_{\sfY_2\txG}} & & }\,.
\qqq
\exdef
\noindent On the next rung of the hierarchy,\ we find
\bedef\label{def:CaEs1g}
Adopt the notation of Def.\,\ref{def:CaEs0g}.\ Assume given a super-3-cocycle $\,\chi_{(3)}\,$ on $\,\txG\,$ representing a class in $\,{\rm CaE}^3(\txG)$.\ A {\bf Cartan--Eilenberg super-1-gerbe} over $\,\txG\,$ of curvature $\,\chi_{(3)}\,$ is a septuple  
\qq\nn
\cG^{(1)}_{\rm CaE}:=\bigl(\sfY\txG,\pi_{\sfY\txG},\b_{(2)},\Lx,\pi_\Lx,\cA_{\Lx\,(1)},\mu_\Lx\bigr)
\qqq
composed of 
\bit
\item a surjective submersion $\,\pi_{\sfY\txG}:\sfY\txG\too\txG\,$ with a structure of a Lie supergroup on its total space (with the binary operation $\,\sfY\txm$) mapped onto that on $\,\txG\,$ by the Lie-supergroup epimorphism $\,\pi_{\sfY\txG}\,$ integrating a Lie-superalgebra epimorphism $\,\sfT\pi_{\sfY\txG}:\sfY\ggt\too\ggt$;
\item a global primitive $\,\b_{(2)}\,$ of the pullback of $\,\chi_{(3)}\,$ to it,\ $\,\pi_{\sfY\txG}^*\chi_{(3)}=\sfd\b_{(2)}$,\ which is LI with respect to the induced left regular action of $\,\sfY\txG\,$ on itself $\,\sfY\ell\equiv\sfY\txm:\sfY\txG\x\sfY\txG\too\sfY\txG$;
\item a CaE super-0-gerbe $\,(\Lx,\pi_\Lx,\cA_{\Lx\,(1)})\,$ over the fibred square $\,\sfY^{[2]}\txG\equiv\sfY\txG\x_\txG\sfY\txG\,$ (endowed with the Lie-supergroup structure induced from the product one on $\,\sfY\txG^{\x 2}$),\ with a principal connection 1-form $\,\cA_{\Lx\,(1)}\,$ of curvature $\,(\pr_2^*-\pr_1^*)\b_{(2)}$;
\item an isomorphism of CaE super-0-gerbes\footnote{Note that pullback along a canonical projection is consistent with the definition of a super-0-gerbe due to its equivariance.} $\,\mu_\Lx\ :\ \pr_{1,2}^*\Lx\ox\pr_{2,3}^*\Lx\xrightarrow{\ \cong\ }\pr_{1,3}^*\Lx\,$ over the fibred cube $\,\sfY^{[3]}\txG\equiv\sfY\txG\x_\txG\sfY\txG\x_\txG\sfY\txG\,$ which satisfies the coherence (associativity) condition $\,\pr_{1,2,4}^*\mu_\Lx\circ(\id_{\pr_{1,2}^*\Lx}\ox\pr_{2,3,4}^*\mu_\Lx)=\pr_{1,3,4}^*\mu_\Lx\circ(\pr_{1,2,3}^*\mu_\Lx\ox \id_{\pr_{3,4}^*\Lx})\,$ over the quadruple fibred product $\,\sfY^{[4]}\txG\equiv\sfY\txG\x_\txG\sfY\txG\x_\txG\sfY\txG\x_\txG\sfY\txG$.
\eit

Given CaE super-1-gerbes $\,\cG^{(1)\,A}_{\rm CaE}=(\sfY_A\txG,\pi_{\sfY_A\txG},\b^A_{(2)},\Lx_A,\pi_{\Lx_A},\cA_{\Lx_A\,(1)},\mu_{\Lx_A})$,\ $A\in\{1,2\}\,$ over a common base $\,\txG$,\ a 1-{\bf isomorphism} between them is a sextuple
\qq\nn
\Phi^{(1)}_{\rm CaE}:=\bigl(\sfY\sfY_{1,2}\txG,\pi_{\sfY\sfY_{1,2}\txG},\txE,\pi_\txE,\cA_{\txE\,(1)},\a_\txE\bigr)\ :\ \cG^{(1)\,1}_{\rm CaE}\xrightarrow{\ \cong\ }\cG^{(1)\,2}_{\rm CaE}
\qqq
composed of 
\bit
\item a surjective submersion $\,\pi_{\sfY\sfY_{1,2}\txG}\ :\ \sfY\sfY_{1,2}\txG\too\sfY_1\txG\x_\txG\sfY_2\txG\equiv\sfY_{1,2}\txG\,$ with a structure of a Lie supergroup on its total space which lifts the product Lie-supergroup structure on the fibred product $\,\sfY_{1,2}\txG\,$ along the Lie-supergroup epimorphism $\,\pi_{\sfY\sfY_{1,2}\txG}$; 
\item a CaE super-0-gerbe $\,(\txE,\pi_\txE,\cA_{\txE\,(1)})\,$ over $\,\sfY\sfY_{1,2}\txG$,\ with a principal $\bC^\x$-connection super-1-form $\,\cA_{\txE\,(1)}\,$ of curvature $\,\pi_{\sfY\sfY_{1,2}\txG}^*(\pr_2^*\b_{(2)}^2-\pr_1^*\b_{(2)}^1)$;
\item an isomorphism of CaE super-0-gerbes $\,\a_\txE\ :\ \pi_{\sfY\sfY_{1,2}\txG}^{\x 2\,*}\pr_{1,3}^*\Lx_1\ox\pr_2^*\txE\xrightarrow{\ \cong\ }\pr_1^*\txE\ox\pi_{\sfY\sfY_{1,2}\txG}^{\x 2\,*}\pr_{2,4}^*\Lx_2\,$ over the fibred product $\,\sfY^{[2]}\sfY_{1,2}\txG=\sfY\sfY_{1,2}\txG\x_\txG\sfY\sfY_{1,2}\txG$,\ subject to the coherence constraint expressed by Axiom (1M2) in \cite[Def.\,1.4]{Waldorf:2007mm},\ regarded as a relation between isomorphisms of CaE super-0-gerbes over the fibred product $\,\sfY^{[3]}\sfY_{1,2}\txG\equiv\sfY\sfY_{1,2}\txG\x_\txG\sfY\sfY_{1,2}\txG\x_\txG\sfY\sfY_{1,2}\txG$.
\eit

Given a pair of 1-isomorphisms $\,\Phi^{(1)\,B}_{\rm CaE}=(\sfY^B\sfY_{1,2}\txG,\pi_{\sfY^B\sfY_{1,2}\txG},\txE_B,\pi_{\txE_B},\cA_{\txE_B\,(1)},$\linebreak $\a_{\txE_B}),\ B\in\{1,2\}\,$ between CaE super-1-gerbes $\,\cG^{(1)\,A}_{\rm CaE}=(\sfY_A \txG,\pi_{\sfY_A \txG},\b_{(2)}^A,\Lx_A,\pi_{\Lx_A},$ $\cA_{\Lx_A\,(1)},\mu_{\Lx_A}),\ A\in\{1,2\}$,\ a 2-isomorphism between the latter is a triple
\qq\nn
\varphi^{(1)}_{\rm CaE}=(\sfY\sfY^{1,2}\sfY_{1,2}\txG,\pi_{\sfY\sfY^{1,2}\sfY_{1,2}\txG},\b)\ :\ \Phi^{(1)\,1}_{\rm CaE}\overset{\cong}{\Longrightarrow}\Phi^{(1)\,2}_{\rm CaE}
\qqq
composed of 
\bit
\item a surjective submersion $\,\pi_{\sfY\sfY^{1,2}\sfY_{1,2}\txG}\ :\ \sfY\sfY^{1,2}\sfY_{1,2}\txG\too\sfY^1\sfY_{1,2}\txG\x_{\sfY_{1,2}\txG}\sfY^2\sfY_{1,2}\txG\equiv\sfY^{1,2}\sfY_{1,2}\txG\,$ with a structure of a Lie supergroup which lifts the product Lie-supergroup structure on the fibred product $\,\sfY^{1,2}\sfY_{1,2}\txG\,$ along the Lie-supergroup epimorphism $\,\pi_{\sfY\sfY^{1,2}\sfY_{1,2}\txG}$;
\item an isomorphism of CaE super-0-gerbes $\,\b\ :\ (\pr_1\circ\pi_{\sfY\sfY^{1,2}\sfY_{1,2}\txG})^*\txE_1\xrightarrow{\ \cong\ }(\pr_2\circ\pi_{\sfY\sfY^{1,2}\sfY_{1,2}\txG})^*\txE_2\,$ subject to the coherence constraint expressed by Axiom (2M) in \cite[Def.\,3]{Waldorf:2007mm},\ regarded as a relation between isomorphisms of CaE super-0-gerbes over $\,\sfY^{[2]}\sfY^{1,2}\sfY_{1,2}\txG$.
\eit
\exdef

The above concrete definitions shall serve as a reference in our subsequent analysis,\ which is why we write them out in all their glory.\ Having done so,\ we pass to examine the details of the geometrisations of the physically distinguished GS super-$(p+2)$-cocycles $\,\chi_{(p+2)}\,$ of \eqref{eq:GS-cocyc} on $\,\txG=\txT$,\ a task that we undertake for $\,p\in\{0,1,2\}\,$ in order to illustrate the general construction.\ The latter has an essentially \emph{non}-algorithmic component,\ studied at great length by de Azc\'arraga {\it et al.}\ in \cite{Chryssomalakos:2000xd,deAzcarraga:2001fi,deAzcarraga:2005jd} ({\it cp.}\ also \cite{Bergshoeff:1995hm}),\ which consists in extracting from a given non-exact CaE super-$(k+2)$-cocycle (with $\,k>0$) the aforementioned cohomologically nontrivial rank-2 $\wedge$-subfactors,\ marking ($H^2$-)stages of its sequential trivialisation.\ Its non-algorithmic nature makes it necessary to meticulously trace a particular path chosen in the space of admissible trivialisations,\ which becomes rather lengthy and relatively (technically) complex already for $\,p=2$,\ so much so that we restrict our construction to the three cases indicated.\ It stands to reason,\ though,\ that the construction works for \emph{all} cases covered by the `brane scan',\ and thus gives rise to a complete physical hierarchy of Green--Schwarz CaE super-$p$-gerbes.

As an instructive prelude to the three higher-supergeometric constructions of immediate physical interest,\ we briefly review -- after \cite{Aldaya:1984gt,Chryssomalakos:2000xd} -- the construction of the super-Minkowski target $\,\txT\,$ itself as a central extension of the abelian Lie supergroup $\,\bR^{0|D_{d,1}}$,\ a.k.a.\ the superpoint,\ with global Gra\ss mann-odd coordinates $\,\{\theta^\a\}^{\a\in\ovl{1,D_{d,1}}}$.\ On the latter,\ we find a family of manifestly LI super-2-cocycles
\qq\nn
\chi_{(2)}^a:=\tfrac{1}{2}\,\vartheta_{\bR^{0|D_{d,1}}}^\a\wedge\ovl\G{}^a_{\a\b}\,\vartheta_{\bR^{0|D_{d,1}}}^\b\,,\qquad a\in\ovl{0,d}\,,
\qqq
written in terms of the canonical LI 1-forms $\,\vartheta_{\bR^{0|D_{d,1}}}^\a(\theta)=\sfd\theta^\a$.\ These do not admit primitives on $\,\bR^{0|D_{d,1}}\equiv\txT_0\,$ invariant with respect to the action $\,\ell^{(0)}_\cdot\equiv\txm_0:\txT_0^{\x 2}\too\txT_0:(\vep^\a,\theta^\a)\longmapsto\theta^\a+\vep^\a$.\ Equivalently,\ the 2-cocycles $\,\om_{(2)}^a=\frac{1}{2}\,\ovl\G{}^a_{\a\b}\,q^\a\wedge q^\b\,,\ a\in\ovl{0,d}\,$ on the supercommutative Lie superalgebra
\qq\nn
\tgt_0=\bigoplus_{\a=1}^{D_{d,1}}\corr{Q_\a}\equiv\bR^{\x D_{d,1}}[1]\,,\qquad\{Q_\a,Q_\b\}=0\,,
\qqq
with $\,q^\a(Q_\b)=\d^\a_{\ \b}$,\ do not trivialise in the relevant cohomology $\,{\rm CE}^2(\tgt_0)$,\ and so we are led to consider the central extension ({\it cp.}\ \eqref{eq:H2ext-expl})\label{p:spoint-ext-smink}
\qq\nn
&\brd0\too\bR^{\x d+1}\too\sfY\tgt_0\xrightarrow{\ \pi_{\sfY\tgt_0}\ }\tgt_0\too\brd0\,,&\cr\cr
&\sfY\tgt_0=\bigoplus_{\a=1}^{D_{d,1}}\corr{Q_\a}\oplus\bigoplus_{a=0}^d\corr{P_a}\equiv\tgt_0\oplus\bR^{\x d+1}\,,\qquad\qquad\pi_{\sfY\tgt_0}=\pr_1\,,&\cr\cr
&\{Q_\a,Q_\b\}_{\sfY\tgt_0}=\{Q_\a,Q_\b\}_{\tgt_0}+\bigl(\om_{(2)}^a\ox P_a\bigr)(Q_\a,Q_\b)=\ovl\G{}^a_{\a\b}\,P_a\,,&\cr\cr
&[Q_\a,P_a]_{\sfY\tgt_0}=0\,,\qquad\qquad[P_a,P_b]_{\sfY\tgt_0}=0\,,&
\qqq
with $\,\pi_{\sfY\tgt_0}^*\om_{(2)}^a=-\d^{(1)}_{\sfY\tgt_0}p^a\,$ for the dual $\,p^a\,$ of $\,P_a$.\ The Lie-superalgebra extension integrates to the central Lie-supergroup extension
\qq\nn
\bd1\too\bR^{\x d+1}\too\sfY\txT_0\xrightarrow{\ \pi_{\sfY\txT_0}\ }\txT_0\too\bd1\,,
\qqq
given by the rank-$(d+1)$ (real) vector bundle
\qq\nn
\pi_{\sfY\txT_0}=\pr_1\ :\ \sfY\txT_0\equiv\txT_0\x\bR^{\x d+1}\too\txT_0\ :\ (\theta^\a,x^a)\longmapsto\theta^\a
\qqq
with fibre coordinates $\,\{x^a\}^{a\in\ovl{0,d}}$.\ The Lie-supergroup structure on the total space $\,\sfY\txT_0\,$ of this surjective submersion is fixed by the demand that the CaE counterparts $\,e^a\in\Om^1(\sfY\txT_0)\,$ of the CE 1-cochains $\,p^a\in\sfY\tgt_0^*\,$ be LI with respect to the left regular action of $\,\sfY\txT_0\,$ on itself.\ Upon expressing the $\,e^a\,$ in terms of the component LI 1-forms $\,\vartheta_{\bR^{\x d+1}}^a\equiv\sfd x^a\,$ on the abelian fibre $\,\bR^{\x d+1}\,$ and the non-LI primitives $\,B_{(1)}^a(\theta)=\frac{1}{2}\,\theta\,\ovl\G{}^a\,\sfd\theta\,$ of the $\,\chi_{(2)}^a\,$ as
\qq\nn
e^a=\pr_2^*\vartheta_{\bR^{\x d+1}}^a+\pr_1^*B_{(1)}^a\,,\qquad\qquad\sfd e^a=\pi_{\sfY\txT_0}^*\chi_{(2)}^a\,,
\qqq
we thus retrieve the familiar group law for $\,\sfY\txT_0$:
\qq
\txm\equiv\sfY\txm_0\ &:&\ \sfY\txT_0\x\sfY\txT_0\too\sfY\txT_0\cr\cr 
&:&\ \bigl(\bigl(\theta_1^\a,x_1^a\bigr),\bigl(\theta_2^\b,x_2^b\bigr)\bigr)\longmapsto\bigl(\theta_1^\a+\theta_2^\a,x_1^a+x_2^a-\tfrac{1}{2}\,\theta_1\,\ovl\G{}^a\,\theta_2\bigr)\,,\label{eq:sact-sMink}
\qqq
which renders the anticipated identity $\,\sfY\txT_0\equiv\txT\,$ manifest,\ and therewith demonstrates the `hidden' nature of the super-Minkowski space:\ $\,{\rm sMink}(d,1|D_{d,1})\,$ is,\ in fact,\ a central extension of the superpoint $\,\bR^{0|D_{d,1}}\,$ associated with the family $\,\{\chi_{(2)}^a\}^{a\in\ovl{0,d}}\,$ of its non-trivial CaE super-2-cocycles.

In the case in hand,\ we do know (and can readily check) that the supermanifold $\,\sfY\txT_0\,$ obtained through the integration of the Lie-superalgebra extension $\,\sfY\tgt_0\,$ detailed above \emph{is} a Lie supergroup.\ In general,\ though,\ we ought to verify the associativity of the binary operation derived from the requirement of left-invariance of the 1-form determined by a quasi-invariant primitive of a CaE 2-cocycle (such as the $\,e^a\,$ for the respective $\,\chi_{(2)}^a$).\ We close the present section by abstracting from the above simple illustration of the universal extension algorithm a simple criterion of associativity,\ which shall be applied throughout our subsequent considerations.\ Thus,\ assume given a CaE 2-cocycle $\,\chi\,$ on a Lie supergroup $\,\txG\,$ (with a binary operation to be denoted by '$\cdot$',\ and with the neutral element $\,1$) with a \emph{trivial} de Rham cohomology in degrees 0,\ 1 and 2.\ Accordingly,\ the 2-cocycle has a de Rham primitive $\,b\in\Om^1(\txG)\,$ which is quasi-$\txG$-invariant,\ and so defines a $\txG$-indexed family of 0-forms $\,\{\D_\vep\}_{\vep\in\txG}$,\ to be referred to as the {\bf quasi-invariance 1-cochain},\ such that $\,\ell_\vep^*b-b=\sfd\D_\vep$.\ We have,\ for arbitrary $\,\vep_1,\vep_2\in\txG$,\ the identity
\qq\nn
\sfd\ell_{\vep_2}^*\D_{\vep_1}=\ell_{\vep_2}^*\bigl(\ell_{\vep_1}^*b-b\bigr)=\bigl(\ell_{\vep_1\cdot\vep_2}^*b-b\bigr)-\bigl(\ell_{\vep_2}^*b-b\bigr)=\sfd\D_{\vep_1\cdot\vep_2}-\sfd\D_{\vep_2}\,,
\qqq
and so we infer \emph{constancy} of maps from the the $\txG^{\x 2}$-indexed family
\qq\nn
f_{\vep_1,\vep_2}:=\D_{\vep_1\cdot\vep_2}-\D_{\vep_2}-\ell_{\vep_2}^*\D_{\vep_1}\,,
\qqq
whence
\qq\nn
f_{\vep_1,\vep_2}=\D_{\vep_1\cdot\vep_2}(1)-\D_{\vep_2}(1)-\D_{\vep_1}(\vep_2)\,.
\qqq
In virtue of Thm.\,\ref{thm:H2ext},\ the 2-cocycle defines a (rank-1) supercentral extension $\,\brd0\too\agt\too\sfY\ggt\xrightarrow{\ \pi_{\sfY\ggt}\ }\ggt\too\brd0\,$ of the tangent Lie superalgebra $\,\ggt\,$ of $\,\txG$,\ which we integrate to the Lie-supergroup level by equipping the cartesian product $\,\txG\x A\equiv\sfY\txG\,$ of $\,\txG\,$ and the 1-(super)dimensional supermanifold $\,A\,$ of the supercommutative Lie group (modelled on $\,\bR\,$ (of appropriate parity) or $\,\bC^\x$,\ and with the tangent Lie superalgebra $\,\agt$) with the binary operation fixed by the requirement of left-invariance of the 1-form 
\qq\nn
e=\pr_2^*\vartheta_A+\pr_1^*b\,,
\qqq
written in terms of the canonical $\agt$-valued LI 1-form on $\,A$,\ which we write as $\,\vartheta_A(\xi)=\sfd\xi\,$ for $\,\xi\,$ a global (\emph{additive}) coordinate on $\,A$.\ The ensuing group law reads
\qq\nn
\sfY\txG\x\sfY\txG\too\sfY\txG\ :\ \bigl((X_1,\xi_1),(X_2,\xi_2)\bigr)\longmapsto\bigl(X_1\cdot X_2,\xi_1+\xi_2-\D_{X_1}(X_2)\bigr)\,,
\qqq
and so its associativity is ensured by the identity $\,\D_{X_1\cdot X_2}(X_3)-\D_{X_2}(X_3)-\D_{X_1}(X_2\cdot X_3)=-\D_{X_1}(X_2)$,\ valid for all $\,X_1,X_2,X_3\in\txG$.\ In its left-hand side,\ we recognise the constant $\,f_{X_1,X_2}$,\ and so we are led to demand that $\,\D_{X_1}(X_2)=-\D_{X_1\cdot X_2}(1)+\D_{X_2}(1)+\D_{X_1}(X_2)$, or,\ equivalently,\ that 
\qq\nn
\forall_{X,Y\in\txG}\ :\ \D_X(1)=\D_Y(1)\,.
\qqq
Besides associativity,\ we need the existence of a (two-sided) unit $\,(1,\xi_e)$,\ which leads to further constraints:
\qq\nn
\xi_e=\D_1(X)=\D_X(1)\,.
\qqq
Finally,\ the existence of a two-sided inverse $\,(X^{-1},\xi^{-1})\,$ for any $\,(X,\xi)\,$ (written in terms of the inverse $\,X^{-1}\,$ on $\,\txG$) requires 
\qq\nn
\D_{X^{-1}}(X)=\D_X\bigl(X^{-1}\bigr)\,,
\qqq
and then
\qq\nn
\xi^{-1}=-\xi+\D_{X^{-1}}(X)+\D_X(1)\,.
\qqq
The simplest way to satisfy the above requirements is stated in
\berop\label{prop:assoconstr}
Adopt the hitherto notation.\ The product supermanifold $\,\txG\x\txA\,$ equipped with the above binary operation is a Lie supergroup with the neutral element $\,(1,\xi_e)=(1,0)\,$ and the inverse $\,(X,\xi)\longmapsto(X^{-1},-\xi+\D_{X^{-1}}(X))\,$ if the quasi-invariance 1-cochain satisfies the conditions
\qq\nn
\D_X(1)=0=\D_1(X)\,,\qquad\qquad\D_{X^{-1}}(X)=\D_X\bigl(X^{-1}\bigr)
\qqq
for all $\,X\in\txG$.
\eerop
\smallskip

\noindent{\bf Remark:} A word is well due at this point regarding the notation used for structural Lie-supergroup operations (the binary operation,\ the inverse {\it etc.}) throughout the paper.\ All Lie supergroups that we shall encounter admit global generators of the respective structure sheaves,\ and so the operations shall be \emph{presented} on the generators.\ Furthermore,\ the $\cS$-point picture shall be adopted \emph{implicitly} throughout,\ resulting in presentations of the familiar type \eqref{eq:sact-sMink} instead of their sheaf-theoretic counterparts
\qq\nn
\txm^*\theta^\a=\theta^\a\ox\bd1+\bd1\ox\theta^\a\,,\qquad\qquad\txm^* x^a=x^a\ox\bd1+\bd1\ox x^a-\tfrac{1}{2}\,\ovl\G{}^a_{\a\b}\,\theta^\a\ox\theta^\b\,,
\qqq
describing the evaluation of the sheaf component $\,\txm^*:\cO_\txT\too\cO_{\txT\x\txT}\cong\cO_\txT\widehat\ox\cO_\txT\,$ on the generators of the structure sheaf $\,\cO_\txT\,$ of $\,\txT$,\ {\it cp.}\ \cite[Ex.\,7.2.4]{Carmeli:2011}.

\subsection{The super-0-gerbe for the super-$0$-brane}

We begin our case-by-case investigation of geometrisations of the GS super-$(p+2)$-cocycles with the distinguished super-2-cocycle $\,\chi_{(2)}\,$ coupling to the superparticle current in $\,{\rm sMink}(9,1|32)$.\ The corresponding nontrivial CE 2-cocycle (written in the previously introduced notation)
\qq\nn
\om_{(2)}=\ovl\G{}_{11\,\a\b}\,q^\a\wedge q^\b
\qqq 
gives rise -- {\it via} Thm.\,\ref{thm:H2ext} -- to a rank-1 central extension 
\qq
&\brd0\too\bR\too\sfY_0\tgt\xrightarrow{\ \pi_{\sfY_0\tgt}\ }\tgt\too\brd0\,,&\label{eq:om2-sparticle-ext}\\ \cr
&\sfY_0\tgt=\tgt\oplus\corr{S}\,,\qquad\qquad\pi_{\sfY_0\tgt}=\pr_1\,,&\cr\cr
&[P_a,P_b]_{\sfY_0\tgt}=0\,,\qquad\qquad[Q_\a,P_a]_{\sfY_0\tgt}=0\,,&\cr\cr
&\{Q_\a,Q_\b\}_{\sfY_0\tgt}=\{Q_\a,Q_\b\}_\tgt+\bigl(\om_{(2)}\ox S\bigr)(Q_\a,Q_\b)=\ovl\G{}^a_{\a\b}\,P_a+2\ovl\G{}_{11\,\a\b}\,S\,,&\cr\cr
&[Q_\a,S]_{\sfY_0\tgt}=0=[P_a,S]_{\sfY_0\tgt}\,,\qquad\qquad[S,S]_{\sfY_0\tgt}=0\,.&\nn
\qqq
with $\,\pi_{\sfY_0\tgt}^*\om_{(2)}=-\d^{(1)}_{\sfY_0\tgt}s\,$ for the dual $\,s\,$ of $\,S$.\ Its integration to the Lie-supergroup level yields
\bethe
The GS super-2-cocycle $\,\chi_{(2)}\,$ determines a CaE super-0-gerbe 
\qq\nn
\cG^{(0)}_{\rm GS}=\bigl(\sfY_0\txT,\pi_{\sfY_0\txT},\b_{(1)}\bigr)
\qqq 
in the sense of Def.\,\ref{def:CaEs0g},\ composed of 
\bit
\item the Lie supergroup $\,\sfY_0\txT=\txT\x\bC^\x\ni(\theta,x,z)\,$ which is mapped epimorphically onto $\,\txT\,$ along $\,\pi_{\sfY_0\txT}=\pr_1\,$ and has the binary operation
\qq\nn
\txm_0^{(1)}\bigl((\theta_1,x_1,z_1),(\theta_2,x_2,z_2)\bigr)=\bigl(\txm\bigl((\theta_1,x_1),(\theta_2,x_2)\bigr),\ee^{\sfi\,\theta_1\,\ovl\G_{11}\,\theta_2}\cdot z_1\cdot z_2\bigr)\,;
\qqq
\item the LI primitive 
\qq\nn
\b_{(1)}=\pr_2^*\vartheta_{\bC^\x}+\pi_{\sfY_0\txT}^*\txB_{(1)}\in\Om^1(\sfY_0\txT)^{\sfY_0\txT}
\qqq
of $\,\pi_{\sfY_0\txT}^*\chi_{(2)}$,\ expressed in terms of the canonical LI 1-form $\,\vartheta_{\bC^\x}(z)=\tfrac{\sfi\,\sfd z}{z}$ on $\,\bC^\x$,\ and a non-LI primitive $\,\txB_{(1)}(\theta,x)=\theta\,\ovl\G{}_{11}\,\si(\theta,x)\,$ of $\,\chi_{(2)}\,$ -- its left-invariance fixes $\,\txm_0^{(1)}$.
\eit
The short exact sequence of Lie supergroups $\,\bd1\too\bC^\x\xrightarrow{(0,\cdot)}\sfY_0\txT\xrightarrow{\pi_{\sfY_0\txT}}\txT\too\bd1\,$ integrates \eqref{eq:om2-sparticle-ext}. 
\ethe
\beroof
With the above choice of the primitive for $\,\chi_{(2)}$,\ we obtain the quasi-invariance 1-cochain $\,\D_{(\theta_1,x_1)}(\theta_2,x_2)=\theta_1\,\ovl\G{}_{11}\,\theta_2$,\ and so the statement follows from Prop.\,\ref{prop:assoconstr}.
\eroof

\noindent The above super-0-gerbe,\ the first from a hierarchy of higher-supergeometric structures associated with the GS super-$(p+2)$-cocycles,\ shall be referred to as the {\bf Green--Schwarz super-0-gerbe over} $\,{\rm sMink}(9,1|32)\,$ in future work.\ Its supersymmetry shall be discussed in Sec.\,\ref{sec:susy-cat}.

\subsection{The super-1-gerbe for the Green--Schwarz superstring}\label{sec:s-1-grb}

On the next level of the hierarchy,\ we have the super-3-cocycle $\,\chi_{(3)}$,\ which determines the WZ term in the DF amplitude for the superstring in $\,{\rm sMink}(d,1|D_{d,1})$.\ Here,\ the point of departure of the cohomological resolution is the CE 3-cocycle $\,\om_{(3)}=\ovl\G{}_{a\,\a\b}\,q^\a\wedge q^\b\wedge p^a$,\ from which we extract a family of nontrivial LI sub-2-cocycles $\,\varpi_{(2)\,\a}=-\ovl\G{}_{a\,\a\b}\,q^\b\wedge p^a\,,\ \a\in\ovl{1,D_{d,1}}$,\ 
whose closedness is ensured by the Fierz identity \eqref{eq:Fierz} for $\,p=1$,
\qq\label{eq:Fierz-1}
\ovl\G{}^a_{\a(\b}\,\bigl(\ovl\G{}_a\bigr)_{\g\d)}=0\,.
\qqq
The 2-cocycles induce a rank-$D_{d,1}$ \emph{super}central Lie-superalgebra extension
\qq
&\brd0\too\bR^{\x D_{d,1}}[1]\too\sfY_1\tgt\xrightarrow{\ \pi_{\sfY_1\tgt}\ }\tgt\too\brd0\,,&\label{eq:omal2-sstring-ext}\\\cr
&\sfY_1\tgt=\tgt\oplus\bigoplus_{\a=1}^{D_{d,1}}\corr{\cZ^\a}\,,\qquad\qquad\pi_{\sfY_1\tgt}=\pr_1\,,&\nn
\qqq
with a single `corrected' bracket of the generators of $\,\tgt$,
\qq\nn
[Q_\a,P_a]_{\sfY_1\tgt}=[Q_\a,P_a]_\tgt+\bigl(\varpi_{(2)\,\b}\ox\cZ^\b\bigr)(Q_\a,P_a)=-\ovl\G{}_{a\,\a\b}\,\cZ^\b\,,
\qqq
and the remaining new brackets trivial due to the supercentrality of the $\,\cZ^\a$.\ This is the Green superalgebra of \cite{Green:1989nn}.\ On it,\ we obtain the identity $\,\pi_{\sfY_1\tgt}^*\varpi_{(2)\,\a}=-\d^{(1)}_{\sfY_1\tgt}z_\a\,$ for the dual $\,z_\a\,$ of $\,Z^\a$,\ and so the 3-cocycle $\,\pi_{\sfY_1\tgt}^*\om_{(3)}\equiv q^\a\wedge\d^{(1)}_{\sfY_1\tgt}z_\a\,$ trivialises owing to the closedness of the $\,q^\a$.\ Upon integration,\ we arrive at
\berop\label{prop:surj-subm-GS1grb}
The GS super-3-cocycle $\,\chi_{(3)}\,$ determines a supercentral extension $\,\bd1\too\bR^{0|D_{d,1}}\xrightarrow{(0,\cdot)}\sfY_1\txT\xrightarrow{\pi_{\sfY_1\txT}}\txT\too\bd1\,$ which integrates \eqref{eq:omal2-sstring-ext}.\ Its total space $\,\sfY_1\txT=\txT\x\bR^{0|D_{d,1}}\ni(\theta^\a,x^a,\xi_\b)\,$ is mapped epimorphically onto $\,\txT\,$ along $\,\pi_{\sfY_1\txT}=\pr_1\,$ and has the binary operation 
\qq\nn
\txm_1^{(1)}\bigl(\bigl(\theta_1,x_1,\xi_1),(\theta_2,x_2,\xi_2)\bigr)&=&\bigl(\txm\bigl((\theta_1,x_1),(\theta_2,x_2)\bigr),\xi_{1\,\a}+\xi_{2\,\a}-\D_{(\theta_1,x_1)\,\a}(\theta_2,x_2)\bigr)\cr\cr
\D_{(\theta_1,x_1)\,\a}(\theta_2,x_2)&=&\ovl\G_{a\,\a\b}\,\theta_1^\b\,x_2^a-\tfrac{1}{6}\,\ovl\G{}^a_{\a\b}\,\bigl(2\theta_1^\b+\theta_2^\b\bigr)\,\theta_1\,\ovl\G{}_a\,\theta_2\,.
\qqq
Furthermore,\ it supports the LI primitive
\qq\nn
\b_{(2)}=\pi_{\sfY_1\txT}^*\si^\a\wedge\varphi_\a\,,\qquad\qquad\varphi_\a=\pr_2^*\vartheta_{\bR^{0|D_{d,1}}\,\a}+\pr_1^*b_{(1)\,\a}\in\Om^1(\sfY_1\txT)^{\sfY_1\txT}
\qqq
of $\,\pi_{\sfY_1\txT}^*\chi_{(3)}$,\ in which the LI primitive $\,\varphi_\a\,$ of the pullback of the LI super-2-cocycle $\,h_{(2)\,\a}=-\ovl\G{}_{a\,\a\b}\,\si^\b\wedge e^a\,$ (the CaE counterpart of $\,\varpi_{(2)\,\a}$) to $\,\sfY_1\txT\,$ is expressed in terms of the component LI 1-forms $\,\vartheta_{\bR^{0|D_{d,1}}\,\a}(\xi)=\sfd\xi_\a\,$ on $\,\bR^{0|D_{d,1}}\,$ and the non-LI primitives $\,b_{(1)\,\a}(\theta,x)=-\ovl\G{}_{a\,\a\b}\,\theta^\b\,(\sfd x^a+\frac{1}{6}\,\theta\,\ovl\G{}^a\,\si(\theta,x))\,$ of the $\,h_{(2)\,\a}\,$ -- its left-invariance fixes $\,\txm_1^{(1)}$.
\eerop
\beroof
The explicit form of $\,b_{(1)\,\a}\,$ follows directly from the identity $\,3\ovl\G{}_{a\,\a\b}\,\si(\theta)^\b\wedge\theta\,\ovl\G{}^a\,\si(\theta)$\linebreak $=\sfd(\ovl\G{}_{a\,\a\b}\,\theta^\b\,\theta\,\ovl\G{}^a\,\si(\theta))\,$ implied by the Fierz identity \eqref{eq:Fierz-1}.\ The group law is subsequently determined by the supersymmetry variation 
\qq\nn
\bigl(\ell_{(\vep,t)}^*b_{(1)\,\a}-b_{(1)\,\a}\bigr)(\theta,x)=\sfd\bigl(-x^a\,\ovl\G{}_{a\,\a\b}\,\vep^\b+\tfrac{1}{6}\,\ovl\G{}^a_{\a\b}\,\bigl(2\vep^\b+\theta^\b\bigr)\,\vep\,\ovl\G{}_a\,\theta\bigr)\,,
\qqq
which is inferred from \Reqref{eq:Fierz-1}.\ The latter also ensures associativity of the binary operation $\,\txm_1^{(1)}$,\ which follows immediately from Prop.\,\ref{prop:assoconstr}.
\eroof

With the candidate surjective submersion of the super-1-gerbe firmly established,\ we may,\ next,\ consider the $\txT$-fibred square thereof,\ which can be presented as the Lie supergroup $\,\sfY_1^{[2]}\txT\equiv\txT\x\bR^{0|D_{d,1}}\x\bR^{0|D_{d,1}}\ni(\theta,x,\xi^1,\xi^2)\equiv(\theta,x,\xi^A)\,$ with the binary operation induced from $\,\txm_1^{(1)}\,$ as
\qq
\txm_1^{(1)\,[2]}\bigl((\theta_1,x_1,\xi^A_1),(\theta_2,x_2,\xi^A_2)\bigr)&=&\bigl(\txm\bigl((\theta_1,x_1),(\theta_2,x_2)\bigr),\xi^A_{1\,\a}+\xi^A_{2\,\a}+\ovl\G_{a\,\a\b}\,\theta_1^\a\,x_2^a\cr\cr
&&\quad-\tfrac{1}{6}\,\ovl\G{}^a_{\a\b}\,\bigl(2\theta_1^\b+\theta_2^\b\bigr)\,\theta_1\,\ovl\G{}_a\,\theta_2\bigr)\,. \label{eq:fibd-mult}
\qqq
On it,\ we find the nontrivial CaE super-2-cocycle (written in the above presentation)
\qq\nn
\cF^1_{(2)}=\bigl(\pr_{1,3}^*-\pr_{1,2}^*\bigr)\b_{(2)}\equiv\pr_1^*\si^\a\wedge\bigl(\pr_{1,3}^*-\pr_{1,2}^*\bigr)\varphi_\a=\pr_1^*\si^\a\wedge\bigl(\pr_3^*-\pr_2^*\bigr)\vartheta_{\bR^{0|D_{d,1}}\,\a}\,.
\qqq
Equivalently,\ we may pass to the fibred-sum Lie superalgebra $\,\sfY_1^{[2]}\tgt:=\sfY_1\tgt\oplus_\tgt\sfY_1\tgt\,$ with the convenient presentation $\,\sfY_1^{[2]}\tgt\equiv\tgt\oplus\bigoplus_{\a=1}^{D_{d,1}}\langle\cZ^\a_{(1)}\rangle\oplus\bigoplus_{\b=1}^{D_{d,1}}\langle\cZ^\b_{(2)}\rangle\,$ and the bracket of the pair $\,(Q_\a,P_a)\,$ replaced by $\,[Q_\a,P_a]_{\sfY_1^{[2]}\tgt}=-\ovl\G{}_{a\,\a\b}\,(\cZ_{(1)}^\b+\cZ_{(2)}^\b)$,\ on which we find the CE counterpart of $\,\cF^1_{(2)}\,$ in the form
\qq\nn
\varpi^{[2]}_{(2)}=q^\a\wedge\bigl(z^{(2)}_\a-z^{(1)}_\a\bigr)
\qqq
Its nontriviality,\ readily established by evaluating it on $\,Q_\a\,$ and the supercentral $\,\cZ_{(2)}^\a$,\ ensures the existence of a central extension 
\qq
&\brd0\too\bR\too\lgt\xrightarrow{\ \pi_\lgt\ }\sfY_1^{[2]}\tgt\too\brd0\,,&\label{eq:F2-sstring-ext}\\ \cr
&\lgt=\sfY_1^{[2]}\tgt\oplus\corr{S}\,,\qquad\qquad\pi_\lgt=\pr_1&\nn
\qqq
with the `corrected' brackets ($A\in\{1,2\}$)
\qq\nn
\{Q_\a,Z_{(A)}^\b\}_\lgt=\{Q_\a,Z_{(A)}^\b\}_{\sfY_1^{[2]}\tgt}+\bigl(\varpi^{[2]}_{(2)}\ox S\bigr)\bigl(Q_\a,Z_{(A)}^\b\bigr)=(-1)^A\,\d_\a^{\ \b}\,S\,.
\qqq
The extension can be integrated,\ whereupon it gives us
\berop\label{prop:bndl-GS1grb}
The super-2-cocycle $\,\cF_{(2)}^1\,$ determines a CaE super-0-gerbe
\qq\nn
\bigl(\Lx,\pi_\Lx,\cA_{\Lx\,(1)}\bigr)
\qqq
composed of
\bit
\item the Lie supergroup $\,\Lx=\sfY_1^{[2]}\txT\x\bC^\x\ni(\theta,x,\xi^A,z)\equiv(y^{12},z)\,$ which is mapped epimorphically onto $\,\sfY_1^{[2]}\txT\,$ along $\,\pi_\Lx=\pr_1\,$ and has the binary operation
\qq\nn
\txm_1^{(2)}\bigl(\bigl(y^{12}_1,z_1\bigr),\bigl(y^{12}_2,z_2\bigr)\bigr)=\bigl(\txm_1^{(1)\,[2]}\bigl(y^{12}_1,y^{12}_2\bigr),\ee^{\sfi\,\theta_1^\a\,\xi^{21}_{2\,\a}}\cdot z_1\cdot z_2\bigr)\,,\qquad\xi^{21}_2=\xi^2_2-\xi^1_2\,;
\qqq
\item the LI primitive
\qq\nn
\cA_{\Lx\,(1)}=\pr_2^*\vartheta_{\bC^\x}+\pr_1^*\txA_{(1)}\in\Om^1(\Lx)^\Lx
\qqq
of $\,\pi_\Lx^*\cF_{(2)}^1$,\ expressed in terms of a non-LI primitive $\,\txA_{(1)}(y^{12})=\theta^\a\,\sfd\xi^{21}_\a\,$ of $\,\cF_{(2)}^1\,$  -- its left-invariance fixes $\,\txm_1^{(2)}$.
\eit
The short exact sequence of Lie supergroups $\,\bd1\too\bC^\x\xrightarrow{(0,\cdot)}\Lx\xrightarrow{\pi_\Lx}\sfY_1^{[2]}\txT\too\bd1\,$ integrates \eqref{eq:F2-sstring-ext}. 
\eerop
\beroof
We have $\,\sfd\txA_{(1)}(y^{12})=\sfd\theta^\a\wedge\sfd\xi^{21}_\a\equiv\si^\a(\theta,x)\wedge(\pr_2^*-\pr_1^*)\vartheta_{\bR^{0|D_{d,1}}}(\xi^1,\xi^2)$,\ and $\,(\sfY_1^{[2]}\ell_{(\vep,t,\z^A)}^*\txA_{(1)}-\txA_{(1)})(y)=\vep^\a\,\sfd\xi^{21}_\a\,$ for $\,\sfY_1^{[2]}\ell_{(\vep,t,\z^A)}\equiv\txm_1^{(1)\,[2]}((\vep,t,\z^A),\cdot)$.\ The rest is a straightforward check (in which we may employ Prop.\,\ref{prop:assoconstr} for the quasi-invariance 1-cochain $\,\D_{y^{12}_1}(y^{12}_2)=\theta_1^\a\,\xi^{21}_{2\,\a}$).
\eroof

The last step in the reconstruction procedure consists in establishing a groupoid structure on the fibres of $\,\Lx$.\ This we attain in
\berop\label{prop:grpd-str-GS1grb}
The principal $\bC^\x$-bundle $\,\pr_{1,2}^*\Lx\ox\pr_{2,3}^*\Lx\,$ over $\,\sfY_1^{[3]}\txT\,$ carries a natural structure of a Lie supergroup.\ The latter is mapped to the Lie supergroup $\,\pr_{1,3}^*\Lx\,$ by the super-0-gerbe isomorphism
\qq\nn
\mu_\Lx\ &:&\ \pr_{1,2}^*\Lx\ox\pr_{2,3}^*\Lx\xrightarrow{\ \cong\ }\pr_{1,3}^*\Lx\,,\cr\cr
&:&\ \bigl(\theta,x,\xi^1,\xi^2,1\bigr)\ox\bigl(\theta,x,\xi^2,\xi^3,z\bigr)\longmapsto\bigl(\theta,x,\xi^1,\xi^3,z\bigr)\,.
\qqq
\eerop
\beroof
The tensor-product bundle $\,\pr_{1,2}^*\Lx\ox\pr_{2,3}^*\Lx\,$ is a bundle associated\footnote{Bundles of this type in the category $\,\sMan\,$ were considered in \cite{Kessler:2019bwp}.} to $\,\pr_{1,2}^*\Lx\,$ through the defining action of the structure Lie supergroup $\,\bC^\x\,$ on $\,\pr_{2,3}^*\Lx\,$ -- this is the so-called contracted product of principal $\bC^\x$-bundles,\ described in \cite[Sec.\,2.1]{Brylinski:1993ab}.\ As such,\ the bundle comes with the Lie-supergroup structure determined by the binary operation with the coordinate presentation
\qq\nn
&&\bigl(\bigl(\theta_1,x_1,\xi^1_1,\xi^2_1,1\bigr)\ox\bigl(\theta_1,x_1,\xi^2_1,\xi^3_1,z_1\bigr),\bigl(\theta_2,x_2,\xi^1_2,\xi^2_2,1\bigr)\ox\bigl(\theta_2,x_2,\xi^2_2,\xi^3_2,z_2\bigr)\bigr)\cr\cr
&\longmapsto&\txm_1^{(2)}\bigl(\bigl(\theta_1,x_1,\xi^1_1,\xi^2_1,1\bigr),\bigl(\theta_2,x_2,\xi^1_2,\xi^2_2,1\bigr)\bigr)\ox\txm_1^{(2)}\bigl(\bigl(\theta_1,x_1,\xi^2_1,\xi^3_1,z_1\bigr),\bigl(\theta_2,x_2,\xi^2_2,\xi^3_2,z_2\bigr)\bigr)\,,
\qqq
which covers the obvious Lie-supergroup structure on $\,\sfY^{[3]}_1\txT$.\ The isomorphic character of $\,\mu_\Lx\,$ now follows from the identity $\,\ee^{\sfi\,\theta_1^\a\,\xi^{21}_{2\,\a}}\,\ee^{\sfi\,\theta_1^\a\,\xi^{32}_{2\,\a}}=\ee^{\sfi\,\theta_1^\a\,\xi^{31}_{2\,\a}}$,\ written for $\,\xi^{BA}_2=\xi^B_2-\xi^A_2,\ (A,B)\in\{(1,2),(2,3),(1,3)\}$.\ That this mapping preserves connections can be read off from the identity 
\qq\nn
\txA_{(1)}\bigl(\theta,x,\xi^1,\xi^2\bigr)+\txA_{(1)}\bigl(\theta,x,\xi^2,\xi^3\bigr)=\txA_{(1)}\bigl(\theta,x,\xi^1,\xi^3\bigr)\,.
\qqq
\eroof
\brem\label{rem:triv-grpd-str}
A super-0-gerbe isomorphism $\,\varphi:\cG^{(0)}_1\xrightarrow{\cong}\cG^{(0)}_2\,$ between super-0-gerbes $\,\cG^{(0)}_A,\ A\in\{1,2\}\,$ of the above (trivial) form shall be referred to as {\bf unital} and denoted as $\,\varphi\equiv\bd1\,$ in what follows.
\erem

The foregoing considerations are summarised in
\bethe\label{thm:GS-1grb}
The GS super-3-cocycle $\,\chi_{(3)}\,$ determines a CaE super-1-gerbe
\qq\nn
\cG^{(1)}_{\rm GS}=\bigl(\sfY_1\txT,\pi_{\sfY_1\txT},\b_{(2)},\Lx,\pi_\Lx,\cA_{\Lx\,(1)},\mu_\Lx\bigr)
\qqq
in the sense of Def.\,\ref{def:CaEs1g},\ with components defined in Props.\,\ref{prop:surj-subm-GS1grb},\ \ref{prop:bndl-GS1grb} and \ref{prop:grpd-str-GS1grb}.
\ethe

\subsection{The super-2-gerbe for the M-theory supermembrane}

The last example of a cocycle in the CaE cohomology whose geometrisation shall be explicited in this paper is the super-4-cocycle $\,\chi_{(4)}$,\ coupling to the super-membrane current in $\,{\rm sMink}(10,1|32)\equiv\txT_{(M)}$,\ {\it cp.}\ \cite{Chryssomalakos:2000xd}.\ In what follows,\ we give the main points of the construction,\ in which we adhere to the description of 2-gerbes given in \cite{Stevenson:2001grb2}.\ The construction develops in conformity with the general logic delineated above,\ and so its details can readily be reproduced by the interested Reader. 

Anticipating subsequent developments,\ we begin with an auxiliary 
\bedef\label{def:full-ext-spoint11}
Adopt the hitherto notation,\ and in particular that of App.\,\ref{app:sISO}.\ The {\bf fully extended 11$d$ superpoint superalgebra} is the central extension 
\qq\nn
&\brd0\too\bR^{\x 528}\too\sfE\tgt_0\xrightarrow{\ \pi_{\sfE\tgt_0}\ }\tgt_0\too\brd0\,,&\cr\cr
&\sfE\tgt_0=\bigoplus_{\a=1}^{32}\corr{Q_\a}\oplus\bigoplus_{\b\leq\g=1}^{32}\corr{Z_{\b\g}\equiv Z_{(\b\g)}}\equiv\tgt_0\oplus\bR^{\x 528}\,,\qquad\qquad\pi_{\sfE\tgt_0}=\pr_1\,,&
\qqq
of the supercommutative Lie superalgebra $\,\tgt_0\,$ for $\,d=10\,$ (and $\,D_{d,1}=32$),\ induced,\ in the sense of Thm.\,\ref{thm:H2ext},\ by the family of CE 2-cocycles
\qq\nn
\varpi_{(2)}^{\a\b}=\tfrac{1}{2}\,q^\a\wedge q^\b
\qqq
on $\,\tgt_0$,\ {\it i.e.},
\qq\nn
&\{Q_\a,Q_\b\}_{\sfE\tgt_0}=\{Q_\a,Q_\b\}_{\tgt_0}+\bigl(\varpi_{(2)}^{\g\d}\ox Z_{\g\d}\bigr)(Q_\a,Q_\b)=Z_{\a\b}\,,&\cr\cr
&[Q_\a,Z_{\b\g}]_{\sfE\tgt_0}=0\,,\qquad\qquad[Z_{\a\b},Z_{\g\d}]_{\sfE\tgt_0}=0\,.&
\qqq
Upon employing the Clifford decomposition \eqref{eq:C32-sym} with its completeness relations \eqref{eq:Cliff11-compl-sym},\ and denoting
\qq\nn
P_a=\tfrac{1}{64}\,\Lx_a^{\a\b}\,Z_{\a\b}\,,\qquad\qquad M_{ab}=\tfrac{1}{64}\,\Lx_{[ab]}^{\a\b}\,Z_{\a\b}\,,\qquad\qquad\cS_{abcde}=\tfrac{1}{64}\,\Lx_{[abcde]}^{\a\b}\,Z_{\a\b}\,,
\qqq
the Lie superalgebra $\,\sfE\tgt_0\,$ is seen to be a central extension 
\qq\nn
&\brd0\too\bR^{\x 55}\x\bR^{\x 462}\too\sfE\tgt_0\xrightarrow{\ p_{2;5}\equiv\pr_1\ }\gt{smink}(10,1|32)\too\brd0\,,&\cr\cr
&\sfE\tgt_0=\gt{smink}(10,1|32)\oplus\bigoplus_{a<b=0}^{10}\corr{M_{ab}\equiv M_{[ab]}}\oplus\bigoplus_{c<d<e<f<g=0}^{10}\corr{\cS_{cdefg}\equiv\cS_{[cdefg]}}\,,&\cr\cr
&\{Q_\a,Q_\b\}_{\sfE\tgt_0}=\ovl\G{}^a_{\a\b}\,P_a+\tfrac{1}{2!}\,\ovl\G{}^{bc}_{\a\b}\,M_{bc}+\tfrac{1}{5!}\,\ovl\G{}^{defgh}_{\a\b}\,\cS_{defgh}&
\qqq
of the super-minkowskian Lie superalgebra.\ The central Lie-supergroup extension 
\qq\nn
&\bd1\too\bR^{\x 528}\too\sfE\sfT_0\xrightarrow{\ \pi_{\sfE\txT_0}\ }\txT_0\too\bd1\,,&\cr\cr
&\pi_{\sfE\txT_0}\equiv\pr_1\ :\ \sfE\txT_0\equiv\txT_0\x\bR^{\x 528}\too\txT_0\ :\ \bigl(\theta^\a,X^{(\b\g)}\bigr)\longmapsto\theta^\a&
\qqq
to which $\,\sfE\tgt_0\,$ integrates is equipped with the binary operation 
\qq\nn
\sfE\txm_0\ &:&\ \sfE\txT_0\x\sfE\txT_0\too\sfE\txT_0\cr\cr 
&:&\ \bigl(\bigl(\theta_1^\a,X_1^{\b\g}),(\theta_2^\d,X_2^{\la\mu})\bigr)\longmapsto\bigl(\theta_1^\a+\theta_2^\a,X_1^{\b\g}+X_2^{\b\g}-\tfrac{1}{4}\,\bigl(\theta_1^\b\,\theta_2^\g+\theta_1^\g\,\theta_2^\b\bigr)\bigr)\,,
\qqq
determined by the left-invariance of the 1-forms
\qq\nn
E^{\a\b}=\pr_2^*\vartheta_{\bR^{\x 528}}^{\a\b}+\pr_1^*b^{\a\b}\,,
\qqq
written in terms of the canonical LI 1-forms $\,\vartheta_{\bR^{\x 528}}^{\a\b}(X)=\sfd X^{\a\b}\,$ on $\,\bR^{\x 528}\,$ and the non-LI primitives $\,b^{\a\b}(\theta)=\frac{1}{4}\,(\theta^\a\,\sfd\theta^\b+\theta^\b\,\sfd\theta^\a)\,$ of the CaE super-2-cocycles $\,\om_{(2)}^{\a\b}(\theta)=\tfrac{1}{2}\,\sfd\theta^\a\wedge\sfd\theta^\b$.\ The extension shall be referred to as the {\bf fully extended 11$d$ superpoint},\ and presented equivalently as the central extension 
\qq\nn
&\bd1\too\bR^{\x 55}\x\bR^{\x 462}\too\sfE\txT_0\xrightarrow{\ \pi_{2;5}\ }{\rm sMink}(10,1|32)\too\bd1\,,&\cr\cr
&\pi_{2;5}\equiv\pr_1\ :\ \sfE\txT_0\equiv{\rm sMink}(10,1|32)\x\bR^{\x 55}\x\bR^{\x 462}\too{\rm sMink}(10,1|32)&\cr\cr 
&\hspace{-2.cm}:\ \bigl(\theta^\a,x^a,\varphi^{bc}\equiv\varphi^{[bc]},\varsigma^{defgh}\equiv\varsigma^{[defgh]}\bigr)\longmapsto\bigl(\theta^\a,x^a\bigr)\,,&\cr\cr
&X^{\a\b}=:\tfrac{1}{32}\,\bigl(x^a\,\Lx_a^{\a\b}+\tfrac{1}{2}\,\varphi^{bc}\,\Lx_{bc}^{\a\b}+\tfrac{1}{5!}\,\varsigma^{defgh}\,\Lx_{defgh}^{\a\b}\bigr)&
\qqq
with 
\qq\nn
&&\sfE\txm_0\bigl(\bigl(\theta^{\a_1}_1,x^{a_1}_1,\varphi^{b_1 c_1}_1,\varsigma^{d_1 e_1 f_1 g_1 h_1}_1\bigr),\bigl(\theta^{\a_2}_2,x^{a_2}_2,\varphi^{b_2 c_2}_2,\varsigma^{d_2 e_2 f_2 g_2 h_2}_2\bigr)\bigr)\cr\cr
&=&\bigl(\theta_1^\a+\theta_2^\a,x_1^a+x_2^a-\tfrac{1}{2}\,\theta_1\,\ovl\G{}^a\,\theta_2,\varphi_1^{bc}+\varphi_2^{bc}-\tfrac{1}{2}\,\theta_1\,\ovl\G{}^{bc}\,\theta_2,\varsigma^{defgh}_1+\varsigma^{defgh}_2-\tfrac{1}{2}\,\theta_1\,\ovl\G{}^{defgh}\,\theta_2\bigr)\,.
\qqq
Accordingly,\ the $\,E^{\a\b}\,$ decompose as
\qq\nn
E^{\a\b}=\tfrac{1}{32}\,\bigl(\widetilde e{}^a\,\Lx_a^{\a\b}+\tfrac{1}{2}\,\widetilde e{}^{bc}\,\Lx_{bc}^{\a\b}+\tfrac{1}{5!}\,\widetilde e{}^{defgh}\,\Lx_{defgh}^{\a\b}\bigr)
\qqq
into LI components
\qq\nn
\widetilde e{}^a(\theta,X)&=&\sfd x^a+\tfrac{1}{2}\,\theta\,\ovl\G{}^a\,\sfd\theta\equiv e^a(\theta,x)\,,\cr\cr
\widetilde e{}^{bc}(\theta,X)&=&\sfd\varphi^{bc}+\tfrac{1}{2}\,\theta\,\ovl\G{}^{bc}\,\sfd\theta=:e^{bc}(\theta,\varphi)\,,\cr\cr
\widetilde e{}^{defgh}(\theta,X)&=&\sfd\varsigma^{defgh}+\tfrac{1}{2}\,\theta\,\ovl\G{}^{defgh}\,\sfd\theta\equiv e^{defgh}(\theta,\varsigma)\,.
\qqq
These correspond,\ in an obvious manner,\ to a hierarchy of (central) extensions
\qq\nn
&&\sfE\txT_0\equiv\bigl({\rm sMink}(10,1|32)\x\bR^{\x 55}\bigr)\x\bR^{\x 462}\xrightarrow{\ \pi_5\equiv\pr_1\ }{\rm sMink}(10,1|32)\x\bR^{\x 55}\cr\cr
&\xrightarrow{\ \pi_2\equiv\pr_1\ }&{\rm sMink}(10,1|32)\equiv\bR^{0|32}\x\bR^{\x 11}\xrightarrow{\ \pr_1\ }\bR^{0|32}\,.
\qqq
\exdef
\noindent The above structures were considered very early on in the supergravity literature,\ {\it cp.}\ \cite{vanHolten:1982mx,West:1998ey}.\ They are an indispensable constitutive element of the construction of the super-2-gerbe presented below.\ In it,\ one may follow two qualitatively different paths:\ The standard one would have as its point of departure essentially the extended superspace of \cite[Sec.\,5.2]{Chryssomalakos:2000xd} with the LI primitive given in Eq.\,(73) {\it ibidem} -- this would be the first level of a higher-supergeometric resolution of the former,\ {\it i.e.},\ the surjective submersion over the super-minkowskian support of $\,\chi_{(4)}\,$ and the curving of the 2-gerbe-to-be over it.\ The associated stepwise extension of $\,{\rm sMink}(10,1|32)\,$ effectively goes through the full extension of the 11$d$ superpoint described above,\ but leads,\ seemingly inevitably,\ to an algebraic anomaly at a late stage of the construction\footnote{The anomaly occurs in the reconstruction of the groupoid structure on the 1-gerbe of the 2-gerbe,\ and consists in the absence of a two-sided inverse for the associative binary operation on the principal $\bC^\x$-bundle of the 1-gerbe 1-isomorphism (a property rarely checked,\ which is also why it was overlooked in the original analysis of \cite{Suszek:2017xlw}).\ This fact is quite intriguing {\it per se},\ and certainly merits further investigation.\ In the light of our findings from Sec.\,\ref{sec:RCorb},\ it might be related to the phenomenon of de-abelianisation of 1-gerbe modules accompanying descent to the non-1-connected orbifolds of the target Lie group $\,{\rm SO}(4n)\,$ with respect to the action of its non-cyclic centre $\,\bZ_2\x\bZ_2$,\ discovered in \cite{Gawedzki:2004tu}.\ We are planning to pursue this intuition in a future work.},\ and the anomaly can be traced back to the `late' completion of the extension $\,\sfE\txT_0$.\ The alternative,\ based on this observation,\ is to \emph{start} with the fully extended 11$d$ superpoint and promote $\,\chi_{(4)}\,$ to the rank of a non-trivial CaE super-4-cocycle on this larger Lie supergroup,\ in which $\,{\rm sMink}(10,1|32)\,$ can be recovered as,\ say,\ the zero section of $\,\pi_{2;5}$.\ A geometrisation organised thus proceeds without anomalies,\ and so we give its details in what follows.

We begin by rewriting,\ with hindsight,\ the super-4-cocycle $\,\chi_{(4)}\,$ on $\,\sfE\txT_0\,$ in the form
\qq\nn
\chi_{(4)}&=&\sfd\bigl(2\la_1\,e_{ab}\wedge e^a\wedge e^b\bigr)+\la_3\,\si\wedge\ovl\G{}_{ab}\,\si\wedge e^a\wedge e^b+2\la_{12}\,e_{ab}\wedge\si\wedge\ovl\G{}^a\,\si\wedge e^b\cr\cr
&&+\bigl[\bigl(2\la_{11}\,e_{ba}\,\ovl\G{}^b_{\a\b}\,+\la_2\,e^b\,\ovl\G{}_{ab\,\a\b}\bigr)\wedge\si^\b\bigr]\wedge\si^\a\wedge e^a
\qqq
in terms of the LI super-1-forms $\,e_{ab}\equiv\eta_{ac}\,\eta_{bd}\,e^{cd}$,\ and subsequently arrange the real parameters $\,\la_1,\la_2,\la_3=1-\la_1-\la_2,\la_{11}\,$ and $\,\la_{12}=\la_1-\la_{11}\,$ in such a way that the $\wedge$-factor in the square bracket in the above expression be closed and hence give rise to a Lie-superalgebra extension of $\,\sfE\tgt_0$,\ and that the super-4-form ensuing from their partial trivialisation of the pullback of $\,\chi_{(4)}\,$ to the corresponding Lie-supergroup extension of $\,\sfE\txT_0\,$ (whose existence shall be verified below) combines with the terms with the parameters $\,\la_3\,$ and $\,\la_{12}\,$ to another (nontrivial) super-4-cocycle.\ It turns out that this problem is not only well-posed but admits a unique solution:
\qq\nn
(\la_1,\la_2,\la_3,\la_{11},\la_{12})=\bigl(\tfrac{1}{3},\tfrac{3}{5},\tfrac{1}{15},\tfrac{3}{10},\tfrac{1}{30}\bigr)\,,
\qqq
for which 
\qq\nn
\chi_{(4)}&=&\sfd\bigl(\tfrac{2}{3}\,e_{ab}\wedge e^a\wedge e^b\bigr)+\tfrac{1}{15}\,\bigl(\si\wedge\ovl\G{}_{ab}\,\si\wedge e^a\wedge e^b+e_{ab}\wedge\si\wedge\ovl\G{}^a\,\si\wedge e^b\bigr)\cr\cr
&&+\tfrac{3}{5}\,\bigl[\bigl(\ovl\G{}^b_{\a\b}\,e_{ba}+\ovl\G{}_{ba\,\a\b}\,e^b\bigr)\wedge\si^\b\bigr]\wedge\si^\a\wedge e^a
\qqq
We justify our choice of the presentation of $\,\chi_{(4)}\,$ and study its consequences at length in
\berop\label{prop:surjsubm-membr}
The GS super-4-cocycle $\,\chi_{(4)}\,$ on $\,\sfE\txT_0\,$ determines supercentral extensions: 
\bit
\item $\bd1\too\bR^{0|352}\xrightarrow{(0,\cdot)}\widehat\txT{}_1\xrightarrow{\widehat\pi{}_1}\sfE\txT_0\too\bd1\,$ whose total space $\,\widehat\txT{}_1=\sfE\txT_0\x\bR^{0|352}\ni(\theta^\a,X^{\b\g},\psi_{a\d})\,$ is mapped epimorphically onto $\,\sfE\txT_0\,$ along $\,\widehat\pi{}_1=\pr_1\,$ and has the binary operation
\qq\nn
\sfE\txm_0^{(1)}\bigl(\bigl(\theta_1,X_1,\psi^1\bigr),\bigl(\theta_2,X_2,\psi^2\bigr)\bigr)=\bigl(\sfE\txm_0\bigl((\theta_1,X_1),(\theta_2,X_2)\bigr),\cr\cr
\psi^1_{a\a}+\psi^2_{a\a}+\bigl(\ovl\G{}^b_{\a\b}\,\varphi^2_{ba}+x_2^b\,\ovl\G{}_{ba\,\a\b}\bigr)\,\theta_1^\b-\tfrac{1}{6}\,G^+_{a;\a\b,\g\d}\,\bigl(2\theta_1^\b+\theta_2^\b\bigr)\,\theta_1^\g\,\theta_2^\d\bigr)\,,
\qqq
written in terms of $\,G^+_{a;\a\b,\g\d}=\ovl\G{}^b_{\a\b}\,\ovl\G_{ba\,\g\d}+\ovl\G{}^b_{\g\d}\,\ovl\G_{ba\,\a\b}$;
\item $\bd1\too\bR^{\x 528}\xrightarrow{(0,\cdot)}\widehat\txT{}_2\xrightarrow{\widehat\pi_2}\widehat\txT{}_1\too\bd1\,$ whose total space $\,\widehat\txT{}_2=\widehat\txT{}_1\x\bR^{\x 528}\ni(\theta^\a,X^{\b\g},\psi_{a\d},\upsilon_{\la\mu}\equiv\upsilon_{(\la\mu)})\,$ is mapped epimorphically onto $\,\widehat\txT{}_1\,$ along $\,\widehat\pi_2=\pr_1\,$ and has the binary operation
\qq\nn
\sfE\txm_0^{(2)}\bigl(\bigl(\theta_1,X_1,\psi^1,\upsilon^1\bigr),\bigl(\theta_2,X_2,\psi^2,\upsilon^2\bigr)\bigr)=\bigl(\sfE\txm_0^{(1)}\bigl(\bigl(\theta_1,X_1,\psi^1\bigr),\bigl(\theta_2,X_2,\psi^2\bigr)\bigr),\cr\cr
\upsilon^1_{\a\b}+\upsilon^2_{\a\b}-\tfrac{1}{16}\,\ovl\G{}^a_{\a\b}\,\theta_1\,\ovl\G{}^b\,\theta_2\cdot\theta_1\,\ovl\G{}_{ab}\,\theta_2\cr\cr
+\tfrac{1}{24}\,\theta_2^\g\,\bigl(\theta_2^\d\,\bigl(2D_{\a\b;\g\d\ep\eta}\,\theta_2^\ep+3\bigl(D_{\a\b;\ep\g\d\eta}-D_{\a\b;\g\ep\d\eta}\bigr)\,\theta_1^\ep\bigr)+6D_{\a\b;\ep\d\g\eta}\,\theta_1^\ep\,\theta_1^\d\bigr)\,\theta_1^\eta\cr\cr
+\tfrac{1}{4}\,x_1^a\,\bigl(\ovl\G_{ab\,\a\b}\,\bigl(2x_2^b-\theta_1\,\ovl\G{}^b\,\theta_2\bigr)+\ovl\G{}^b_{\a\b}\,\bigl(2\varphi^2_{ab}-\theta_1\,\ovl\G_{ab}\,\theta_2\bigr)\bigr)+\tfrac{1}{4}\,x_2^b\,G^+_{b;\a\b,\g\d}\,\theta_1^\g\,\theta_2^\d\cr\cr
+\bigl(\varphi^2_{ab}\,\ovl\G{}^a_{\a\g}\,\ovl\G{}^b_{\b\d}+\tfrac{1}{2}\,x_2^a\,G^-_{a;\a\g,\b\d}\bigr)\,\theta_1^\g\,\theta_1^\d+\bigl(\tfrac{1}{4}\,\ovl\G{}^a_{\a\b}\,\psi^2_{a\g}+\ovl\G{}^a_{\a\g}\,\psi^2_{a\b}+\ovl\G{}^a_{\b\g}\,\psi^2_{a\a}\bigr)\,\theta_1^\g\,,
\qqq
written in terms of $\,G^-_{a;\a\b,\g\d}=\ovl\G{}^b_{\a\b}\,\ovl\G_{ba\,\g\d}-\ovl\G{}^b_{\g\d}\,\ovl\G_{ba\,\a\b}\,$ and $\,2D_{\a\b;\g\d\vep\eta}=\ovl\G{}^a_{\a\d}\,G^+_{a;\b\g,\ep\eta}+\ovl\G{}^a_{\b\d}\,G^+_{a;\a\g,\ep\eta}$,
\eit
altogether composing a Lie-supergroup epimorphism 
\qq\nn
\widehat\pi{}_{1,2}\equiv\widehat\pi{}_1\circ\widehat\pi{}_2\ :\ \widehat\txT{}_2\too\sfE\txT_0\,
\qqq 
The extensions provide the following CaE resolutions:
\bit
\item the LI primitives 
\qq\nn
\si^1_{a\a}=\pr_2^*\vartheta_{\bR^{0|352}\,a\a}+\pr_1^*b^1_{a\a}\,,\qquad\qquad\sfd\si^1_{a\a}=\widehat\pi{}_1^*h^1_{a\a}
\qqq 
for the super-2-cocycles $\,h^1_{a\a}=(\ovl\G{}^b_{\a\b}\,e_{ba}+\ovl\G{}_{ba}\,e^b)\wedge\si^\b$,\ written in terms of the canonical LI 1-forms $\,\vartheta_{\bR^{0|352}\,a\a}(\psi)=\sfd\psi_{a\a}\,$ on $\,\bR^{0|352}\,$ and the non-LI primitives $\,b^1_{a\a}(\theta,X)=(\ovl\G{}^b_{\a\b}\,(e_{ab}(\theta,\varphi)-\frac{1}{3}\,\theta\,\ovl\G{}_{ab}\,\si(\theta,x))+\ovl\G{}_{ab\,\a\b}\,(e^b(\theta,x)-\frac{1}{3}\,\theta\,\ovl\G{}^b\,\si(\theta,x)))\,\theta^\b\,$ of the $\,h^1_{a\a}\,$ on $\,\sfE\txT_0$;
\item the LI primitives 
\qq\nn
\si^2_{\a\b}=\pr_2^*\vartheta_{\bR^{\x 528}\,\a\b}+\pr_1^*b^2_{\a\b}\,,\qquad\qquad\sfd\si^2_{\a\b}=\widehat\pi{}_2^*h^2_{\a\b}
\qqq
for the super-2-cocycles $\,h^2_{\a\b}=(\frac{1}{4}\,\ovl\G{}^a_{\a\b}\,\si^1_{a\g}+\ovl\G{}^a_{\a\g}\,\si^1_{a\b}+\ovl\G{}^a_{\b\g}\,\si^1_{a\a})\wedge\widehat\pi{}_1^*\si^\g$ $-\frac{1}{2}\,\widehat\pi{}_1^*((\ovl\G{}^a_{\a\b}\,e_{ab}+\ovl\G_{ab\,\a\b}\,e^a)\wedge e^b)$,\ written in terms of the canonical LI 1-forms $\,\vartheta_{\bR^{\x 528}\,\a\b}(\upsilon)=\sfd\upsilon_{\a\b}\,$ on $\,\bR^{\x 528}\,$ and the non-LI primitives $\,b^2_{\a\b}(\theta,X,\psi)=-(\frac{1}{4}\,\ovl\G{}^a_{\a\b}\,\si^1_{a\g}+\ovl\G{}^a_{\a\g}\,\si^1_{a\b}+\ovl\G{}^a_{\b\g}\,\si^1_{a\a})(\theta,X,\psi)\,\theta^\g+\frac{1}{2}\,\ovl\G{}^a_{\a\b}\,e_{ab}(\theta,\varphi)\,x^b-(\ovl\G{}^a_{\a\g}\,\ovl\G{}^b_{\b\d}\,e_{ab}(\theta,\varphi)+\frac{1}{2}\,G^-_{a;\a\g,\b\d}\,e^a(\theta,x))\,\theta^\g\,\theta^\d+\frac{1}{4}\,D_{\a\b;\g\d\vep\eta}\,\theta^\g\,\theta^\d\,\theta^\vep\,\si^\eta(\theta,x)-\frac{1}{4}\,x^a\,(2\ovl\G{}_{ab\,\a\b}\,e^b(\theta,x)+G^+_{a;\a\b,\g\d}\,\theta^\g\,\si^\d(\theta,x))\,$ of the $\,h^2_{\a\b}\,$ on $\,\widehat\txT{}_1$\,,
\eit
which fix the respective binary operations (in the previous sense).

Consequently,\ the Lie supergroup $\,\widehat\txT{}_2\,$ supports the LI primitive 
\qq\nn
\widetilde\b{}_{(3)}=\tfrac{2}{3}\,\widehat\pi{}_{1,2}^*\bigl(e_{ab}\wedge e^a\wedge e^b\bigr)-\tfrac{3}{5}\,\widehat\pi{}_2^*\bigl(\si^1_{a\a}\wedge\widehat\pi{}_1^*\bigl(e^a\wedge\si^\a\bigr)\bigr)-\tfrac{2}{15}\,\si_{\a\b}^2\wedge\widehat\pi{}_{1,2}^*\bigl(\si^\a\wedge\si^\b\bigr)
\qqq
of the pullback $\,\widehat\pi{}_{1,2}^*\chi_{(4)}$.
\eerop
\beroof
The closedness of the $\,h^1_{a\a}\,$ is ensured by the Fierz identity \eqref{eq:Fierz} for $\,p=2$,\ which also yields the primitives $\,b^1_{a\a}$.\ A supersymmetry variation of the latter quickly yields the expression $\,b^1_{a\a}(\sfE\txm_0((\vep,Y),(\theta,X)))-b^1_{a\a}(\theta,X)=\sfd[\vep^\b\,(\ovl\G{}^b_{\a\b}\,\varphi_{ab}+x^b\,\ovl\G_{ab\,\a\b}-\frac{1}{3}\,G^+_{a;\a\b,\g\d})\,\vep^\g\,\theta^\d]-\frac{1}{6}\,\eta^{(\vep)}_{a\a}(\theta)$,\ in which the last term $\,\eta^{(\vep)}_{a\a}(\theta)=(G^+_{a;\a\b,\g\d}+2G^+_{a;\a\g,\b\d})\,\vep^\b\,\theta^\g\,\sfd\theta^\d\,$ is closed (by construction),\ and hence exact.\ Its primitive is established through application of the standard homotopy formula,\ which gives us $\,\eta^{(\vep)}_{a\a}=\sfd F^{(\vep)}_{a\a}\,$ with $\,F^{(\vep)}_{a\a}(\theta)=G^+_{a;\a\g,\b\d}\,\vep^\b\,\theta^\g\,\theta^\d$.\ From this,\ we read off the complete quasi-invariance 1-cochains
\qq\nn
\D_{a\a\,(\theta_1,X_1)}\bigl(\theta_2,X_2\bigr)=\bigl(\ovl\G{}^b_{\a\b}\,\varphi^2_{ab}+x_2^b\,\ovl\G_{ab\,\a\b}-\tfrac{1}{3}\,G^+_{a;\a\b,\g\d}\,\theta_1^\g\,\theta_2^\d\bigr)\,\theta_1^\b-\tfrac{1}{6}\,G^+_{a;\a\g,\b\d}\,\theta_1^\b\,\theta_2^\g\,\theta_2^\d\,,
\qqq
and thus infer the first part of the thesis upon invoking Prop.\,\ref{prop:assoconstr}.\ We attain partial trivialisation:
\qq\nn
\widehat\pi{}_1^*\chi_{(4)}&=&\sfd\bigl[\tfrac{2}{3}\,\widehat\pi{}_1^*\bigl(e_{ab}\wedge e^a\wedge e^b\bigr)-\tfrac{3}{5}\,\si^1_{a\a}\wedge\widehat\pi{}_1^*\bigl(e^a\wedge\si^\a\bigr)\bigr]-\tfrac{2}{15}\,h^2_{\a\b}\wedge\widehat\pi{}_1^*\bigl(\si^\a\wedge\si^\b\bigr)\,,
\qqq
with the $\,h^2_{\a\b}\,$ as in the statement of the proposition.

In the second extension,\ simple manipulations,\ involving the Fierz identity \eqref{eq:Fierz} for $\,p=2$,\ lead to the expression for the (partially trivialised) $\,h^2_{\a\b}(\theta,X,\psi)=\sfd[-(\frac{1}{4}\,\ovl\G{}^a_{\a\b}\,\si^1_{a\g}+\ovl\G{}^a_{\a\g}\,\si^1_{a\b}+\ovl\G{}^a_{\b\g}\,\si^1_{a\a})(\theta,X,\psi)\,\theta^\g+\tfrac{1}{2}\,\ovl\G{}^a_{\a\b}\,e_{ab}(\theta,\varphi)\,x^b-(\ovl\G{}^a_{\a\d}\,\ovl\G{}^b_{\b\g}\,e_{ab}(\theta,\varphi)+\tfrac{1}{2}\,G^-_{a;\a\d,\b\g}\,e^a(\theta,x))\,\theta^\d\,\theta^\g-\frac{1}{2}\,\ovl\G{}_{ab\,\a\b}\,x^a\,e^b(\theta,x)$ $-\tfrac{1}{4}\,G^+_{a;\a\b,\g\d}\,x^a\,\theta^\g\,\si^\d(\theta,x)]+\eta_{\a\b}(\theta)$,\ with the closed (by construction) and hence exact remainder $\,8\eta_{\a\b}(\theta)=2(2\ovl\G{}^a_{\a\d}\,\ovl\G{}^b_{\b\g}\,\ovl\G_{ab\,\ep\z}+G^-_{a;\a\d,\b\g}\,\ovl\G{}^a_{\ep\z}\bigr)\,\theta^\d\,\theta^\g\,\sfd\theta^\ep\wedge\sfd\theta^\z+(\ovl\G{}^a_{\a\b}\,\theta\,\ovl\G_{ab}\,\sfd\theta+\ovl\G_{ab\,\a\b}\,\theta\,\ovl\G{}^a\,\sfd\theta)\wedge\theta\,\ovl\G{}^b\,\sfd\theta$,\ whose further trivialisation with the help of the homotopy formula brings the result stated in the proposition.\ Analogous calculations for the supersymmetry variation of the thus obtained primitives $\,b^2_{\a\b}\,$ of the $\,h^2_{\a\b}\,$ yield the binary operation with the quasi-invariance 1-cochains
\qq\nn
\D_{\a\b\,(\theta_1,X_1,\psi_1)}(\theta_2,X_2,\psi_2)=\tfrac{1}{16}\,\ovl\G{}^a_{\a\b}\,\theta_1\,\ovl\G{}^b\,\theta_2\cdot\theta_1\,\ovl\G{}_{ab}\,\theta_2\cr\cr
-\tfrac{1}{4}\,x_1^a\,\bigl(\ovl\G_{ab\,\a\b}\,\bigl(2x_2^b-\theta_1\,\ovl\G{}^b\,\theta_2\bigr)+\ovl\G{}^b_{\a\b}\,\bigl(2\varphi^2_{ab}-\theta_1\,\ovl\G_{ab}\,\theta_2\bigr)\bigr)-\tfrac{1}{4}\,x_2^b\,G^+_{b;\a\b,\g\d}\,\theta_1^\g\,\theta_2^\d\cr\cr
-\bigl(\varphi^2_{ab}\,\ovl\G{}^a_{\a\g}\,\ovl\G{}^b_{\b\d}+\tfrac{1}{2}\,x_2^a\,G^-_{a;\a\g,\b\d}\bigr)\,\theta_1^\g\,\theta_1^\d-\bigl(\tfrac{1}{4}\,\ovl\G{}^a_{\a\b}\,\psi^2_{a\g}+\ovl\G{}^a_{\a\g}\,\psi^2_{a\b}+\ovl\G{}^a_{\b\g}\,\psi^2_{a\a}\bigr)\,\theta_1^\g\cr\cr
-\tfrac{1}{24}\,\theta_2^\g\,\bigl(\theta_2^\d\,\bigl(2D_{\a\b;\g\d\ep\eta}\,\theta_2^\ep+3\bigl(D_{\a\b;\ep\g\d\eta}-D_{\a\b;\g\ep\d\eta}\bigr)\,\theta_1^\ep\bigr)+6D_{\a\b;\ep\d\g\eta}\,\theta_1^\ep\,\theta_1^\d\bigr)\,\theta_1^\eta
\qqq
which satisfy the conditions of Prop.\,\ref{prop:assoconstr}.
\eroof

The two-tier extension discussed above is sufficient for the trivialisation of the GS super-4-cocycle $\,\chi_{(4)}$,\ and so could well serve as the point of departure of the essentially algorithmic realisation of the Murray scheme of Diag.\,\eqref{diag:Murray},\ to wit,\ as the source of the definition of the surjective submersion of the super-2-gerbe.\ With hindsight,\ and using the intrinsic freedom of the choice of the latter inscribed in the definition of the higher-geometric object,\ we propose to move one step farther,\ which yields a \emph{convenient} surjective submersion of $\,\sfE\txT_0\,$ to be employed in the next stage of geometrisation of the super-4-cocycle along the lines of Sec.\,\ref{sec:s-1-grb}. 
\berop\label{prop:surj-subm-GS2}
The family of CaE super-2-cocycles $\,h_3^{a\a}=\widehat\pi{}_{1,2}^*(8(\d^\a_{\ \g}\,e^a-\ovl\G{}^a_{\b\g}\,E^{\a\b})\wedge\si^\g)\,$ on $\,\widehat\txT{}_2\,$ determines a central extension 
\qq\nn
\bd1\too\bR^{0|352}\xrightarrow{\ (0,\cdot)\ }\widehat\txT{}_3\xrightarrow{\ \widehat\pi{}_3\ }\widehat\txT{}_2\too\bd1
\qqq
whose total space $\,\widehat\txT{}_3\equiv\widehat\txT{}_2\x\bR^{0|352}\ni(\theta^\a,X^{\b\g},\psi_{a\d},\upsilon_{\la\mu},\z^{b\nu})\,$ is mapped epimorphically onto $\,\widehat\txT{}_2\,$ along $\,\widehat\pi{}_3\equiv\pr_1\,$ and has the binary operation 
\qq\nn
&&\sfE\txm_0^{(3)}\bigl(\bigl(\theta_1,X_1,\psi^1,\upsilon^1,\z^1\bigr),\bigl(\theta_2,X_2,\psi^2,\upsilon^2,\z^2\bigr)\bigr)\cr\cr
&=&\bigl(\sfE\txm_0^{(2)}\bigl(\bigl(\theta_1,X_1,\psi^1,\upsilon^1\bigr),\bigl(\theta_2,X_2,\psi^2,\upsilon^2\bigr)\bigr),\z^{a\a}_1+\z^{a\a}_2-\D_{(\theta_1,X_1,\psi_1,\upsilon_1)}(\theta_2,X_2,\psi_2,\upsilon_2)\bigr)\,,
\qqq
where
\qq\nn
\D_{(\theta_1,X_1,\psi_1,\upsilon_1)}(\theta_2,X_2,\psi_2,\upsilon_2)=-8\theta_1^\a\,x_2^a+2\bigl(2\theta_1^\a+\theta_2^\a\bigr)\,\theta_1\,\ovl\G{}^a\,\theta_2+8\theta_1^\b\,\ovl\G{}^a_{\b\g}\,X_2^{\a\g}\,.
\qqq
It supports the LI primitives
\qq\nn
\si_3^{a\a}=\pr_2^*\vartheta_{\bR^{0|352}}+\pr_1^*b^{a\a}_3\,,\qquad\qquad\sfd\si_3^{a\a}=\widehat\pi{}_3^*h^{a\a}_3
\qqq
for the super-2-cocycles ,\ written in terms of the canonical LI 1-forms $\,\vartheta_{\bR^{0|352}}^{a\a}(\z)=\sfd\z^{a\a}\,$ on $\,\bR^{0|352}\,$ and the non-LI primitives $\,b^{a\a}_3(\theta,X,\psi,\upsilon)=-8\theta^\g\,(\d^\a_{\ \g}\,\sfd x^a-\ovl\G{}^a_{\b\g}\,E^{\a\b})\equiv\unl b{}^{a\a}_3(\theta,X)\,$ of the $\,h^{a\a}_3\,$ on $\,\widehat\txT{}_2$.
\eerop
\beroof
Closedness of the $\,h_3^{a\a}\,$ is straightforward to check,\ as is to derive the form of their quasi-invariant primitives $\,b^{a\a}_3$.\ A supersymmetry variation of the latter produces the quasi-invariance 1-cochains $\,\D_{(\theta_1,X_1,\psi_1,\upsilon_1)}$,\ whose properties imply the statement of the proposition in virtue of Prop.\,\ref{prop:assoconstr}. 
\eroof

Altogether,\ then,\ we obtain the surjective submersion 
\qq\label{eq:surj-subm_M}
\pi_{\sfY\sfE\txT_0}\equiv\widehat\pi{}_{1,2}\circ\widehat\pi{}_3\ :\ \sfY\sfE\txT_0\equiv\widehat\txT{}_3\too\sfE\txT_0\, 
\qqq
such that
\qq\label{eq:curv_M}
\pi_{\sfY\sfE\txT_0}^*\chi{}_{(4)}=\sfd\b_{(3)}\,,\qquad\qquad\b_{(3)}=\widehat\pi{}_3^*\widetilde\b{}_{(3)}\,,
\qqq
which defines the trivial super-2-gerbe over $\,\sfY\sfE\txT_0$,\ {\it i.e.},\ the surjective submersion and the curving of the supermembrane super-2-gerbe,\ respectively.

On the Lie supergroup $\,\sfY^{[2]}\sfE\txT_0\equiv\sfY\sfE\txT_0\x_{\sfE\txT_0}\sfY\sfE\txT_0$,\ we find
\qq\nn
\D^{(1)}\b_{(3)}&\equiv&\bigl(\pr_2^*-\pr_1^*\bigr)\b_{(3)}=\sfd\bigl(\tfrac{4}{15}\,\xcZ_{\a\b}\wedge\pr_1^*\pi_{\sfY\sfE\txT_0}^*E^{\a\b}\bigr)-\tfrac{1}{15}\,\xcY_{a\a}\wedge\pr_1^*\pi_{\sfY\sfE\txT_0}^*h^{a\a}_3
\qqq
where we have used the shorthand notation for the LI 1-forms
\qq\nn
\xcY_{a\a}=\D^{(1)}\widehat\pi{}_3^*\widehat\pi{}_2^*\si^1_{a\a}\,,\qquad\qquad\xcZ_{\a\b}=\D^{(1)}\widehat\pi{}_3^*\si_{\a\b}^2\,.
\qqq
Taking into account the defining property of the judicious extension of Prop.\,\ref{prop:surj-subm-GS2} and the closedness of the $\,\xcY_{a\a}$,\ we may cast the above in the form 
\qq\nn
\D^{(1)}\b_{(3)}=\sfd\bigl(\tfrac{4}{15}\,\xcZ_{\a\b}\wedge\pr_1^*\pi_{\sfY\sfE\txT_0}^*E^{\a\b}+\tfrac{1}{15}\,\xcY_{a\a}\wedge\pr_1^*\si^{a\a}_3\bigr)\,.
\qqq
Triviality of the latter in the CaE cohomology of the Lie supergroup $\,\sfY^{[2]}\sfE\txT_0\,$ has a straightforward higher-(super)geometric implication:
\berop\label{prop:GS2grb-1grb}
The super-3-cocycle $\,\widehat\chi{}_{(3)}=\D^{(1)}\b_{(3)}\,$ defines a \emph{trivial} CaE super-1-gerbe 
\qq\nn
\widehat\cG{}^{(1)}=\bigl(\sfY^{[2]}\sfE\txT_0,\id_{\sfY^{[2]}\sfE\txT_0},\widehat\b{}_{(2)},\widehat\Lx\equiv\sfY^{[2]}\sfE\txT_0\x\bC^\x,\pr_1,\widehat\cA{}_{\widehat\Lx\,(1)}=\pr_2^*\vartheta_{\bC^\x},\mu_{\widehat\Lx}\equiv\bd1\bigr)
\qqq
of the curvature $\,\widehat\chi{}_{(3)}$,\ with the LI curving
\qq\nn
\widehat\b{}_{(2)}=\tfrac{4}{15}\,\xcZ_{\a\b}\wedge\pr_1^*\pi_{\sfY\sfE\txT_0}^*E^{\a\b}+\tfrac{1}{15}\,\xcY_{a\a}\wedge\pr_1^*\si^{a\a}_3\,.
\qqq
Its (trivial) principal $\bC^\x$-bundle $\,\widehat\Lx\ni(y^1,y^2,z)\,$ carries the \emph{product} Lie-supergroup structure,\ with the binary operation
\qq\nn
\sfE\txm_0^{(4)}\bigl(\bigl(y^1_1,y^2_1,z_1\bigr),\bigl(y^1_2,y^2_2,z_2\bigr)\bigr)=\bigl(\sfE\txm_0^{(3)}\bigl(y^1_1,y^1_2\bigr),\sfE\txm_0^{(3)}\bigl(y^2_1,y^2_2\bigr),z_1\cdot z_2\bigr)\,.
\qqq
\eerop
\beroof
Obvious.
\eroof

Next,\ we identify the groupoid structure $\,\cM_{\widehat\cG{}^{(1)}}:\pr_{1,2}^*\widehat\cG{}^{(1)}\ox\pr_{2,3}^*\widehat\cG{}^{(1)}\cong\pr_{1,3}^*\widehat\cG{}^{(1)}\,$ on $\,\widehat\cG{}^{(1)}\,$ over the fibred-product Lie supergroup $\,\sfY^{[3]}\sfE\txT_0$,\ using the canonical projections $\,\pr_{i,j}:\sfY^{[3]}\sfE\txT_0\too\sfY^{[2]}\sfE\txT_0,\ (i,j)\in\{(1,2),(2,3),(1,3)\}\,$ on the latter.\ In view of the triviality of $\,\widehat\cG{}^{(1)}$,\ the identification boils down to the comparison of the two pullbacks:\ $\,(\pr_{1,2}^*+\pr_{2,3}^*)\widehat\b{}_{(2)}\,$ and $\,\pr_{1,3}^*\widehat\b{}_{(2)}\,$ of the curving of the super-1-gerbe to $\,\sfY^{[3]}\sfE\txT_0$,\ whose result is given in
\berop\label{prop:GS2grb-1prod-bndl}
The super-2-cocycle $\,\cF_{(2)}=(\pr_{1,3}^*-\pr_{1,2}^*-\pr_{2,3}^*)\widehat\b{}_{(2)}\,$ on $\,\sfY^{[3]}\sfE\txT_0\,$ determines a CaE super-0-gerbe
\qq\nn
\bigl(\widehat\cE,\pi_{\widehat\cE},\cA_{\widehat\cE\,(1)}\bigr)
\qqq
composed of
\bit
\item a Lie supergroup $\,\widehat\cE\equiv\sfY^{[3]}\sfE\txT_0\x\bC^\x\ni(y^1,y^2,y^3,z),\ y^A=(\theta,X,\psi^A,\upsilon^A,\z_A)\,$ which is mapped epimorphically onto $\,\sfY^{[3]}\sfE\txT_0\,$ along $\,\pi_{\widehat\cE}\equiv\pr_1\,$ and has the binary operation
\qq\nn
&&\sfE\txm_0^{(5)}\bigl(\bigl(y^1_1,y^2_1,y^3_1,z_1\bigr),\bigl(y^1_2,y^2_2,y^3_2,z_2\bigr)\bigr)\cr\cr
&=&\bigl(\sfE\txm_0^{(3)}\bigl(y^1_1,y^1_2\bigr),\sfE\txm_0^{(3)}\bigl(y^2_1,y^2_2\bigr),\sfE\txm_0^{(3)}\bigl(y^3_1,y^3_2\bigr),\ee^{\frac{\sfi}{15}\,\z^{a\a}_{1\,21}\,\psi^{32}_{2\,\a\a}}\cdot z_1\cdot z_2\bigr)\,,
\qqq
where we have used the usual shorthand notation:\ $\,\z^{a\a}_{1\,21}\equiv\z^{a\a}_{1\,2}-\z^{a\a}_{1\,1}\,$ and $\,\psi^{32}_{2\,\a\a}\equiv\psi^3_{2\,\a\a}-\psi^2_{2\,\a\a}$;
\item the LI primitive 
\qq\nn
\cA_{\widehat\cE\,(1)}=\pr_2^*\vartheta_{\bC^\x}+\pr_1^*\widehat\txa{}_{(1)}\in\Om^1\bigl(\widehat\cE\bigr)^{\widehat\cE}
\qqq
of $\,\pi_{\widehat\cE}^*\cF^2_{(2)}$,\ expressed in terms of the non-LI primitive $\,\widehat\txa{}_{(1)}(y^1,y^2,y^3)=\frac{1}{15}\,\z^{a\a}_{21}\,\sfd\psi^{32}_{\a\a}\,$ -- its left-invariance fixes $\,\sfE\txm_0^{(5)}$.
\eit

Over the product Lie supergroup $\,\sfY^{[3]}\sfE\txT_0\x_{\sfY^{[3]}\sfE\txT_0}\sfY^{[3]}\sfE\txT_0\cong\sfY^{[3]}\sfE\txT_0$,\ with its canonical projections $\,\pr_A:\sfY^{[3]}\sfE\txT_0\x_{\sfY^{[3]}\sfE\txT_0}\sfY^{[3]}\sfE\txT_0\too\sfY^{[3]}\sfE\txT_0,\ A\in\{1,2\}\,$ and $\,\pr_{i,j}^{[2]}\equiv(\pr_{i,j}\x\pr_{i,j}):\sfY^{[3]}\sfE\txT_0\x_{\sfY^{[3]}\sfE\txT_0}\sfY^{[3]}\sfE\txT_0\too\sfY^{[2]}\sfE\txT_0\,$,\ there exists a super-0-gerbe isomorphism
\qq\nn
\a_{\widehat\cE}\equiv\bd1\ :\ \pr_{1,2}^{[2]\,*}\widehat\Lx\ox\pr_{2,3}^{[2]\,*}\widehat\Lx\ox\pr_2^*\widehat\cE\xrightarrow{\ \cong\ }\pr_1^*\widehat\cE\ox\pr_{1,3}^{[2]\,*}\widehat\Lx\,.
\qqq
\eerop
\beroof
The proof bases on the identity $\,\widehat\b{}_{(2)}(y^1,y^2)=\tfrac{1}{15}\,\sfd\z^{a\a}_1\wedge\sfd\psi^{21}_{a\a}+\Xi(y^2)-\Xi(y^1)$,\ written in terms of the super-2-form $\,15\Xi(\theta,X,\psi,\upsilon)=4\si^2_{a\a}(\theta,X,\psi,\upsilon)\wedge E^{\a\b}(\theta,X)+\sfd\psi_{a\a}\wedge\unl b{}_3^{a\a}(\theta,X)$,\ with the $\,\unl b{}_3^{a\a}\,$ as in Prop.\,\ref{prop:surj-subm-GS2}.\ From it,\ the expression for $\,\widehat\txa{}_{(1)}\,$ follows readily.\ A supersymmetry variation of the latter yields the quasi-invariance 1-cochain $\,\D_{(y^1_1,y^2_1,y^3_1)}(y^1_2,y^2_2,y^3_2)=\tfrac{1}{15}\,\z^{a\a}_{1\,21}\,\psi^{32}_{2\,\a\a}$,\ and so the first part of the statement follows from Prop.\,\ref{prop:assoconstr}.\ The second part is now a simple consequence of the triviality of $\,\widehat\Lx$.
\eroof

Putting the above results together,\ we arrive at the anticipated
\berop\label{prop:GS2grb-1prod}
The super-0-gerbe of Prop.\,\ref{prop:GS2grb-1prod-bndl} determines a super-1-gerbe 1-isomorphism
\qq\nn
\cM_{\widehat\cG{}^{(1)}}\ :\ \pr_{1,2}^*\widehat\cG{}^{(1)}\ox\pr_{2,3}^*\widehat\cG{}^{(1)}\xrightarrow{\ \cong\ }\pr_{1,3}^*\widehat\cG{}^{(1)}
\qqq
over $\,\sfY^{[3]}\sfE\txT_0$,\ in the sense of Def.\,\ref{def:CaEs1g},\ as
\qq\nn
\cM_{\widehat\cG{}^{(1)}}=\bigl(\sfY^{[3]}\sfE\txT_0,\id_{\sfY^{[3]}\sfE\txT_0},\widehat\cE,\pi_{\widehat\cE},\cA_{\widehat\cE\,(1)},\a_{\widehat\cE}\bigr)\,.
\qqq
\eerop
\beroof
Follows readily from Props.\,\ref{prop:GS2grb-1grb} and \ref{prop:GS2grb-1prod-bndl}.
\eroof

The above 1-isomorphism are readily shown to satisfy
\berop\label{prop:GS2grb-1assoc}
There exists a super-1-gerbe 2-isomorphism,\ in the sense of Def.\,\ref{def:CaEs1g},
\qq\nn
\alxydim{@C=4.cm@R=2cm}{\pr_{1,2}^*\widehat\cG{}^{(1)}\ox\pr_{2,3}^*\widehat\cG{}^{(1)}\ox\pr_{3,4}^*\widehat\cG{}^{(1)} \ar[r]^{\qquad\pr_{1,2,3}^*\cM_{\widehat\cG{}^{(1)}}\ox\id_{\pr_{3,4}^*\widehat\cG{}^{(1)}}} \ar[d]_{\id_{\pr_{1,2}^*\widehat\cG{}^{(1)}}\ox\pr_{2,3,4}^*\cM_{\widehat\cG{}^{(1)}}} & \pr_{1,3}^*\widehat\cG{}^{(1)}\ox\pr_{3,4}^*\widehat\cG{}^{(1)} \ar[d]^{\pr_{1,3,4}^*\cM_{\widehat\cG{}^{(1)}}} \ar@{=>}[dl]|{\ \mu_{\widehat\cG{}^{(1)}}\ } \\ \pr_{1,2}^*\widehat\cG{}^{(1)}\ox\pr_{2,4}^*\widehat\cG{}^{(1)} \ar[r]_{\qquad\pr_{1,2,4}^*\cM_{\widehat\cG{}^{(1)}}} & \pr_{1,4}^*\widehat\cG{}^{(1)} }
\qqq
over $\,\sfY^{[4]}\sfE\txT_0$,\ the latter coming with the (obvious) canonical projections $\,\pr_{i,j}:\sfY^{[4]}\sfE\txT_0\too\sfY^{[2]}\sfE\txT_0,\ i<j\in\{1,2,3,4\}\,$ and $\,\pr_{a,b,c}:\sfY^{[4]}\sfE\txT_0\too\sfY^{[3]}\sfE\txT_0,\ a<b<c\in\{1,2,3,4\}$),\ with data
\qq\nn
\mu_{\widehat\cG{}^{(1)}}&=&\bigl(\sfY^{[4]}\sfE\txT_0,\id_{\sfY^{[4]}\sfE\txT_0},\bd1\bigr)\,.
\qqq
\eerop
\beroof
Upon taking into account Prop.\,\ref{prop:GS2grb-1grb} and recalling the structure of the identity 1-isomorphism $\,\id_{\widehat\cG{}^{(1)}}\,$ of the (super-)1-gerbe $\,\widehat\cG{}^{(1)}\,$ ({\it cp.}\ \cite[Sec.\,1.1]{Waldorf:2007mm}),\ the statement of the proposition follows straightforwardly from the identity $\,\widehat\txa{}_{(1)}(y^1,y^2,y^3)+\widehat\txa{}_{(1)}(y^1,y^3,y^4)=\widehat\txa{}_{(1)}(y^2,y^3,y^4)+\widehat\txa{}_{(1)}(y^1,y^2,y^4)\,$ satisfied by the base component of the principal $\bC^\x$-connection $\,\cA_{\widehat\cE\,(1)}\,$ of $\,\widehat\cE\,$ (unitality of $\,\mu_{\widehat\cG{}^{(1)}}$),\ and an analogous identity $\,\D_{(y^1_1,y^2_1,y^3_1)}(y^1_2,y^2_2,y^3_2)+\D_{(y^1_1,y^3_1,y^4_1)}(y^1_2,y^3_2,y^4_2)=\D_{(y^2_1,y^3_1,y^4_1)}(y^2_2,y^3_2,y^4_2)+\D_{(y^1_1,y^2_1,y^4_1)}(y^1_2,y^2_2,y^4_2)\,$ obeyed by the quasi-invariance 1-cochain of the Lie supergroup $\,\widehat\cE\,$ (homomorphicity of $\,\mu_{\widehat\cG{}^{(1)}}$).
\eroof

The crowning of our efforts comes in the form of
\bethe
The GS super-4-cocycle $\,\chi_{(4)}\,$ on $\,\sfE\txT_0\,$ determines a CaE super-2-gerbe
\qq\nn
\cG^{(2)}_{\rm GS}=\bigl(\sfY\sfE\txT_0,\pi_{\sfY\sfE\txT_0},\b_{(3)},\widehat\cG{}^{(1)},\cM_{\widehat\cG{}^{(1)}},\mu_{\widehat\cG{}^{(1)}}\bigr)\,,
\qqq
with components defined in Eqs.\,\eqref{eq:surj-subm_M} and \eqref{eq:curv_M},\ and in Props.\,\ref{prop:GS2grb-1grb},\ \ref{prop:GS2grb-1prod} and \ref{prop:GS2grb-1assoc}.
\ethe

\subsection{Supersymmetry categorified}\label{sec:susy-cat}

Geometrisation of the GS $p$-brane super-$(p+2)$-cocycles,\ while certainly interesting in its own right as a purely mathematical phenomenon,\ serves the important physical purpose delineated in the Introduction:\ It paves the way to prequantisation of the corresponding GS super-$\si$-models and,\ {\it i.a.},\ to a rigorous study of their prequantisable symmetries.\ Having completed a systematic reconstruction of the CaE super-$p$-gerbes for the cases:\ $\,p\in\{0,1,2\}\,$ in the previous section,\ we may now make the first structural step on that way which consists in identifying a gerbe-theoretic manifestation of the constitutive symmetry of the GS super-$\si$-model,\ {\it i.e.},\ its \emph{rigid} supersymmetry induced from the supergeometric action $\,\ell\equiv\txm\,$ of \Reqref{eq:sact-sMink} of the supersymmetry Lie supergroup $\,\txT\,$ on the target supermanifold $\,\txT$,\ and in examining special consequences of our choice  of the geometrisation scheme in the present work for the ensuing higher-geometric realisation of supersymmetry.\ In what follows,\ we carry out at length the analysis for the GS superstring super-1-gerbe $\,\cG^{(1)}_{\rm GS}\,$ of Thm.\,\ref{thm:GS-1grb} for the sake of concreteness,\ with the understanding that the general conclusions derived from that analysis extend straightforwardly to the other cases of interest.\medskip

Given that the existence of a comomentum (1-)form for the action $\,\ell\,$ is ensured by the quasi-$\tgt$-invariance of the global primitive $\,\b_{(2)}\,$ of the curvature $\,\chi_{(3)}\,$ of $\,\cG^{(1)}_{\rm GS}$,\ we may begin our supersymmetry analysis of the GS super-$\si$-model for the superstring in $\,\txT\,$ directly on the higher-geometric level.\ Thus,\ following the logic recapitulated on p.\,\pageref{p:rig-qmorph},\ we look for a $\txT$-indexed (invariably in the $\cS$-point picture) family of (1-)gerbe 1-isomorphisms of the type \eqref{eq:sym-inv-1isos},
\qq\nn
\Phi_{(\vep,t)}\ :\ \ell_{(\vep,t)}^*\cG^{(1)}_{\rm GS}\xrightarrow{\ \cong\ }\cG^{(1)}_{\rm GS}\,,\qquad(\vep,t)\in\txT\,.
\qqq
The point of departure for its identification is the reconstruction of the pullback gerbe $\,\ell_{(\vep,t)}^*\cG^{(1)}_{\rm GS}\,$ over $\,\txT$.\ Taking into account the existence of the extension $\,\sfY_1\ell\equiv\txm_1^{(1)}\,$ defined in Prop.\,\ref{prop:surj-subm-GS1grb},\ we may -- in the light of Prop.\,\ref{prop:pullbck-alt} -- \emph{choose} the surjective submersion of the latter gerbe judiciously in the form 
\qq\nn
\ell_{(\vep,t)}^*\sfY_1\txT:=\sfY_1\txT\,,\qquad\qquad\pi_{\ell_{(\vep,t)}^*\sfY_1\txT}:=\pi_{\sfY_1\txT}
\qqq
by picking up the invertible covering map $\,\widehat\ell{}_{(\vep,t)}:=\sfY_1\ell_{(\vep,t,0)}\,$ for $\,\ell_{(\vep,t)}$.\ We then obtain the attendant curving
\qq\nn
\widehat\ell{}_{(\vep,t)}^*\b_{(2)}\equiv\sfY_1\ell_{(\vep,t,0)}^*\b_{(2)}=\b_{(2)}\,,
\qqq
owing to the left-invariance of $\,\b_{(2)}$.\ Over the base-fibred square $\,\ell_{(\vep,t)}^*\sfY_1\txT\x_\txT\ell_{(\vep,t)}^*\sfY_1\txT\equiv\sfY_1^{[2]}\txT\,$ of this distinguished surjective submersion,\ we may subsequently \emph{choose},\ once more invoking Prop.\,\ref{prop:pullbck-alt},\ the pullback bundle in the form
\qq\nn
\widehat\ell{}_{(\vep,t)}^{[2]\,*}\Lx:=\Lx\,,\qquad\qquad\pi_{\widehat\ell{}_{(\vep,t)}^{[2]\,*}\Lx}:=\pi_\Lx
\qqq
by picking up the covering map $\,\widehat{\widehat\ell}{}^{[2]}_{(\vep,t)}:=\Lx\ell_{((\vep,t,0),(\vep,t,0),1)}\equiv\txm_1^{(2)}(((\vep,t,0),$ \linebreak $(\vep,t,0),1),\cdot)\,$ for $\,\widehat\ell{}_{(\vep,t)}^{[2]}$,\ whereby we get the connection 1-form
\qq\nn
\widehat{\widehat\ell}{}^{[2]\,*}_{(\vep,t)}\cA_{\Lx\,(1)}\equiv\Lx\ell_{((\vep,t,0),(\vep,t,0),1)}^*\cA_{\Lx\,(1)}=\cA_{\Lx\,(1)}\,.
\qqq
It is now readily seen that the groupoid structure of the pullback 1-gerbe is identical with the (unital) one of $\,\cG^{(1)}_{\rm GS}$.\ Altogether,\ then,\ we arrive at the identity
\qq\label{eq:GS1grb-selfinv}
\ell_{(\vep,t)}^*\cG^{(1)}_{\rm GS}=\cG^{(1)}_{\rm GS}\,,
\qqq
which yields
\qq\label{eq:GS1grb-selfinv-1iso}
\Phi_{(\vep,t)}\equiv\id_{\cG^{(1)}_{\rm GS}}\,.
\qqq
We may summarise our analysis by stating
\berop
The CaE super-1-gerbe $\,\cG^{(1)}_{\rm GS}\,$ of Thm.\,\ref{thm:GS-1grb} admits a trivial realisation of supersymmetry through the identity 1-isomorphisms $\,\{\Phi_{(\vep,t)}\equiv\id_{\cG^{(1)}_{\rm GS}}\}_{(\vep,t)\in\txT}$.
\eerop

The last proposition captures the key structural meaning of the chosen geometrisation scheme of the GS super-$(p+2)$-cocycles,\ of direct relevance to the underlying physics in the quantum r\'egime and in conformity with our expectations.

\section{Equivariance and descent to the Rabin--Crane superorbifold}\label{sec:RCorb}

The foregoing analysis yields the sought-after geometrisation,\ in the sense of Murray {\it et al.},\ of the physically relevant Cartan--Eilenberg super-$(p+2)$-cocycles on $\,\txT$,\ and promotes the supersymmetry $\,\ell\,$ of the super-target to a fully fledged quantomorphism of the superfield theory through consistent categorification:\ The supersymmetry is now implemented by 1-isomorphisms of the higher-geometric objects and as such it is sure to transgress to the prequantised theory in the form of its automorphisms.\ Being a result of algebraic relations in the tangent of a target supermanifold with a completely trivial topology,\ admitting only {\it ex post} integration to the Lie-supergroup level,\ the geometrisation seems to merely formally \emph{imitate} the topological mechanisms known from the un-graded geometric category in which the Murray diagram \eqref{diag:Murray} can be understood as a presentation of a coherent resolution\footnote{The meaning of the term `resolution' is best exemplified by the pullback from the base $\,\bS^2\,$ of the Hopf fibration $\,\bS^3\too\bS^2\,$ to its total space,\ which maps the dual of the generator of $\,H_2(\bS^2)\,$ to a trivial de Rham 2-cocycle,\ dual to a 2-surface in $\,\bS^3\,$ that bounds.} of a homology class in $\,M\,$ with the dual $\,[\chi]$.\ In the light of the long-known interplay between the nontrivial topology (for $\,p>0$) of the elementary objects of the (super)field theory of interest (loops,\ membranes {\it etc.}) and the topology of the target in the un-graded setting,\ manifesting itself,\ {\it e.g.},\ through induction of (winding-)charge extensions of symmetry algebras,\ this situation may leave a physicist and a geometer with a feeling of unsatisfaction.\ In this closing section of the present work,\ we intend to disperse this feeling by elaborating and `categorifying' -- with the help of the concept of a gauge-symmetry defect worked out in \cite{Suszek:2012ddg,Suszek:2013} -- a beautiful and beautifully simple idea,\ conceived by Rabin and Crane in \cite{Rabin:1984rm,Rabin:1985tv} on the basis of the previous superfield-theoretical findings due to Kosteleck\'y and Rabin reported in \cite{Kostelecky:1983qu},\ which enables us to regard the geometrisation as a resolution of a nontrivial topology.\ A natural point of departure here is the identification of the `supergeometric/topological content' of the supersymmetric refinement of the de Rham cohomology of $\,\txT\,$ -- an outwardly ill-posed problem,\ first taken up in \cite{Rabin:1985tv},\ which admits an astonishingly well-defined and straightforward solution.\medskip

Returning to the question originally asked by Rabin and Crane almost forty years after the works of Rothstein \cite{Rothstein:1986ax},\ which introduced a rather strict axiomatic framework for the \emph{differential} geometry of supermanifolds and scrupulously pointed out pathologies (as seen from the thus established vantage point) inherent in a number of supergeometric constructions considered at that time,\ calls for a careful statement of the meaning that shall be attached to the concept of a `supergeometric content' in what follows.\ The stage for it is set by the folllowing useful analogy:\ When considering group actions $\,\la:\G\x M\too M\,$ on smooth manifolds $\,M$,\ we encounter situations in which the set of orbits $\,M//\G\,$ does \emph{not} carry the structure of a smooth manifold,\ and so dividing out the action takes us out of the original geometric category.\ Indeed,\ unless the action is free and proper,\ there is no natural choice of such a structure on $\,M//\G$,\ and various pathologies are known to arise.\ These do not prevent us from contemplating `descent' of differential-geometric objects (including higher ones,\ such as $p$-gerbes) and even field theories along the projection $\,\pi_\sim:M\too M//\G$ -- this is the fundamental idea behind the gauging of rigid configurational symmetries in the lagrangean field theory,\ which gives rise to \emph{effective} models of dynamics with spaces of internal degrees of freedom modelled on orbifolds,\ orientifiolds and more general Lie-group orbispaces ({\it cp.}\ \cite{Gawedzki:2012fu}).\ In the former case,\ the general idea is to distinguish a class of objects on $\,M\,$ which \emph{would be} given by (or,\ more generally,\ isomorphic to) pullbacks along $\,\pi_\sim\,$ of objects of the same kind on $\,M//\G\,$ \emph{if} the space of orbits \emph{were} smooth -- in this manner,\ we arrive at $\G$-equivariant objects over $\,M$,\ further coupled to principal $\G$-connections in the standard model of the \emph{differential} geometry of $\,M//\G$,\ {\it cp.}\ \cite{Meinrenken:2006ecm} and \cite{Gawedzki:2010rn,Gawedzki:2012fu}.\ In the latter case,\ such distinguished structures on the typical fibre $\,M\,$ of the covariant configuration bundle of the field theory are used to extend the class of admissible field configurations to include those piecewise continuous ones (in $\,M$) which \emph{would} project along $\,\pi_\sim\,$ to smooth ones in $\,M//\G$,\ their discontinuities being determined by $\,\la\,$ -- this is the so-called $\G$-twisted sector of \cite{Dixon:1985jw,Runkel:2008gr,Frohlich:2009gb,Suszek:2012ddg,Suszek:2013},\ represented -- \emph{via} pullback along local reference gauges $\,\si_i\in\G(\sfP_\txG\rstr_{\cO_i})\,$ over some ($\sfP_\txG$-)trivialising cover $\,\{\cO_i\}_{i\in I}\,$ of the spacetime of the field theory -- by global sections $\,\G(\sfP_\txG\x_\la M)\cong{\rm Hom}_\txG(\sfP_\txG,M)\,$ of bundles associated (through $\,\la$) with \emph{non}-trivial principal $\G$-bundles $\,\sfP_\txG\,$ over that spacetime.\ In any event,\ the idea of modelling structures on $\,M//\G\,$ by those on $\,M$,\ suitably constrained resp.\ enhanced,\ becomes particularly attractive in the absence of a smooth structure on $\,M//\G\,$ as we then have no direct means of making sense of the differential calculus resp.\ physics on the space of orbits.\ As mentioned before,\ this is a field-theoretic avatar of the homotopy quotient,\ first contemplated by Cartan in \cite{Cartan:1950mix}.\ We shall have more to say on the $\G$-twisted sector of the ensuing field theory when we come to discuss a categorification of the Rabin--Crane construction.\ Meanwhile,\ we employ the elementary principle of `\emph{would-be} descent' for tensors to specify what we mean by a `supergeometric content' of $\,\Om^\bullet(\txT)^\txT$.\ 

Thus,\ following Rabin and Crane,\ we look for a \emph{discrete} subgroup $\,\G\subset\txT\,$ whose action $\,\la\equiv\ell\rstr_{\G\x\txT}\,$ on $\,\txT\,$ has the property $\,\Om^\bullet(\txT)^\G\equiv\Om^\bullet(\txT)^\txT$,\ which would permit us to regard the Cartan--Eilenberg cohomology $\,{\rm CaE}^\bullet(\txT)\,$ as a (`would-be') model of the de Rham cohomology of the space $\,\txT//\G\,$ of $\G$-orbits in $\,\txT$,\ the latter space \emph{not} being required to carry a supermanifold structure in the sense of Rothstein,\ but instead being described in terms of (the nerve of) the action groupoid $\,\G\lx_\la\txT$.\ The rationale behind looking for discrete subgroups $\,\G\,$ is twofold:\ For these,\ $\G$-invariant forms are automatically basic $\G$-equivariant \emph{and} $\,\txT//\G\,$ has a tame \emph{local} structure,\ identical with that of $\,\txT$.\ As argued before,\ all this is fairly standard \emph{conceptually},\ and so the actual challenge consists in finding $\,\G\subset\txT\,$ with the properties specified -- as it happens,\ this calls for a fundamental change of the paradigm in which we place our supergeometric constructions\ldots 

The language of our hitherto supergeometric considerations has been,\ implicitly,\ the one of the sheaf-theoretic Berezin--Le{\"i}tes--Kostant approach to supermanifolds in general \cite{Berezin:1975,Kostant:1975} and the Kostant approach to Lie supergroups in particular \cite{Kostant:1975,Carmeli:2011},\ of which,\ however,\ no subtle aspects have ever entered the explicit constructions -- indeed,\ the latter have been phrased mostly in terms of global generators of the relevant structure sheaves,\ and -- more generally -- in the fairly concrete and tractable $\cS$-point picture.\ Therefore,\ it is now perfectly legitimate to switch,\ with hindsight,\ to the concrete Rogers--DeWitt language of \cite{Rogers:1980,DeWitt:1984,Rogers:2007},\ bearing in mind its intrinsic limitations expounded in \cite{Rothstein:1986ax}.\ Thus,\ from now onwards,\ we shall realise the target supermanifold $\,\txT\equiv{\rm sMink}(d,1|D_{d,1})\,$ as a supermanifold\footnote{Note that $\,\txT_L$ can be thought of as the (structured) set of $B_L$-points in the Berezin--Le{\"i}tes--Kostant supermanifold $\,\txT$,\ {\it cp.}\ \cite{Rothstein:1986ax,Sachse:2008}.\ These are the ones naturally probed by the super-$\si$-model superfields in Freed's approach of \cite{Freed:1999} adopted in this work.} 
\qq\nn
\txT_L\equiv B_{L,0}^{\x d+1}\x B_{L,1}^{\x D_{d,1}}\equiv B_L^{d+1,D_{d,1}}
\qqq
modelled on a Gra\ss mann algebra $\,B_L=B_{L,0}\oplus B_{L,1}\,$ with $\,L(\geq D_{d,1})\,$ generators $\,\b^i,\ i\in\ovl{1,L}$,\ belonging to a nested family $\,\{B_L\}_{L\in\bN^\x}\,$ of Gra\ss mann algebras (with $\,B_{L_1}\emb B_{L_2}:\b^i\longmapsto\b^i,\ i\in\ovl{1,L_1}\,$ for $\,L_1<L_2$) over which a direct limit is to be taken in the end for any construction of relevance that we consider (and similarly for the extensions of $\,\txT$).\ (While this is not going to play any practical r\^ole in our discussion,\ we assume $\,\txT_L\,$ to be endowed with a topology induced from the norm topology on $\,B_L\,$ which makes it a Banach algebra,\ {\it cp.}\ \cite{Rogers:1980},\ and we think of the direct limit as approximating the Jadczyk--Pilch supermanifold $\,B_\infty^{d+1,D_{d,1}}\,$ of \cite{Jadczyk:1980xp},\ based on the countably infinitely generated Banach--Gra\ss mann algebra $\,B_\infty\,$ and known to satisfy the Rothstein axioms,\ {\it cp.}\ \cite[Sec.\,1]{Rothstein:1986ax}.)\ Accordingly,\ the (global) coordinates $\,\{\theta^\a\}^{\a\in\ovl{1,D_{d,1}}}\,$ and $\,\{x^a\}^{a\in\ovl{0,d}}\,$ on $\,\txT_L\,$ admit basis expansions
\qq\nn
\theta^\a=\sum_{k=0}^{E(\frac{L-1}{2})}\,\theta^\a_{i_1 i_2\ldots i_{2k+1}}\,\b^{i_1 i_2\ldots i_{2k+1}}\,,\qquad\qquad x^a=x^a_0\,\bd1+\sum_{k=1}^{E(\frac{L}{2})}\,x^a_{i_1 i_2\ldots i_{2k}}\,\b^{i_1 i_2\ldots i_{2k}}
\qqq
in the Gra\ss mann basis $\,\{\bd1\}\cup\{\b^{i_1 i_2\ldots i_k}\equiv\b^{i_1}\b^{i_2}\cdots\b^{i_k}\}_{1\leq i_1<i_2<\ldots<i_k\leq L,\ k\in\ovl{1,L}}$,\ and smooth differential 1-forms (of interest) are functional-linear combinations
\qq\nn
\om(\theta,x)=\om_\a(\theta,x)\,\sfd\theta^\a+\om_a(\theta,x)\,\sfd x^a
\qqq
with the coordinate differentials $\,\sfd\theta^\a\,$ and $\,\sfd x^a\,$ inheriting expansions from the respective coordinates,\ and with the $\,\om_\a\,$ and the $\,\om_a\,$ having (terminating) expansions in the odd coordinates with (body-smooth) functional coefficients admitting a further Taylor expansion in the {\bf soul} components $\,s(x^a)\equiv x^a-x^a_0\,\bd1\,$ of the even coordinates:
\qq\nn
\om_\a(\theta,x)&=&\sum_{k=0}^{E(\frac{D_{d,1}}{2})}\,\varsigma_{\a,\a_1\a_2\ldots\a_{2k}}(x)\,\theta^{\a_1}\,\theta^{\a_2}\,\cdots\,\theta^{\a_{2k}}\,,\cr\cr
\varsigma_{\a,\a_1\a_2\ldots\a_{2k}}(x)&=&\sum_{l=0}^{E(\frac{L}{2})}\,s\bigl(x^{a_1}\bigr)\,s\bigl(x^{a_2}\bigr)\cdots s\bigl(x^{a_l}\bigr)\,\p^l_{a_1 a_2\ldots a_l}\varsigma_{\a,\a_1\a_2\ldots\a_{2k}}(x_0)
\qqq
and 
\qq\nn
\om_a(\theta,x)&=&\sum_{k=0}^{E(\frac{D_{d,1}-1}{2})}\,\varsigma_{a,\a_1\a_2\ldots\a_{2k+1}}(x)\,\theta^{\a_1}\,\theta^{\a_2}\,\cdots\,\theta^{\a_{2k+1}}\,,\cr\cr
\varsigma_{a,\a_1\a_2\ldots\a_{2k+1}}(x)&=&\sum_{l=0}^{E(\frac{L}{2})}\,s\bigl(x^{a_1}\bigr)\,s\bigl(x^{a_2}\bigr)\cdots s\bigl(x^{a_l}\bigr)\,\p^l_{a_1 a_2\ldots a_l}\varsigma_{a,\a_1\a_2\ldots\a_{2k+1}}(x_0)
\qqq
for $\,\om\,$ odd,\ and
\qq\nn
\om_\a(\theta,x)&=&\sum_{k=0}^{E(\frac{D_{d,1}-1}{2})}\,\varpi_{\a,\a_1\a_2\ldots\a_{2k+1}}(x)\,\theta^{\a_1}\,\theta^{\a_2}\,\cdots\,\theta^{\a_{2k+1}}\,,\cr\cr
\varpi_{\a,\a_1\a_2\ldots\a_{2k+1}}(x)&=&\sum_{l=0}^{E(\frac{L}{2})}\,s\bigl(x^{a_1}\bigr)\,s\bigl(x^{a_2}\bigr)\cdots s\bigl(x^{a_l}\bigr)\,\p^l_{a_1 a_2\ldots a_l}\varpi_{\a,\a_1\a_2\ldots\a_{2k+1}}(x_0)
\qqq
and 
\qq\nn
\om_a(\theta,x)&=&\sum_{k=0}^{E(\frac{D_{d,1}}{2})}\,\varpi_{a,\a_1\a_2\ldots\a_{2k}}(x)\,\theta^{\a_1}\,\theta^{\a_2}\,\cdots\,\theta^{\a_{2k}}\,,\cr\cr
\varpi_{a,\a_1\a_2\ldots\a_{2k}}(x)&=&\sum_{l=0}^{E(\frac{L}{2})}\,s\bigl(x^{a_1}\bigr)\,s\bigl(x^{a_2}\bigr)\cdots s\bigl(x^{a_l}\bigr)\,\p^l_{a_1 a_2\ldots a_l}\varpi_{a,\a_1\a_2\ldots\a_{2k}}(x_0)
\qqq
for $\,\om\,$ even.\ Under the above decomposition,\ the binary operation $\,\txm\,$ of $\,\txT\,$ gives rise to a family of bilinear mappings on the components $\,\theta^\a_{i_1 i_2\ldots i_{2k+1}}\,$ and $\,x^a_0,x^a_{i_1 i_2\ldots i_{2k}}$,\ and we readily identify -- for each $\,L\,$ -- the distinguished \emph{discrete} subgroup\footnote{This requires an appropriate choice of the Majorana representation of $\,\Cliff(\bR^{d,1})$,\ {\it cp.}\ \cite{Kostelecky:1983qu}.} $\,\G_{{\rm KR}(L)}\subset\txT_L\,$ generated ($\txm$-)multiplicatively by the $\bZ$-linear span of the basis $\,\b^{i_1 i_2\ldots i_k}\,$ of the \emph{soul} of $\,\txT_L$.\ These subgroups,\ which are readily seen to nest ($\G_{{\rm KR}(L_1)}\subset\G_{{\rm KR}(L_2)}\,$ for $\,L_1<L_2$) and hence survive in the direct limit,\ were first encountered in the study of the supersymmetric lattice field theory reported by Kosteleck\'y and Rabin \cite{Kostelecky:1983qu},\ and so we shall refer to $\,\G_{{\rm KR}(L)}\,$ as the {\bf Kosteleck\'y--Rabin} (KR) ({\bf discrete supersymmetry}) {\bf group} ({\bf of rank} $\,L$).\ They have the simple yet remarkable property: 
\qq\nn
\Om^1(\txT_L)^{\txT_L}\equiv\Om^1(\txT_L)^{\G_{{\rm KR}(L)}}\,,
\qqq
first noted in \cite{Rabin:1985tv}.\ We now present a direct proof of this fact,\ in which we readily recognise a consequence of the polynomial dependence of the 1-forms on the soul components of the coordinates.\ Let us begin with the analysis of an odd 1-form,\ imposing the requirement of invariance \emph{at a topological point} $\,(\theta^\a,x^a)=(0,x_0^a)\,$ under a shift $\,(0,x_0^a)\longmapsto(0,x_0^a+n^a_{\unl i\unl j}\,\b^{\unl i\unl j})\,$ with $\,n^a_{\unl i\unl j}\in\bZ\,$ non-zero \emph{only for a fixed pair} $\,(\unl i,\unl j)\in\ovl{1,L}^{\x 2}$.\ Due to the nilpotency of $\,\b^{\unl i\unl j}$,\ we then obtain $\,\om(0,x_0+n_{\unl i\unl j}\,\b^{\unl i\unl j})-\om(0,x_0)=\p_a\varsigma_\a(x_0)\,n^a_{\unl i\unl j}\,\b^{\unl i\unl j}\,\sfd\theta^\a$,\ and so conclude (for $\,L\geq 3$,\ which we assume henceforth) that $\,\varsigma_\a\in\bR\,$ (are necessarily constant).\ In the next step,\ we carry out the invariance analysis for $\,(0,x_0^a)\longmapsto(\nu^\a_{\unl i}\,\b^{\unl i},x_0^a)\,$ with $\,\nu^\a_{\unl i}\in\bZ\,$ non-zero \emph{only for a fixed} $\,\unl i\in\ovl{1,L}$.\ Reasoning as before,\ we establish $\,\om(\nu_{\unl i}\,\b^{\unl i},x_0)-\om(0,x_0)=\varsigma_{a,\a}(x_0)\,\nu^\a_{\unl i}\,\b^{\unl i}\,\sfd x^a$,\ and so demand $\,\varsigma_{a,\a}\equiv 0$.\ Repeating the analysis for $\,(0,x_0^a)\longmapsto(\nu^\a_{\unl i{}_1}\,\b^{\unl i{}_1}+\nu^\a_{\unl i{}_2}\,\b^{\unl i{}_2},x_0^a)\,$ with $\,\nu^\a_{\unl i{}_1},\nu^\a_{\unl i{}_2}\in\bZ\,$ non-zero \emph{only for a fixed pair} $\,(\unl i{}_1,\unl i{}_2)\in\ovl{1,L}^{\x 2}$,\ we readily nullify $\,\varsigma_{\a,\b_1\b_2}\equiv 0$.\ It is by now clear that a stepwise increase of the number of (rank-1) components of the odd translation (invariably at a topological point) leads to the removal of \emph{all} $\theta$-dependent components of $\,\om$,\ and so at the end,\ we get the desired result $\,\om(\theta,x)\equiv\varsigma_\a\,\si^\a(\theta,x),\ \varsigma_\a\in\bR$.\ An essentially identical analysis can be carried out for an even 1-form,\ with a minor departure from the previously observed scenario:\ For $\,(0,x_0^a)\longmapsto(0,x_0^a+n^a_{\unl i\unl j}\,\b^{\unl i\unl j})\,$ as before,\ we obtain $\,\om(0,x_0+n_{\unl i\unl j}\,\b^{\unl i\unl j})-\om(0,x_0)=\p_b\varpi_a(x_0)\,n^b_{\unl i\unl j}\,\b^{\unl i\unl j}\,\sfd x^\a$,\ whence $\,\varpi_a\in\bR$,\ and so $\,(0,x_0^a)\longmapsto(\nu^\a_{\unl i}\,\b^{\unl i},x_0^a)\,$ yields $\,\om(\nu_{\unl i}\,\b^{\unl i},x_0)-\om(0,x_0)=(\varpi_{\a,\b}(x_0)-\frac{1}{2}\,\varpi_a\,\ovl\G{}^a_{\a\b})\,\nu^\b_{\unl i}\,\b^{\unl i}\,\sfd\theta^\a$,\ implying $\,\varpi_{\a,\b}=\frac{1}{2}\,\varpi_a\,\ovl\G{}^a_{\a\b}$;\ upon subtracting from $\,\om\,$ the thus obtained $\txT_L$-invariant (and hence,\ in particular,\ $\G_{{\rm KR}(L)}$-invariant) component $\,\varpi_a\,e^a$,\ we readily nullify the remaining $\theta$-dependent terms in the same manner as in the previous case -- thus, altogether,\ $\,\om(\theta,x)\equiv\varpi_a\,e^a(\theta,x),\ \varpi_a\in\bR$,\ as desired.\ In this manner,\ the LI 1-forms on $\,\txT\,$ furnish a model of the de Rham calculus on the {\bf Rabin--Crane} (RC) {\bf superorbifold}
\qq\nn
\sfT//\G_{\rm KR}\equiv\varinjlim_{L}\,\txT_L//\G_{{\rm KR}(L)}\,,
\qqq
as first put forward in \cite{Rabin:1985tv}.\ In the remainder of this section,\ we lift Rabin's idea to the fully fledged geometrisations of the distinguished Green--Schwarz classes in the supersymmetric refinement of the de Rham cohomology given in Sec.\,\ref{sec:dAzcladder}.\ Thereby,\ we arrive at a deeper interpretation of the geometrisation itself in terms of descent to the RC superorbifold,\ and -- as a byproduct -- at an effective definition of a superorbifold super-$\si$-model.\ As we proceed,\ we keep the rank label $\,L\,$ implicit in all our constructions,\ with the understanding that they are to be carried out for a fixed (finite) $\,L$,\ whereupon the direct limit over the nested family of constructions is to be taken.\medskip

The logic of our subsequent considerations,\ which we restrict to the case $\,p=1\,$ for the sake of concreteness and an easy reference to extensive literature,\ is organised largely by the content of Theorem 5.3 of \cite{Gawedzki:2010rn} ({\it cp.}\ also \cite[Sec.\,8.2]{Gawedzki:2012fu}) which states the existence of an equivalence between the bicategory of gerbes over the base $\,M/\G\,$ of a principal $\G$-bundle $\,M\too M/\G\,$ and the bicategory of those gerbes over its total space $\,M\,$ which are equipped with a distinguished (flat) $\G$-equivariant structure,\ a particular simplicial gerbe over Segal's nerve of the action groupoid $\,\G\lx_\la M\,$ (described at length\footnote{In what follows,\ we shall be dealing with a simpler structure relevant for $\,\G\,$ discrete,\ and so we dispense with the detailed definitions,\ referring the Reader to the original papers instead.} {\it ibid.}) that lifts the action $\,\la:\G\x M\too M\,$ coherently to the gerbe $\,\cG\,$ over $\,M$.\ While fundamental,\ the theorem is also quite restrictive in its raw form in that it presupposes a high degree of regularity of the action $\,\la$,\ ensuring the existence of a smooth structure on the space $\,M/\txG\,$ of its orbits,\ and a rather rigid behaviour of the gerbe $\,\cG\,$ (and so,\ in particular,\ of its curvature) under symmetry transformations,\ reflected in the existence of a \emph{flat} $\cG$-bi-module $\,\Upsilon\ :\ \la^*\cG\cong\pr_2^*\cG\ox\cI^{(1)}_{\rho=0}\,$ (an element of the said $\G$-equivariant structure) over the arrow manifold $\,\G\x M\,$ of $\,\G\lx_\la M$.\ A universal construction which serves to circumnavigate these restrictions is that of the homotopy quotient of the $\G$-space $\,M$,\ whose Cartan--Borel pullbacks $\,\sfP_\G\x_\la M\,$ to the spacetime of a field theory (from the classifying space $\,B\G$) were mentioned before as a natural source of the $\G$-twisted sector in the worldheet-orbifold field theory.\ As we want to understand both:\ the descent of the super-gerbes to $\,T//\G_{\rm KR}\,$ \emph{and} the emergence of the superorbifold $\si$-model,\ we place our discussion directly in the image of that pullback.\ Here,\ the pullback from over $\,M\,$ to $\,\sfP_\G\x M\,$ of a gerbe $\,\cG\,$ with a not necessarily flat ($\rho\not\equiv 0$) $\G$-equivariant structure receives a trivial tensor correction (with a global curving,\ determined by $\,\rho$) dependent on the principal $\G$-connection on $\,\sfP_\G$,\ whereupon it descends,\ in conformity with the Theorem,\ to the total space of the associated bundle $\,\sfP_\G\x_\la M\,$ and,\ in so doing,\ defines a $\si$-model with the symmetry $\,\G\,$ gauged according to the universal gauge principle elaborated in \cite{Gawedzki:2010rn,Gawedzki:2012fu}.\ In \cite{Suszek:2012ddg,Suszek:2013},\ the latter principle was translated into a hands-on construction of a gauge-symmetry defect in the $\si$-model with the target space $\,M$,\ at whose edges (defect lines) the field of the $\si$-model was allowed to have discontinuities given by a (point-dependent) action of the symmetry group $\,\G$,\ the ensuing defect field theory being rendered well-defined,\ in keeping with the general principles worked out in \cite{Runkel:2008gr},\ by the placement of the data of the $\G$-equivariant structure on $\,\cG\,$ at edges (1-cells of the simplicial gerbe) and vertices (its 2-cells) of the gauge-symmetry defect.\ In this construction,\ the universal ideas of the homotopy quotient as a model of the space of orbits and of equivariantisation of a differential structure as a prerequisite of the descent of the latter to the former were brought together in a coherent (and intuitive) manner.\ In the setting of immediate interest,\ which is that of the KR group acting on ($B_L$-modelled) super-minkowskian space,\ a tremendous simplification takes place:\ The (super)symmetry group $\,\G_{\rm KR}\,$ is \emph{discrete},\ and so the $\G_{\rm KR}$-equivariance boils down to $\G_{\rm KR}$-\emph{invariance} with the additional structure described on p.\,\pageref{p:jump-inv-equiv}.\ This observation sets the stage for the analysis that follows below.

After the above conceptual preparations,\ we may finally investigate the super-1-gerbe of Sec.\,\ref{sec:s-1-grb} in the context of descent $\,\txT\searrow\txT//\G_{\rm KR}$.\ We begin by noting the existence of the relevant $\G_{\rm KR}$-indexed family of gerbe 1-isomorphisms
\qq\nn
\Phi_\nu\equiv\id_{\cG^{(1)}_{\rm GS}}\ :\ \cG^{(1)}_{\rm GS}\xrightarrow{\ \cong\ }\ell_{\nu^{-1}}^*\cG^{(1)}_{\rm GS}\,,\qquad\nu\in\G_{\rm KR}\,,
\qqq
a discrete subfamily of the $\txT$-indexed family \eqref{eq:GS1grb-selfinv-1iso} of 1-isomorphisms reconstructed in Sec.\,\ref{sec:susy-cat}.\ On the next level,\ we look for a $\G_{\rm KR}^{\x 2}$-indexed family of gerbe 2-isomorphisms of the type \eqref{eq:discr-sym-equiv-2isos},
\qq\nn
\varphi_{\nu_1,\nu_2}\ :\ \ell_{\nu_1^{-1}}^*\Phi_{\nu_2}\circ\Phi_{\nu_1}\overset{\cong}{\Longrightarrow}\Phi_{\nu_1\cdot\nu_2}\,.
\qqq
Invoking (strict 2-)functoriality of the pullback ({\it cp.}\ \cite[Sec.\,1.4]{Waldorf:2007mm}) and taking into account identity \eqref{eq:GS1grb-selfinv},\ we readily rewrite the above as $\,\varphi_{\nu_1,\nu_2}\ :\ \id_{\cG^{(1)}_{\rm GS}}\circ\id_{\cG^{(1)}_{\rm GS}}\overset{\cong}{\Longrightarrow}\id_{\cG^{(1)}_{\rm GS}}$,\ whereupon it becomes clear that the sought-after 2-isomorphisms coincide with the (left,\ say) unit 2-isomorphism $\,\varphi_{\nu_1,\nu_2}\equiv\la_{\id_{\cG^{(1)}_{\rm GS}}}\,$ from the definition of the bicategory of gerbes given in \cite[Sec.\,1]{Waldorf:2007mm}.\ From a suitable specialisation of the construction of the unit 2-isomorphisms explicited in \cite[Sec.\,1.2]{Waldorf:2007mm},\ we infer that the data of the principal $\bC^\x$-bundle isomorphism of $\,\la_{\id_{\cG^{(1)}_{\rm GS}}}\,$ can be expressed entirely in terms of those of the groupoid structure of the 1-gerbe $\,\cG^{(1)}_{\rm GS}$,\ and so 
\qq\nn
\varphi_{\nu_1,\nu_2}\equiv\bd1
\qqq
in virtue of Prop.\,\ref{prop:grpd-str-GS1grb} ({\it cp.}\ Rem.\,\ref{rem:triv-grpd-str}).\ Clearly,\ the corresponding associator 2-cocycle\footnote{Strictly speaking,\ the associator 2-cocycle was defined for an \emph{abelian} group,\ and so one ought to specialise the coherence identity (5.1) of \cite{Gawedzki:2010rn} for the 2-cell of a $\G$-equivariant structure instead in our situation.\ Be as it may,\ the conclusion holds for the exact same reason as the one given above.} of \cite[Eq.\,(2.102)]{Runkel:2008gr} is unital,\ and hence cohomologically trivial.\ The triple
\qq\label{eq:KR-equiv-str}
\bigl(\cG^{(1)}_{\rm GS},\{\Phi_\nu\}_{\nu\in\G_{\rm KR}},\{\varphi_{\nu_1,\nu_2}\}_{(\nu_1,\nu_2)\in\G_{\rm KR}^{\x 2}}\bigr)
\qqq
defines a flat and hence \emph{descendable} $\G_{\rm KR}$-structure on the GS super-1-gerbe $\,\cG^{(1)}_{\rm GS}$.

Given the $\G_{\rm KR}$-structure \eqref{eq:KR-equiv-str},\ we may close the present section with a word on the superorbifold $\si$-model,\ first contempleted by Rabin in his insightful article \cite{Rabin:1985tv}.\ Thus,\ following the prototypical construction of the $Z(\txG)$-orbifold WZW $\si$-model (given for the centre $\,Z(\txG)\,$ of the target Lie group $\,\txG$) delineated in \cite{Runkel:2008gr} (to which we refer the Reader for the technical details),\ we define the superorbifold $\si$-model to be a superfield theory with topological defects (in the sense of \cite{Runkel:2008gr,Suszek:2022lpf,Suszek:2022gtmp}),\ forming an arbitrarily fine mesh,\ to whose edges we pull back,\ in the manner discussed in detail in \cite[Secs.\,2.3.1 \& 2.4.1]{Runkel:2008gr},\ the data of the 1-isomorphisms $\,\Phi_\nu\,$ ({\it i.e.},\ essentially those of the bundle $\,\Lx\,$ of the GS super-1-gerbe from Prop.\,\ref{prop:bndl-GS1grb} ({\it cp.}\ \cite[Sec.\,1.1]{Waldorf:2007mm})),\ and to whose vertices we pull back,\ in the manner discussed in detail in \cite[Secs.\,2.5.1 \& 2.8.1]{Runkel:2008gr},\ the (trivial) data of 2-isomorphisms induced from the elementary ones $\,\varphi_{\nu_1,\nu_2}\,$ along the lines of \cite[Sec.\,2.8]{Runkel:2008gr}.\ The thus constructed $\si$-model is to be thought of as a superfield theory with the supertarget given by the RC superorbifold $\,T//\G_{\rm KR}$.

\section{Conclusions and outlook}

The present study puts forward a complete proposal of a \emph{concrete} geometrisation scheme for a family of classes in the supersymmetry-invariant de Rham cohomology of the super-minkowskian spacetime $\,{\rm sMink}(d,1|D_{d,1})$,\ of direct relevance to the supersymmetric dynamics of super-$p$-branes captured by the Green--Schwarz super-$\si$-models.\ The geometrisation yields distinguished ($p$-)gerbe objects,\ in the sense of Murray {\it et al.},\ in the category of Lie supergroups,\ which have been dubbed {\bf Cartan--Eilenberg super-$p$-gerbes} by the Author.\ In the case of the $M$-theory supermembrane,\ the geometrisation appears to favour as its base the fully extended 11$d$ superpoint over $\,{\rm sMink}(d,1|D_{d,1})$.\ The study also provides an interpretation of the particular scheme of geometrisation proposed in terms of a canonical categorification of the supersymmetries present in the form of identity $p$-gerbe isomorphisms,\ and of the resultant descent of the super-$p$-gerbes to an orbifold of (the set of superpoints in) their super-minkowskian base with respect to the action of a discrete subgroup $\,\G_{\rm KR}\,$ of the supersymmetry group,\ effectively `topologising' the supersymmetric refinement of the trivial de Rham cohomology.\ The descent is effected by an extension of the aforementioned $p$-gerbe isomorphisms to a fully fledged $\G_{\rm KR}$-equivariant structure on the super-$p$-gerbes.\ As such,\ it affords a novel definition of a (super-)$\si$-model with the super-orbifold as the supertarget,\ based on the construction of a $\G_{\rm KR}$-jump defect carrying the data of the $\G_{\rm KR}$-equivariant structure.

The geometrisation scheme hinges on the classic 1-1 correspondence between classes in the second cohomology group in the Chevalley--Eilenberg cohomology of the tangent Lie superalgebra of the supersymmetry group and equivalence classes of supercentral extensions of that Lie superalgebra.\ It uses amply a hands-on technique of a ($H^2$-)stepwise resolution of physically distinguished CE super-$(p+2)$-cocycles and of integration of the ensuing Lie-superalgebra extensions to the Lie-supergroup level,\ originally considered by de Azc\'arraga {\it et al.}\ The extended Lie supergroups which it produces are subsequently taken as elementary ingredients in a construction of the super-$p$-gerbes (carried out explicitly for $\,p\in\{0,1,2\}$) along the lines of the standard un-graded geometrisation scheme for de Rham cocycles,\ due to Murray and Stevenson.\ The interpretation,\ on the other hand,\ builds on the original ideas of Rabin and Crane and employs the discrete Kosteleck\'y--Rabin supersymmetry group,\ known from early studies of lattice supersymmetric field theory.\ The descent is realised in what is to be viewed as an adaptation of the Cartan--Borel model of a homotopy quotient associated with a (not necessarily free and proper) group action on a (super)manifold,\ worked out in full generality in the un-graded geometric setting of the (gauged) two-dimensional $\si$-model with the WZ term by Gaw\k{e}dzki,\ Waldorf and the Author.\ Its transcription into an effective definition of a super-orbifold $\si$-model draws upon former studies of worldsheet orbifolds (including continuous gauge-symmetry orbifolds) by Dixon {\it et al.},\ Fr\"ohlich {\it et al.},\ Runkel and the Author.

Being \emph{concrete} and \emph{tractable},\ the higher-supergeometric constructions advanced in the present work pave the way to a variety of applications and extensions,\ in particular those listed in the Introduction,\ {\it i.e.},\ to a higher-supergeometric study of $\kap$-symmetry,\ to a systematic reconstruction of full (weak) $(p+1)$-categories for super-$\si$-models with defects (in particular,\ the maximally supersymmetric ones),\ and to an in-depth investigation of the higher geometry behind the {\.I}n{\"o}n{\"u}--Wigner-type asymptotic relations between super-$\si$-models with curved supertargets and the flat super-minkowskian ones considered herein.\ The latter line of research promises to shed some light on (the higher geometry of) the celebrated AdS/CFT correspondence,\ largely successful physically but still elusive mathematically.\ Last but not least,\ the intricate supergeometry of Rabin--Crane super-orbifolds and the associated superstring dynamics certainly deserves a separate study.\ In this latter context,\ the r\^ole of the fully extended 11$d$ superpoint in the geometrisation of the super-4-cocycle for the $M$-theory supermembrane awaits an elucidation.\ We shall return to these topics in a future work.\\[-19pt]

\section*{Appendices}

\appendix

\stoptocwriting

\section{Mappings and pullbacks}\label{app:maps}

Let $\,\cM\,$ and $\,\cN_A,\ A\in\{1,2\}\,$ be supermanifolds,\ and let $\,f_A:\cM\too\cN_A\,$ be supermanifold mappings (morphisms).\ We then define a supermanifold mapping
\qq\nn
(f_1,f_2)\ :\ \cM\too\cN_1\x\cN_2
\qqq
as the unique one fixed by the conditions $\,\pr_A\circ(f_1,f_2)=f_A,\ A\in\{1,2\}$,\ written in terms of the canonical projections $\,\pr_A:\cN_1\x\cN_2\too\cN_A$.

Similarly,\ given supermanifolds $\,\cM_A,\ A\in\{1,2\}\,$ and $\,\cN_B,\ B\in\{1,2\}\,$ alongside supermanifold mappings $\,F_C:\cM_C\too\cN_C,\ C\in\{1,2\}$,\ we define a supermanifold mapping
\qq\nn
F_1\x F_2\ :\ \cM_1\x\cM_2\too\cN_1\x\cN_2
\qqq
as the unique one fixed by the conditions $\,\pr_A\circ(F_1\x F_2)=F_A\circ\pr_A,\ A\in\{1,2\}$.

Let,\ next,\ $\,\cM_A,\ A\in\{1,2\}\,$ and $\,\cN\,$ be supermanifolds,\ and let $\,f_A:\cM_A\too\cN\,$ be supermanifold mappings.\ The {\bf pullback of} $\,(\cM_2,\cN,f_2)\,$ {\bf along} $\,f_1\,$ (or the {\bf pullback of} $\,\cM_2\,$ {\bf along} $\,f_1$,\ for short) is a triple $\,(f_1^*\cM_2,p_1,p_2)\,$ composed of a supermanifold $\,f_1^*\cM_2\,$ and supermanifold mappings $\,p_A:f_1^*\cM_2\too\cM_A\,$ which close the cospan of $\,(f_1,f_2)\,$ to a commutative diagram
\qq\nn
\alxydim{@C=.75cm@R=1.cm}{ &  f_1^*\cM_2 \ar[dl]_{p_1} \ar[dr]^{p_2} & \\ \cM_1 \ar[dr]_{f_1} & & \cM_2 \ar[dl]^{f_2} \\ & \cN &}
\qqq
and are universal for this property in the following sense:\ Given any triple $\,(\cS,\si_1,\si_2)\,$ closing the cospan of $\,(f_1,f_2)\,$ to a commutative diagram analogous to the one above,\ there exists a \emph{unique} supermanifold mapping $\,F:\cS\too f_1^*\cM_2\,$ which renders the following diagram commutative
\qq\nn
\alxydim{@C=.75cm@R=1.cm}{ & \cS \ar@{-->}[d]^{F} \ar@/_1.6pc/[ddl]_{\si_1} \ar@/^1.6pc/[ddr]^{\si_2} & \\ &  f_1^*\cM_2 \ar[dl]_{p_1} \ar[dr]^{p_2} & \\ \cM_1 \ar[dr]_{f_1} & & \cM_2 \ar[dl]^{f_2} \\ & \cN &}
\qqq
Whenever $\,f_2\,$ is a surjective submersion,\ the corresponding pullback $\,f_1^*\cM_2\,$ exists and has a natural model given by the ($\cN$-){\bf fibred product} of the $\,\cM_A\,$ ({\it cp.}\ \cite[Prop.\,3.2.11]{Kessler:2019bwp} and \cite[Sec.\,2.4.9]{Voronov:2014}),\ given by the triple
\qq\nn
\bigl(f_1^*\cM_2,p_1,p_2\bigr)=\bigl(\cM_1{}_{f_1}\hspace{-1pt}\x_{f_2}\hspace{-1pt}\cM_2,\pr_1\rstr_{\cM_1\x\cM_2},\pr_2\rstr_{\cM_1\x\cM_2}\bigr)
\qqq
in which $\,\cM_1{}_{f_1}\hspace{-1pt}\x_{f_2}\hspace{-1pt}\cM_2\,$ is a sub-supermanifold of the cartesian product $\,\cM_1\x\cM_2\,$ on which (the restrictions of) the canonical projections $\,\pr_A:\cM_1\x\cM_2\too\cM_A\,$ (which we shall write without the restriction symbol henceforth) satisfy the desired identity $\,f_1\circ\pr_1\rstr_{\cM_1{}_{f_1}\hspace{-1pt}\x_{f_2}\hspace{-1pt}\cM_2}=f_2\circ\pr_2\rstr_{\cM_1{}_{f_1}\hspace{-1pt}\x_{f_2}\hspace{-1pt}\cM_2}$.\ Clearly,\ the previously discussed map $\,F\,$ takes the form $\,F=(\si_1,\si_2)\,$ here.\ This is the model which shall implicitly be employed most of the time in the present work.\ An important alternative model,\ which we encounter in our considerations in Sec.\,\ref{sec:susy-cat},\ is described in
\berop\label{prop:pullbck-alt}
Let $\,\cN\,$ and $\,(\cM_A,f_A),\ A\in\{1,2\}\,$ be as above.\ Assume that $\,\cN=\cM_1\,$ and $\,f_1\,$ is an automorphism of $\,\cM_1\,$ in $\,\sMan$.\ If there exists an automorphism $\,\widehat f{}_1\,$ of $\,\cM_2\,$ covering $\,f_1\,$ in the sense of the identity:\ $\,f_2\circ\widehat f{}_1=f_1\circ f_2$,\ then $\,(\cM_2,f_2,\widehat f{}_1)\,$ is a pullback of $\,\cM_2\,$ along $\,f_1$.
\eerop
\beroof
The mapping $\,\mu:=\widehat f{}_1^{-1}\circ\pr_2:\cM_1{}_{f_1}\hspace{-1pt}\x_{f_2}\hspace{-1pt}\cM_2\too\cM_2\,$ satisfies the identities
\qq\nn
&\bigl(f_2,\widehat f{}_1\bigr)\circ\mu=\bigl(f_1^{-1}\circ f_2\circ\pr_2,\pr_2\bigr)=\bigl(\pr_1,\pr_2\bigr)\equiv\id_{\cM_1{}_{f_1}\hspace{-1pt}\x_{f_2}\hspace{-1pt}\cM_2}\,,&\cr\cr
&\mu\circ\bigl(f_2,\widehat f{}_1\bigr)=\id_{\cM_2}\,,&
\qqq
and so it inverts the (unique) mapping $\,(f_2,\widehat f{}_1):\cM_2\too\cM_1{}_{f_1}\hspace{-1pt}\x_{f_2}\hspace{-1pt}\cM_2\,$ into the terminal object $\,\cM_1{}_{f_1}\hspace{-1pt}\x_{f_2}\hspace{-1pt}\cM_2\,$ considered previously.\ Hence,\ for any triple $\,(\cS,\si_1,\si_2)\,$ as above,\ we see that the mapping $\,F_\cS:=\mu\circ(\si_1,\si_2)\,$ is -- in virtue of the invertibility of $\,\widehat f{}_1\,$ -- the \emph{unique} one with the properties
\qq\nn
\widehat f{}_1\circ F_\cS=\si_2\,,\qquad\qquad f_2\circ F_\cS=f_1^{-1}\circ f_2\circ\si_2=\si_1\,.
\qqq
\eroof

Given a supermanifold $\,\cM\,$ and two surjective submersions $\,\pi_{\sfY_A\cM}\ :\ \sfY_A\cM\too\cM,\ A\in\{1,2\}\,$ with $\,\cM\,$ as the common base,\ we shall write the \emph{standard} pullback of one of them along the other one as
\qq\nn
\sfY_1\cM\x_\cM\hspace{-1pt}\sfY_2\cM\equiv\sfY_1\cM{}_{\pi_{\sfY_1\cM}}\hspace{-1pt}\x_{\pi_{\sfY_2\cM}}\hspace{-1pt}\sfY_2\cM\,.
\qqq
When fibring a given surjective submersion with itself,\ we shall use the by now standard notation 
\qq\nn
\sfY^{[n]}\cM\equiv\underbrace{\sfY\cM\x_\cM\sfY\cM\x_\cM\cdots\x_\cM\sfY\cM}_{n\ {\rm times}}\,.
\qqq

\section{Rudiments of the Lie-superalgebra cohomology}\label{app:LieAlgCohom}

In this appendix,\ we collect basic facts concerning the Lie-superalgebra cohomology which prove useful in an algebraic description of supertargets and of their differential geometry.\ In our exposition and discussion,\ we adopt the conventions of the original articles: \cite{Berezin:1970} by Berezin and Ka\v c,\ and \cite{Leites:1975} by Le{\"i}tes. 

We begin with the basic
\bedef\label{def:LSA}
A \textbf{Lie superalgebra} (LSA) over field $\,\bK\,$ is a pair $\,\left(\ggt,[\cdot,\cdot\}_\ggt\right)\,$ composed of a $\bK$-linear space $\,\ggt\,$ endowed with a $\bZ/2\bZ$-grading $\,|\cdot|_\ggt\,$ which induces a decomposition $\,\ggt=\ggt^{(0)}\oplus\ggt^{(1)}\,$ into a direct sum of homogeneous components,\ $\,|\cdot|_\ggt\rstr_{\ggt^{(n)}}=n$,\ and of a \textbf{Lie superbracket} $\,[\cdot,\cdot]_\ggt:\ggt\x\ggt\too\ggt$,\ which preserves the grading,\ $\,|[X_1,X_2]_\ggt|_\ggt\equiv|X_1|_\ggt+|X_2|_\ggt\mod 2\,$ and has the symmetry property $\,[X_1,X_2]_\ggt=-(-1)^{|X_1|_\ggt\cdot|X_2|_\ggt}[X_2,X_1]_\ggt\,$ (both written for arbitrary homogeneous $\,X_1,X_2\in\ggt$).\ The bracket has a vanishing \textbf{super-Jacobiator} (evaluated on arbitrary homogeneous elements $\,X_1,X_2,X_3\in\ggt$)
\qq
{\rm sJac}_\ggt(X_1,X_2,X_3)&:=&(-1)^{|X_1|_\ggt\cdot|X_3|_\ggt}\,[[X_1,X_2]_\ggt,X_3]_\ggt+(-1)^{|X_3|_\ggt\cdot|X_2|_\ggt}\,[[X_3,X_1]_\ggt,X_2]_\ggt\cr\cr
&+&(-1)^{|X_2|_\ggt\cdot|X_1|_\ggt}\,[[X_2,X_3]_\ggt,X_1]_\ggt=0\,.\label{eq:sJac}
\qqq
Given two LSAs $\,\left(\ggt_A,[\cdot,\cdot]_{\ggt_A}\right),\ A\in\{1,2\}$,\ an \textbf{LSA morphism} between them is a $\bK$-linear map $\,\chi\ :\ \ggt_1\too\ggt_2\,$ which preserves the $\bZ/2\bZ$-grading, $\,|\cdot|_{\ggt_2}\circ\chi=|\cdot|_{\ggt_1}$,\ and the Lie superbracket,\ $\,\chi\circ[\cdot,\cdot]_{\ggt_1}=[\cdot,\cdot]_{\ggt_2}\circ(\chi\x\chi)$.
\exdef 
\noindent Next,\ we have
\bedef
A (\textbf{left}) \textbf{$\ggt$-module} is a pair $\,(V,\ell_\cdot)\,$ composed of a $\bK$-linear superspace with a decomposition $\,V=V^{(0)}\oplus V^{(1)}\,$ into homogeneous components induced by the $\bZ/2\bZ$-grading $\,|\cdot|_V$,\ and endowed with a left $\ggt$-action $\,\ell_\cdot\ :\ \ggt\x V\too V\ :\ (X,v)\longmapsto X\lact v\,$ consistent with the $\bZ/2\bZ$-gradings,\ $\,|X\lact v|_V\equiv|X|_\ggt+|v|_V\mod 2$,\ and such that for any two homogeneous elements $\,X_1,X_2\in\ggt\,$ and $\,v\in V$, 
\qq\nn
[X_1,X_2\}_\ggt\lact v=X_1\lact(X_2\lact v)-(-1)^{|X_1|_\ggt\cdot|X_2|_\ggt}\,X_2\lact(X_1\lact v)\,.
\qqq
\exdef
\noindent The object of our main interest is introduced in
\bedef
Let $\,\left(\ggt,[\cdot,\cdot]_\ggt\right)\,$ and $\,\left(\agt,[\cdot,\cdot]_\agt\right)\,$ be two LSAs over a base field $\,\bK$.\ A \textbf{supercentral extension of $\,\ggt\,$ by $\,\agt\,$} is an LSA $\,\left(\widetilde\ggt,[\cdot,\cdot]_{\widetilde\ggt}\right)\,$ over $\,\bK\,$ described by the short exact sequence of LSAs
\qq\label{eq:LSASES}
\brd0\too\agt\xrightarrow{\ \jmath_\agt\ }\widetilde\ggt\xrightarrow{\ \pi_\ggt\ }\ggt\too\brd0\,,
\qqq 
written in terms of an LSA monomorphism $\,\jmath_\agt\,$ and of an LSA epimorphism $\,\pi_\ggt$,\ and such that $\,\jmath_\agt(\agt)\subset\zgt(\widetilde\ggt)\,$ (the (super)centre of $\,\widetilde\ggt$). Hence, in particular, $\,\agt\,$ is necessarily supercommutative, that is $\,[\cdot,\cdot]_\agt\equiv 0$.

Whenever $\,\pi_\ggt\,$ admits a \textbf{section} that is an LSA homomorphism, \textit{i.e.}, there exists $\,\si\in\Hom_{\rm sLie}(\ggt,\widetilde\ggt)\,$ such that $\,\pi_\ggt\circ\si=\id_\ggt$,\ the central extension is said to \textbf{split}.

An equivalence of central extensions $\,\widetilde\ggt_A, A\in\{1,2\}\,$ of $\,\ggt\,$ by $\,\agt\,$ is represented by a commutative diagram 
\qq\nn
\alxydim{@C=.75cm@R=.5cm}{ & & \widetilde\ggt_1 \ar[dd]^{\cong} \ar[dr] & & \\ \brd0 \ar[r] & \agt \ar[ur] \ar[dr] & & \ggt \ar[r] & \brd0 \\ & & \widetilde\ggt_2 \ar[ur] & & }\,,
\qqq
in which the vertical arrow is an LSA isomorphism.
\exdef
\noindent In close analogy with the purely Gra\ss mann-even case, equivalence classes of central extensions of LSAs are neatly captured by the cohomology of the latter. The relevant cohomology is specified in 
\bedef\label{def:LSAcohom}
Let $\,\left(\ggt,[\cdot,\cdot\}_\ggt\right)\,$ be an LSA over field $\,\bK\,$ and let $\,(V,\ell_\cdot)\,$ be a $\ggt$-module. A \textbf{$p$-cochain on $\,\ggt\,$ with values in} $\,V\,$ (also termed a \textbf{$p$-form on $\,\ggt\,$ with values in $\,V$}) is a $p$-linear map $\,\underset{\tx{\ciut{(p)}}}{\varphi}:\ggt^{\x p}\too V\,$ that is totally super-skewsymmetric, \textit{i.e.}, for any homogeneous elements $\,X_i\in\ggt,\ i\in\ovl{1,p}\,$ and for $\,j\in\ovl{1,p-1}$,\ it satisfies
\qq\nn
\underset{\tx{\ciut{(p)}}}{\varphi}(X_1,X_2,\ldots,X_{j-1},X_{j+1},X_j,X_{j+2},\ldots,X_p)=(-1)^{|X_j|_\ggt\cdot|X_{j+1}|_\ggt+1}\,\underset{\tx{\ciut{(p)}}}{\varphi}(X_1,X_2,\ldots,X_p)\,.
\qqq
Such maps form a $\bZ/2\bZ$-graded \textbf{group of $p$-cochains on $\,\ggt\,$ with values in} $\,V$,\ $\,C^p(\ggt,V)=C^p_0(\ggt,V)\oplus C^p_1(\ggt,V)$,\ with the respective gradations $\,|\cdot|_p\,$ such that $\,\underset{\tx{\ciut{(p)}}}{\varphi}(X_1,X_2,\ldots,X_p)\in V_{\sum_{i=1}^p\,|X_i|_\ggt+|\underset{\tx{\ciut{(p)}}}{\varphi}|_p\mod 2}$.

The family of these groups indexed by $\,p\in\bN\,$ forms a semi-bounded complex 
\qq\nn
C^\bullet(\ggt,V)\ :\ C^0(\ggt,V)\xrightarrow{\ \d_\ggt^{(0)}\ }C^1(\ggt,V)\xrightarrow{\ \d_\ggt^{(1)}\ }\cdots\xrightarrow{\ \d_\ggt^{(p-1)}\ }C^p(\ggt,V)\xrightarrow{\ \d_\ggt^{(p)}\ }\cdots
\qqq 
with the coboundary operators $\,\d_\ggt^{(p)}:C^p_n(\ggt,V)\too C^{p+1}_n(\ggt,V)$,\ written for homogeneous elements $\,X,X_i\in\ggt,\ i\in\ovl{1,p+1}\,$ and $\,\underset{\tx{\ciut{(p)}}}{\varphi}\in C^p(\ggt,V)$:
\qq\nn
&\bigl(\d_\ggt^{(0)}\underset{\tx{\ciut{(0)}}}{\varphi}\bigr)(X):=(-1)^{|X|_\ggt\cdot|\underset{\tx{\ciut{(0)}}}{\varphi}|_0}\,X\lact\underset{\tx{\ciut{(0)}}}{\varphi}\,,&\cr\cr
&\bigl(\d_\ggt^{(p)}\underset{\tx{\ciut{(p)}}}{\varphi}\bigr)(X_1,X_2,\ldots,X_{p+1}):=\sum_{i=1}^{p+1}\,(-1)^{|X_i|_\ggt\,|\underset{\tx{\ciut{(p)}}}{\varphi}|_p+S(|X_i|_\ggt)}\,X_i\lact\underset{\tx{\ciut{(p)}}}{\varphi}(X_1,X_2,\underset{\widehat i}{\ldots},X_{p+1})&\cr\cr
&+\sum_{1\leq i<j\leq p+1}\,(-1)^{S(|X_i|_\ggt)+S(|X_j|_\ggt)+|X_i|_\ggt\cdot|X_j|_\ggt}\,\underset{\tx{\ciut{(p)}}}{\varphi}([X_i,X_j\},X_1,X_2,\underset{\widehat{i,j}}{\ldots},X_{p+1})\,,&
\qqq
where $\,S(|X_i|_\ggt):=|X_i|_\ggt\cdot\sum_{j=1}^{i-1}\,|X_j|_\ggt+i-1$.\ We distinguish the \textbf{group of $p$-cocycles} $\,Z^p(\ggt,V):=\ker\,\d_\ggt^{(p)}$,\ and the \textbf{group of $p$-coboundaries} $\,B^p(\ggt,V):=\im\,\d_\ggt^{(p-1)}$.\ The $\bZ/2\bZ$-graded homology groups of the complex $\,(C^\bullet(\ggt,V),\d_\ggt^{(\bullet)})\,$ are called the \textbf{cohomology groups of $\,\ggt\,$ with values in} $\,V\,$ and denoted by
\qq\nn
H^p(\ggt,V):=H^p_0(\ggt,V)\oplus H^p_1(\ggt,V)\,,\qquad H^p_n(\ggt,V):=\frac{\ker\,\d_\ggt^{(p)}\rstr_{C^p_n(\ggt,V)}}{\im\,\d_\ggt^{(p-1)}\rstr_{C^{p-1}_n(\ggt,V)}}\,.
\qqq
\exdef

The classic correspondence between elements of $\,H^2(\ggt,\agt)\,$ and equivalence classes of supercentral extensions of $\,\ggt\,$ by a trivial supercommutative $\ggt$-module LSA $\,\agt\,$ is made precise in
\bethe\label{thm:H2ext}
Let $\,\left(\ggt,[\cdot,\cdot\}_\ggt\right)\,$ be an LSA, and let $\,\agt\,$ be a supercommutative LSA. An equivalence class of supercentral extensions $\,\left(\widetilde\ggt,[\cdot,\cdot\}_{\widetilde\ggt}\right)\,$ of $\,\ggt\,$ by $\,\agt\,$ canonically determines a class in $\,H^2_0(\ggt,\agt)$,\ which vanishes iff the short exact sequence determined by the extensions splits.\ Conversely,\ a class in $\,H^2_0(\ggt,\agt)\,$ canonically induces an equivalence class of supercentral extensions $\,\left(\widetilde\ggt,[\cdot,\cdot\}_{\widetilde\ggt}\right)\,$ of $\,\ggt\,$ by $\,\agt$.\ The extensions split iff the former class vanishes.
\ethe

\noindent Let us conclude the purely algebraic part of our exposition with the following simple reinterpretation which translates beautifully into differential (super)geometry {\it via} the standard correspondence between the CE and CaE cohomologies.\ The extension $\,\widetilde\ggt\,$ of $\,\ggt\,$ by $\,\agt\,$ determined by an arbitrary even 2-cocycle $\,\Th\in Z^2_0(\ggt,\agt)\,$ can be modelled on the $\bK$-linear space $\,\agt\oplus\ggt\,$ with the canonical projection $\,\pi_\ggt:\agt\oplus\ggt\too\ggt:(A,X)\longmapsto X$,\ and with a $\bZ/2\bZ$-grading induced from that of its direct summands.\ On it,\ we have the manifestly super-skewsymmetric 2-linear map
\qq\label{eq:H2ext-expl}
[\cdot,\cdot]_\Th\ :\ \widetilde\ggt^{\x 2}\too\widetilde\ggt\ :\ \bigl((A_1,X_1),(A_2,X_2)\bigr)\longmapsto\bigl(\Th(X_1,X_2),[X_1,X_2]_\ggt\bigr)\qquad
\qqq
which is readily verified to satisfy the Jacobi identity.\ Its existence ensures the trivialisation of the pullback 2-cocycle $\,\widetilde\Th:=\pi_\ggt^*\Th:\widetilde\ggt^{\x 2}\too\agt:((A_1,X_1),(A_2,X_2))\longmapsto\Th(X_1,X_2)$,\ given by
\qq\label{eq:sLieAlg-H2triv}
\widetilde\Th=\d_{\widetilde\ggt}^{(1)}\widetilde\mu\,,\qquad\widetilde\mu:=-\pi_\agt\ :\ \widetilde\ggt\too\agt\ :\ (A,X)\longmapsto-A\,.
\qqq
In the main text,\ we use the less cumbersome notation
\qq\nn
[A_1+X_1,A_2+X_2]_\Th=[X_1,X_2]_\ggt+\Th(X_1,X_2)\,.
\qqq

\section{Conventions for $\,\Cliff(\bR^{\x d+1})\,$ and $\,{\rm sISO}(d,1|D_{d,1})$}\label{app:sISO}

The point of departure of our discussion is the $(d+1)$-dimensional Minkowski space $\,\bR^{d,1}\equiv(\bR^{\x d+1},\eta)\,$ with the translationally invariant metric $\,\eta=\eta_{ab}\,E^a\ox E^b,\ (\eta_{ab})=\diag(-1,\underbrace{1,1,\ldots,1}_{d\ \tx{times}})$,\ the latter being written in terms of the components of the left-invariant (LI) Maurer--Cartan 1-form $\,E=E^a\ox P_a\,$ on the Lie group $\,{\rm Mink}(d,1)\,$ with values in its abelian Lie algebra $\,\gt{mink}(d,1)=\bigoplus_{a=0}^d\corr{P_a}$.\ The Clifford algebra $\,\Cliff(\bR^{d,1})=\corr{\ \G_a\equiv\eta_{ab}\,\G^b\ \vert\ a\in\ovl{0,d}\ }\,$ of $\,\bR^{d,1}\,$ contains the spin group $\,{\rm Spin}(d,1)$,\ the universal cover of the connected component $\,{\rm SO}_0(d,1)\,$ of the identity of the Lorentz group $\,{\rm SO}(d,1)\equiv{\rm SO}(\bR^{d,1})$,
\qq\nn
\bd1\too\bZ/2\bZ\too{\rm Spin}(d,1)\xrightarrow{\ \pi_{\rm Spin}\ }{\rm SO}_0(d,1)\too\bd1\,,
\qqq
and we pick up a vector space $\,S_{d,1}\,$ which carries a Majorana-spinor representation $\,S\ :\ {\rm Spin}(d,1)\too{\rm End}(S_{d,1})\,$ such that the charge-conjugation matrix $\,C\,$ and the $C$-contracted generators $\,\ovl\G{}_{a_1 a_2\ldots a_k}\equiv C\G{}_{a_1 a_2\ldots a_k}:=C\G_{[a_1}\G_{a_2}\cdots\G_{a_k]}\,$ have symmetry properties\footnote{The assumptions made constrain the admissible values of $\,(d,p)$,\ placing them within the so-called `brane scan' of \cite{Achucarro:1987nc}.} 
\qq\nn
C^{\rm T}=-C\,,\qquad\qquad\ovl\G{}^{\rm T}_a=\ovl\G{}_a
\qqq
\emph{and} -- for a given $\,p\in\ovl{1,10}\,$ --
\qq\nn
\ovl\G{}^{\rm T}_{a_1 a_2\ldots a_p}=\ovl\G{}_{a_1 a_2\ldots a_p}\,,
\qqq
as well as the {\bf Fierz identities}
\qq\label{eq:Fierz}
\ovl\G{}^{a_1}_{(\a\b}\bigl(\ovl\G_{a_1 a_2\ldots a_p}\bigr)_{\g\d)}=0\,.
\qqq
In the special case $\,(d,p)=(10,2)$,\ we span the endomorphism algebra of the Majorana-spinor representation on the above elements of the basis of $\,\Cliff(\bR^{10,1})\,$ as
\qq\nn
\bC(32)&=&\corr{\ovl\G{}^a}\oplus\bigoplus_{b_1<b_2\in\ovl{0,10}}\corr{\ovl\G{}^{b_1 b_2}}\oplus\bigoplus_{c_1<c_2<c_3<c_4<c_5\in\ovl{0,10}}\corr{\ovl\G{}^{c_1 c_2 c_3 c_4 c_5}}\cr\cr
&&\oplus\corr{C}\oplus\bigoplus_{d_1<d_2<d_3\in\ovl{0,10}}\corr{\ovl\G{}^{d_1 d_2 d_3}}\oplus\bigoplus_{e_1<e_2<e_3<e_4\in\ovl{0,10}}\corr{\ovl\G{}^{e_1 e_2 e_3 e_4}}\,,
\qqq
with the first three summands giving a convenient decomposition of the subalgebra of symmetric matrices,
\qq\label{eq:C32-sym}
\bC(32)^{\rm sym}=\corr{\ovl\G{}^a}\oplus\bigoplus_{b_1<b_2\in\ovl{0,10}}\corr{\ovl\G{}^{b_1 b_2}}\oplus\bigoplus_{c_1<c_2<c_3<c_4<c_5\in\ovl{0,10}}\corr{\ovl\G{}^{c_1 c_2 c_3 c_4 c_5}}\,,\qquad\qquad
\qqq
{\it cp.}\ \cite{vanHolten:1982mx}.\ Upon denoting elements of the Clifford basis jointly as $\,\ovl\G{}^I,\ I\in\{a,[b_1,b_2],[c_1,c_2,c_3,c_4,c_5]\}\,$ and their inverses as $\,\ovl\G{}^{I\,-1}_{\a\b}\equiv\Lx_I^{\a\b}$,\ we may write out the completeness relations 
\qq\label{eq:Cliff11-compl-sym}
\ovl\G{}^I_{\a\b}\,\Lx_I^{\g\d}=32\,\d^{(\g}_\a\,\d^{\d)}_\b\,,\qquad\qquad\ovl\G{}^I_{\a\b}\,\Lx_J^{\a\b}=32\,\d^I_{\ J}\,,
\qqq
{\it cp.}\ \cite{Kamimura:2003rx}.\ Finally,\ for $\,(d,p)=(9,0)$,\ we consider the anticentral volume element $\,\G_{11}=\G^0\G^1\cdots\G^9\,$ with the properties
\qq\nn
\bigl(C\G_{11}\bigr)^{\rm T}=\ovl\G{}_{11}\equiv C\G_{11}\,,\qquad\qquad\G_{11}^2=\id_{S_{D_{d,1}}}\,.
\qqq
 
Given these,\ we consider the associated super-Poincar\'e (super)group
\qq\nn
&{\rm sISO}(d,1|D_{d,1})=\bigl(\widetilde{{\rm ISO}}(d,1)\equiv\bR^{\x d+1}\rx_L{\rm Spin}(d,1),\cO_{{\rm sISO}(d,1|D_{d,1})}\bigr)\,,&\cr\cr
&\cO_{{\rm sISO}(d,1|D_{d,1})}=C^\infty(\cdot,\bR)\circ\pr_1\ox C^\infty(\cdot,\bR)\circ\pr_2\ox\bigwedge\hspace{-3pt}{}^\bullet\,S_{d,1}\,,&
\qqq 
with the crossed product in the definition of its body $\,\widetilde{{\rm ISO}}(d,1)\,$ determined by the vector realisation 
\qq\nn
\alxydim{@C=1cm@R=.5cm}{L & :\ & {\rm Spin}(d,1) \ar[rr] \ar[dr]_{\pi_{\rm Spin}} & & \End\bigl(\bR^{d,1}\bigr) \\ & & & {\rm SO}_0(d,1) \ar@{^{(}->}[ur] & }
\qqq
of the spin group,\ and the structure sheaf $\,\cO_{{\rm sISO}(d,1|D_{d,1})}\,$ written in terms of the structure sheaf $\,C^\infty(\cdot,\bR)\circ\pr_1\ox C^\infty(\cdot,\bR)\circ\pr_2\equiv\cO_{\widetilde{{\rm ISO}}(d,1)}\,$ of the body.\ The binary operation 
\qq\nn
\widehat\txm\ :\ {\rm sISO}(d,1|D_{d,1})\x{\rm sISO}(d,1|D_{d,1})\too{\rm sISO}(d,1|D_{d,1})
\qqq
of the Lie-supergroup structure on $\,{\rm sISO}(d,1|D_{d,1})\,$ is customarily described,\ in the $\cS$-point picture,\ in terms of the anticommuting generators $\,\theta^\a,\ \a\in\ovl{1,D_{d,1}}\,$ of $\,\bigwedge\hspace{-3pt}{}^\bullet\,S_{d,1}$,\ the global (cartesian-coordinate) generators $\,x^a,\ a\in\ovl{0,d}\,$ of the structure sheaf of $\,\bR^{\x d+1}$,\ and local (Lie-algebra) coordinates $\,\phi^{ab}\equiv\phi^{[ab]}\,$ on $\,{\rm Spin}(d,1)\,$ as
\qq
&&\widehat\txm\bigl(\bigl(\theta_1^\a,x_1^a,\phi_1^{bc}\bigr),\bigl(\theta_2^\a,x_2^a,\phi_2^{bc}\bigr)\bigr)\cr
&=&\bigl(\theta_1^\a+S(\phi_1)^\a_{\ \b}\,\theta_2^\b,x_1^a+L(\phi_1)^a_{\ b}\,x_2^b-\tfrac{1}{2}\,\theta_1\,\ovl\G{}^a\,S(\phi_1)\,\theta_2,\bigl(\phi_1\star\phi_2\bigr)^{ab}\bigr)\,,\label{eq:sISOgl}
\qqq
where $\,\star\,$ represents the standard binary operation on the spin group.\ In these coordinates,\ the basis LI vector fields on $\,{\rm sISO}(d,1|D_{d,1})\,$ take the form
\qq\nn
Q_\a(\theta,x,\phi)&=&S(\phi)^\b_{\ \a}\,\bigl(\tfrac{\vec\p\ }{\p\theta^\b}+\tfrac{1}{2}\,\theta^\g\,C_{\g\d}\,\G^{a\,\d}_{\ \ \ \b}\,\tfrac{\p\ }{\p x^a}\bigr)=:S(\phi)^\b_{\ \a}\,\unl Q{}_\b(\theta,x)\,,\cr\cr 
P_a(\theta,x,\phi)&=&L(\phi)^b_{\ a}\,\tfrac{\p\ }{\p x^b}=:L(\phi)^b_{\ a}\,\unl P{}_b(\theta,x)\,,\cr\cr 
J_{ab}(\theta,x,\phi)&=&\tfrac{\sfd\ }{\sfd t}\rstr_{t=0}\phi\star t\phi_{ab}\,,\qquad\qquad\bigl(\phi_{ab}\bigr)^{cd}=\d_a^{\ c}\,\d_b^{\ d}-\d_a^{\ d}\,\d_b^{\ c}\,.
\qqq
These obey the superalgebra
\qq\nn
&\{Q_\a,Q_\b\}=\ovl\G{}_{\a\b}^a\,P_a\,,\qquad\qquad[P_a,P_b]=0\,,\qquad\qquad[Q_\a,P_a]=0\,,&\cr\cr
&[J_{ab},Q_\a]=\tfrac{1}{2}\,\bigl(Q\,\G_{ab}\bigr){}_\a=\tfrac{1}{2}\,\G_{ab}{}^\b_{\ \a}\,Q_\b\,,\qquad\qquad[J_{ab},P_c]=\eta_{bc}\,P_a-\eta_{ac}\,P_b\,,&\cr\cr
&[J_{ab},J_{cd}]=\eta_{ad}\,J_{bc}-\eta_{ac}\,J_{bd}+\eta_{bc}\,J_{ad}-\eta_{bd}\,J_{ac}\,,&
\qqq
called the {\bf super-Poincar\'e} ({\bf super}){\bf algebra} and decomposing \emph{reductively} as
\qq\nn
\gt{siso}(d,1|D_{d,1})&=&\bigoplus_{\a=1}^{D_{d,1}}\,\corr{Q_\a}\oplus\bigoplus_{a=0}^d\,\corr{P_a}\oplus\bigoplus_{a<b=0}^d\,\corr{J_{ab}=-J_{ba}}\cr\cr
&=&\gt{smink}(d,1|D_{d,1})\oplus\gt{spin}(d,1)
\qqq
into $\,\gt{spin}(d,1)={\rm Lie}({\rm Spin}(d,1))\,$ and the {\bf super-minkowskian Lie superalgebra} 
\qq\nn
\gt{smink}(d,1|D_{d,1})=\bigoplus_{\a=1}^{D_{d,1}}\,\corr{Q_\a}\oplus\bigoplus_{a=0}^d\,\corr{P_a}\,.
\qqq
The cotangent sheaf $\,\cT^*{\rm sISO}(d,1|D_{d,1})\,$ of $\,{\rm sISO}(d,1|D_{d,1})\,$ is globally generated by the LI 1-forms dual to the $\,Q_\a,\ P_a\,$ and $\,J_{bc}$,\ with the coordinate presentation
\qq\nn
\theta_{\rm L}^\a(\theta,x,\phi)&=&S(\phi)^{-1\,\a}_{\ \ \ \ \b}\,\sfd\theta^\b=:S(\phi)^{-1\,\a}_{\ \ \ \ \b}\,\si^\b(\theta,x)\,,\cr\cr
\theta_{\rm L}^a(\theta,x,\phi)&=&L(\phi)^{-1\,a}_{\ \ \ \ b}\,\bigl(\sfd x^b+\tfrac{1}{2}\,\theta\,\ovl\G{}^b\,\sfd\theta\bigr)=:L(\phi)^{-1\,a}_{\ \ \ \ b}\,e^b(\theta,x)\,,\cr\cr
\theta_{\rm L}^{ab}(\theta,x,\phi)&=&L(\phi)^{-1\,a}_{\ \ \ \ c}\,\sfd L(\phi)^c_{\ d}\,\eta^{-1\,db}\,.
\qqq
Through the ensuing (super-)Maurer--Cartan equations
\qq
&\sfd q^\a=-\tfrac{1}{4}\,j^{ab}\wedge\bigl(\G_{ab}\,q\bigr)^\a\,,\qquad\qquad\sfd p^a=\tfrac{1}{2}\,q\wedge\ovl\G{}^a\,q-\eta_{bc}\,j^{ab}\wedge p^c\,,&\cr\cr
&\sfd j^{ab}=-\eta_{cd}\,j^{ac}\wedge j^{bd}\,,&\label{eq:sMCeqs}
\qqq
they generate the Cartan--Eilenberg cochain complex of $\,{\rm sISO}(d,1|D_{d,1})$.\ Its cohomology,\ 
\qq\nn
H_{\rm dR}^\bullet\bigl({\rm sISO}(d,1|D_{d,1}),\bR\bigr)^{{\rm sISO}(d,1|D_{d,1})}\equiv{\rm CaE}^\bullet\bigl({\rm sISO}(d,1|D_{d,1})\bigr)\,,
\qqq
the supersymmetric refinement of the de Rham cohomology of $\,{\rm sISO}(d,1|D_{d,1})$,\ is termed the Cartan--Eilenberg cohomology of $\,{\rm sISO}(d,1|D_{d,1})$.\ By the $\bZ/2\bZ$-graded version of the classic Lie-algebraic result,\ it is isomorphic with the Chevalley--Eilenberg cohomology of the Lie superalgebra $\,\gt{siso}(d,1|D_{d,1})\,$ with values in the trivial $\gt{siso}(d,1|D_{d,1})$-module $\,\bR$,
\qq\nn
{\rm CaE}^\bullet\bigl({\rm sISO}(d,1|D_{d,1})\bigr)\cong H^\bullet\bigl(\gt{siso}(d,1|D_{d,1}),\bR\bigr)\,.
\qqq
Among nontrivial classes in $\,{\rm CaE}^\bullet({\rm sISO}(d,1|D_{d,1}))$,\ we find that of the GS $(p+2)$-cocycle
\qq\label{eq:GS_p_coc}
\widehat\chi{}_{(p+2)}=\theta_{\rm L}^\a\wedge\bigl(\ovl\G{}_{a_1 a_2\ldots a_p}\bigr)_{\a\b}\,\theta_{\rm L}^\b\wedge\theta_{\rm L}^{a_1}\wedge\theta_{\rm L}^{a_2}\wedge\cdots\wedge\theta_{\rm L}^{a_p}
\qqq
for $\,p>0$,\ and that of the GS 2-cocycle
\qq\label{eq:GS_2_coc}
\widehat\chi{}_{(2)}=\theta_{\rm L}^\a\wedge\bigl(\ovl\G{}_{11}\bigr)_{\a\b}\,\theta_{\rm L}^\b
\qqq
for $\,(d,p)=(10,0)$.\ The closedness of the former follows directly from \eqref{eq:Fierz}.\ Besides them,\ there exists on $\,{\rm sISO}(d,1|D_{d,1})\,$ a manifestly LI lift of the minkowskian metric
\qq\nn
\widehat\txg=\eta_{ab}\,\theta^a_{\rm L}\ox\theta^b_{\rm L}\,.
\qqq

The GS $(p+2)$-cocycles are $\gt{spin}(d,1)$-horizontal and ${\rm Spin}(d,1)$-invariant,\ and hence ${\rm Spin}(d,1)$-basic.\ Therefore,\ they descend to the super-Minkowski space ($e^{a_1 a_2\ldots a_p}\equiv e^{a_1}\wedge e^{a_2}\wedge\cdots\wedge e^{a_p}$)
\qq\nn
\widehat\chi{}_{(p+2)}=\pi^*\chi_{(p+2)}\,,\qquad\qquad\chi_{(p+2)}=\left\{ \barr{cl} \si^\a\wedge\bigl(\ovl\G{}_{a_1 a_2\ldots a_p}\bigr)_{\a\b}\,\si^\b\wedge e^{a_1 a_2\ldots a_p} & \ \tx{if}\ p\in\ovl{1,10}\\ \si^\a\wedge\bigl(\ovl\G{}_{11}\bigr)_{\a\b}\,\si^\b & \ \tx{if}\ p=0\earr \right.\,,
\qqq
engendering the nontrivial (GS) $(p+2)$-cocycles in the Cartan--Eilenberg cohomology of the {\bf super-minkowskian Lie supergroup}
\qq\nn
{\rm sMink}(d,1|D_{d,1})=\bigl({\rm Mink}(d,1)\equiv\bR^{\x d+1},\cO_{{\rm sMink}(d,1|D_{d,1})}=C^\infty(\cdot,\bR)\ox\bigwedge\hspace{-3pt}{}^\bullet\,S_{d,1}\bigr)\,,
\qqq 
with the binary operation induced from $\,\widehat\txm\,$ as
\qq\nn
\txm\bigl(\bigl(\theta_1^\a,x_1^a\bigr),\bigl(\theta_2^\a,x_2^a\bigr)\bigr)=\pi\circ\widehat\txm\bigl(\bigl(\theta_1^\a,x_1^a,0\bigr),\bigl(\theta_2^\a,x_2^a,0\bigr)\bigr)=\bigl(\theta_1^\a+\theta_2^\a,x_1^a+x_2^a-\tfrac{1}{2}\,\theta_1\,\ovl\G{}^a\,\theta_2\bigr)\,,
\qqq
where $\,\pi\ :\ {\rm sISO}(d,1|D_{d,1})\too{\rm sMink}(d,1|D_{d,1})\,$ is the quotient map.\ For the same reason,\ the pseudo-metric $\,\widehat\txg\,$ descends as
\qq\nn
\widehat\txg=\pi^*\txg\,,\qquad\qquad\txg=\eta_{ab}\,e^a\ox e^b\,.
\qqq

\section{A proof of Prop.\,\ref{prop:surjsubm-membr}}\label{app:surjsubm-membr}

The first extension forms part of the full one discussed in Def.\,\ref{def:full-ext-spoint11} and yields the partially trivialised super-4-cocycle
\qq\nn
\pi_{\txT^1_{(M)}}^*\chi_{(4)}&=&\tfrac{2}{3}\,\bigl(\ovl\G{}^a_{\a\b}\,e^1_{ab}+\ovl\G{}_{ab\,\a\b}\,\pi_{\txT^1_{(M)}}^*e^a\bigr)\wedge\pi_{\txT^1_{(M)}}^*\bigl(e^b\wedge\si^\a\wedge\si^\b\bigr)\cr\cr
&&+\sfd\bigl(\tfrac{2}{3}\,e^1_{ab}\wedge\pi_{\txT^1_{(M)}}^*\bigl(e^a\wedge e^b\bigr)\bigr)\,.
\qqq
In the case of the second one,\ the closedness of the $\,h^2_{a\a}\,$ is ensured by the Fierz identity \eqref{eq:Fierz} for $\,p=2$,\ which also yields the primitives $\,b^2_{a\a}$.\ A supersymmetry variation of the latter quickly yields the expression
\qq\nn
&&b^2_{a\a}\bigl(\txm_2^{(1)}\bigl((\vep,y,\phi),(\theta,x,\varphi)\bigr)\bigr)-b^2_{a\a}(\theta,x,\varphi)\cr\cr
&=&\sfd\bigl[\bigl(\ovl\G{}^b_{\a\b}\,\varphi_{ab}+x^b\,\ovl\G_{ab\,\a\b}-\tfrac{1}{3}\,\bigl(\ovl\G{}^b_{\a\b}\,\vep\,\ovl\G_{ab}\,\theta+\ovl\G_{ab\,\a\b}\,\vep\,\ovl\G{}^b\,\theta\bigr)\bigr)\,\vep^\b\bigr]+\eta^{(\vep)}_{a\a}(\theta)
\qqq
in which the last term 
\qq\nn
\eta^{(\vep)}_{a\a}(\theta)=\tfrac{1}{6}\,\bigl(\ovl\G{}^b_{\a\b}\,\ovl\G_{ab\,\g\d}+\ovl\G{}^b_{\g\d}\,\ovl\G_{ab\,\a\b}+2\ovl\G{}^b_{\a\g}\,\ovl\G_{ab\,\b\d}+2\,\ovl\G{}^b_{\b\d}\,\ovl\G_{ab\,\a\g}\bigr)\,\vep^\b\,\theta^\g\,\sfd\theta^\d
\qqq
is closed (by construction),\ and hence exact.\ Its primitive is established through application of the standard homotopy formula,\ which gives us
\qq\nn
\eta^{(\vep)}_{a\a}=\sfd F^{(\vep)}_{a\a}\,,\qquad\qquad F^{(\vep)}_{a\a}(\theta)=\tfrac{1}{6}\,\bigl(\ovl\G{}^b_{\a\g}\,\ovl\G_{ab\,\b\d}+\ovl\G{}^b_{\b\d}\,\ovl\G_{ab\,\a\g}\bigr)\,\vep^\b\,\theta^\g\,\theta^\d\,.
\qqq
From this,\ we read off the complete quasi-invariance 1-cochain(s)
\qq\nn
\D_{(\theta_1,x_1,\varphi^1)\,a\a}\bigl(\theta_2,x_2,\varphi^2\bigr)&=&-\bigl(\ovl\G{}^b_{\a\b}\,\varphi^2_{ab}+x_2^b\,\ovl\G_{ab\,\a\b}-\tfrac{1}{3}\,\bigl(\ovl\G{}^b_{\a\b}\,\theta_1\,\ovl\G_{ab}\,\theta_2+\ovl\G_{ab\,\a\b}\,\theta_1\,\ovl\G{}^b\,\theta_2\bigr)\bigr)\,\theta_1^\b\cr\cr
&&-\tfrac{1}{6}\,\bigl(\ovl\G{}^b_{\a\g}\,\ovl\G_{ab\,\b\d}+\ovl\G{}^b_{\b\d}\,\ovl\G_{ab\,\a\g}\bigr)\,\theta_1^\b\,\theta_2^\g\,\theta_2^\d\,,
\qqq
and thus infer associativity of $\,\txm_2^{(2)}\,$ upon invoking Prop.\,\ref{prop:assoconstr}.\ We attain further trivialisation
\qq\nn
\pi_{\txT^2_{(M)}}^*\pi_{\txT^1_{(M)}}^*\chi_{(4)}&=&-\tfrac{2}{15}\,h^3_{\a\b}\wedge\pi_{\txT^2_{(M)}}^*\pi_{\txT^1_{(M)}}^*\bigl(\si^\a\wedge\si^\b\bigr)\cr\cr
&&+\sfd\bigl[\pi_{\txT^2_{(M)}}^*\bigl(\tfrac{2}{3}\,e^1_{ab}\wedge\pi_{\txT^1_{(M)}}^*\bigl(e^a\wedge e^b\bigr)\bigr)-\tfrac{3}{5}\,\si^2_{a\a}\wedge\pi_{\txT^2_{(M)}}^*\pi_{\txT^1_{(M)}}^*\bigl(e^a\wedge\si^\a\bigr)\bigr]\,,
\qqq
with the $\,h^3_{\a\b}\,$ as in the statement of the proposition.

In the third extension,\ simple manipulations,\ involving the Fierz identity \eqref{eq:Fierz} for $\,p=2$,\ lead to the expression for the (partially trivialised) $\,h^3_{\a\b}$:
\qq\nn
h^3_{\a\b}(\theta,x,\varphi,\psi)&=&\sfd\bigl[-(\tfrac{1}{4}\,\ovl\G{}^a_{\a\b}\,\si^2_{a\g}+\ovl\G{}^a_{\a\g}\,\si^2_{a\b}+\ovl\G{}^a_{\b\g}\,\si^2_{a\a})(\theta,x,\varphi,\psi)\,\theta^\g+\tfrac{1}{2}\,\ovl\G{}^a_{\a\b}\,e^1_{ab}(\theta,x,\varphi)\,x^b\cr\cr
&&-(\ovl\G{}^a_{\a\d}\,\ovl\G{}^b_{\b\g}\,e^1_{ab}(\theta,x,\varphi)+\tfrac{1}{2}\,(\ovl\G{}^a_{\a\d}\,\ovl\G{}_{ab\,\b\g}+\ovl\G{}^a_{\b\d}\,\ovl\G{}_{ab\,\a\g})\,e^b(\theta,x))\,\theta^\d\,\theta^\g\cr\cr
&&-\tfrac{1}{2}\,\ovl\G{}_{ab\,\a\b}\,x^a\,e^b(\theta,x)-\tfrac{1}{4}\,(\ovl\G{}^a_{\a\b}\,\ovl\G{}_{ab\,\g\d}+\ovl\G{}^a_{\g\d}\,\ovl\G{}_{ab\,\a\b})\,x^b\,\theta^\g\,\si^\d(\theta,x)\bigr]+\eta_{\a\b}(\theta)\,,
\qqq
with the closed (by construction) and hence exact remainder
\qq\nn
8\eta_{\a\b}(\theta)&=&2\bigl(2\ovl\G{}^a_{\a\d}\,\ovl\G{}^b_{\b\g}\,\ovl\G_{ab\,\ep\z}+\bigl(\ovl\G{}^a_{\a\d}\,\ovl\G_{ab\,\b\g}+\ovl\G{}^a_{\b\d}\,\ovl\G_{ab\,\a\g}\bigr)\,\ovl\G{}^b_{\ep\z}\bigr)\,\theta^\d\,\theta^\g\,\sfd\theta^\ep\wedge\sfd\theta^\z\cr\cr
&&+\bigl(\ovl\G{}^a_{\a\b}\,\theta\,\ovl\G_{ab}\,\sfd\theta+\ovl\G_{ab\,\a\b}\,\theta\,\ovl\G{}^a\,\sfd\theta\bigr)\wedge\theta\,\ovl\G{}^b\,\sfd\theta
\qqq
whose further trivialisation with the help of the homotopy formula brings the result stated in the proposition.\ Analogous calculation for the supersymmetry variation of the thus obtained primitives $\,b^3_{\a\b}\,$ of the $\,h^3_{\a\b}\,$ yields the binary operation with the quasi-invariance 1-cochain(s)
\qq\nn
&&\D_{(\theta_1,x_1,\varphi_1,\psi_1)\,\a\b}(\theta_2,x_2,\varphi_2,\psi_2)\cr\cr
&=&-\tfrac{1}{4}\,x_1^a\,\bigl(\ovl\G_{ab\,\a\b}\,\bigl(2x_2^b-\theta_1\,\ovl\G{}^b\,\theta_2\bigr)+\ovl\G{}^b_{\a\b}\,\bigl(2\varphi^2_{ab}-\theta_1\,\ovl\G_{ab}\,\theta_2\bigr)\bigr)-\tfrac{1}{4}\,x_2^b\,\bigl(\ovl\G{}^a_{\a\b}\,\ovl\G_{ab\,\g\d}+\ovl\G{}^a_{\g\d}\,\ovl\G_{ab\,\a\b}\bigr)\,\theta_1^\g\,\theta_2^\d\cr\cr
&&-\bigl(\varphi^2_{ab}\,\ovl\G{}^a_{\a\g}\,\ovl\G{}^b_{\b\d}+\tfrac{1}{2}\,x_2^b\,\bigl(\ovl\G{}^a_{\a\g}\,\ovl\G_{ab\,\b\d}+\ovl\G{}^a_{\b\g}\,\ovl\G_{ab\,\a\d}\bigr)\bigr)\,\theta_1^\g\,\theta_1^\d-\bigl(\tfrac{1}{4}\,\ovl\G{}^a_{\a\b}\,\psi^2_{a\g}+\ovl\G{}^a_{\a\g}\,\psi^2_{a\b}+\ovl\G{}^a_{\b\g}\,\psi^2_{a\a}\bigr)\,\theta_1^\g\cr\cr
&&+\tfrac{1}{24}\,\theta_2^\g\,\bigl(\theta_2^\d\,\bigl(2\D_{\a\b;\g\d\ep\eta}\,\theta_2^\ep+3\bigl(\D_{\a\b;\ep\g\d\eta}-\D_{\a\b;\g\ep\d\eta}\bigr)\,\theta_1^\ep\bigr)+6\D_{\a\b;\ep\d\g\eta}\,\theta_1^\ep\,\theta_1^\d\bigr)\,\theta_1^\eta\cr\cr
&&+\tfrac{1}{16}\,\ovl\G{}^a_{\a\b}\,\theta_1\,\ovl\G{}^b\,\theta_2\cdot\theta_1\,\ovl\G{}_{ab}\,\theta_2\,,
\qqq
which satisfy $\,\D_{(\theta_1,x_1,\varphi_1,\psi_1)\,\a\b}(0,0,0,0)=0$,\ whence associativity of the operation follows by Prop.\,\ref{prop:assoconstr}.

Verification of the identity $\,\pi_{\sfY_2\txT_{(M)}}^*\chi_{(4)}=\sfd\widetilde\b{}_{(4)}\,$ is now a matter of a simple calculation using the previously established structural (Maurer--Cartan) equations and the Fierz identity.

\resumetocwriting

\end{document}